\documentclass[11pt,a4paper]{article}
\usepackage{jheppub}
\usepackage{graphicx} 
\usepackage{appendix}
\usepackage{slashed}
\usepackage{braket}
\usepackage[normalem]{ulem}

\usepackage{xcolor}

\definecolor{acolor}{rgb}{0.1,.6,0}

\definecolor{bcolor}{rgb}{0.0,0.6,0.6}

\newcommand*{\hbb}{\mathbb{H}}
\newcommand*{\ibb}{\mathbb{I}}

\newcommand*{\acal}{\mathcal{A}}
\newcommand*{\bcal}{\mathcal{B}}
\newcommand*{\tcal}{\mathcal{T}}
\newcommand*{\dcal}{\mathcal{D}}
\newcommand*{\ical}{\mathcal{I}}
\newcommand*{\Tr}{\mathrm{Tr}}
\newcommand*{\bnd}{{\partial\gamma}}
\newcommand*{\avg}[1]{\langle #1 \rangle}
\newcommand*{\sw}{\mathcal{S}}
\newcommand{\ind}[1]{\vec{\textbf{#1}}}
\newcommand*{\core}{\Pi_{\text{core}}}
\newcommand*{\coloneqq}{\mathrel{\vcenter{\baselineskip0.5ex \lineskiplimit0pt \hbox{\scriptsize.}\hbox{\scriptsize.}}} =}
\newcommand*{\revcoloneqq}{=\mathrel{\vcenter{\baselineskip0.5ex \lineskiplimit0pt \hbox{\scriptsize.}\hbox{\scriptsize.}}} }

	\title{\Large Holographic properties of superposed quantum geometries}
	\author[a,b,c]{Eugenia Colafranceschi}
	\emailAdd{ecolafra@uwo.ca}
	\affiliation[a]{School of Mathematical Sciences, University of Nottingham, University Park Campus, Nottingham NG7 2RD, United Kingdom}
 \affiliation[b]{Department of Physics, University of California, Santa Barbara, CA 93106, USA}
 \affiliation[c]{Department of Physics and Astronomy, Western University, N6A 3K7, London ON, Canada}
	
	\author[d,e]{Simon Langenscheidt}
	\emailAdd{s.langenscheidt@physik.lmu.de}
	\affiliation[d]{Arnold Sommerfeld Center for Theoretical Physics, Ludwig-Maximilians-Universit\"at München, Theresienstrasse 37, 80333 M\"unchen, Germany}
        \affiliation[e]{Munich Center for Quantum Science and Technology (MCQST), Schellingstrasse 4, 80799 M\"unchen, Germany}
	
	\author[d,e]{Daniele Oriti}
	\emailAdd{daniele.oriti@physik.lmu.de}

\date{August 2023}

\abstract{
We study the holographic properties of a class of quantum geometry states characterized by a superposition of discrete geometric data, in the form of generalised tensor networks. This class specifically includes spin networks, the kinematic states of lattice gauge theory and discrete quantum gravity. We employ an algebraic, operatorial definition of holography based on quantum information channels, an approach which is particularly valuable in settings, such as the one we consider, where the relevant Hilbert space of states does not factorize into subsystem Hilbert spaces due to gauge invariance. We apply random tensor network techniques (successfully used in the AdS/CFT context) to analyse information transport properties of the bulk-to-boundary and boundary-to-boundary maps associated with this superposition of quantum geometries, and produce typicality results about the average over the geometric data colouring the fixed graph structure. In this context, one naturally obtains a nontrivial area operator encoding the dominant contribution to entropy calculations. Among our main results is the requirement that one can only isometrically map a bulk region onto boundaries with fixed total area. We furthermore inquire about similar state-induced mappings between segments of the boundary, and discuss related conditions for isometric behaviour. These generalisations make further steps towards quantum gravity implementations of tensor network holography.  
}

\begin{document}

\maketitle

\section{Introduction}
	Since the discovery of the Bekenstein-Hawking entropy formula for black holes~\cite{PhysRevD.7.2333,Hawking:1975vcx}, holography has taken center stage in the exploration of the overlap of quantum, gravitational and thermodynamic properties of spacetime. Indeed, according to the conventional microstate counting interpretation of entropy, the Bekenstein-Hawking formula signifies that the number of degrees of freedom associated with a black hole does not scale with its \textit{volume} (as for most systems), but instead scales with the \textit{area} of the surface bounding that volume. This implies a form of (informational) \textit{holography}: information on the degrees of freedom of the system is encoded on the boundary of the region of space  it occupies, from where it may be recovered. Although discovered in the context of semiclassical gravity, it is generally believed that holography calls for a quantum gravity explanation.
    In fact, this holographic behaviour is found for entanglement entropy in a number of quantum many-body systems and often (for local Hamiltonians) it characterizes ground states, distinguishing them from the vast majority of their quantum states~\cite{eisertAreaLawsEntanglement2010}. The suggestion, therefore, is that it may also provide a similar characterizing role for quantum gravitational systems.
	
	Following the discovery of the Bekenstein-Hawking area law for black hole entropy, other forms of holography 
	have been related to gravity and used to gain access to its quantum properties. The most explored example is of course the AdS/CFT correspondence~\cite{witten1998anti,maldacenaLargeLimitSuperconformal1999}, and its generalizations~\cite{Araujo_Regado_2023}.
	What all of these more recent examples have in common, and different from the original black hole case, is that they describe a relation between a \textit{bulk} gravitational theory and degrees of freedom on an \textit{asymptotic boundary}, apparently describing the same physics. However, a different type of holography for finite regions of space/spacetime is suggested to exist in a variety of contexts, besides the original black hole one. This finite-distance holographic behaviour is signaled too by entropy bounds. For example, recent work in classical gravity suggests that corner charges of general relativity provide an encoding of bulk information~\cite{donnellyGravitationalEdgeModes2021,freidelEdgeModesGravity2020,freidelExtendedCornerSymmetry2021}, which applies to any finite region of space with boundary. In the already mentioned condensed matter context, holographic properties refer to finite regions; indeed, the ground states of local Hamiltonians on lattice systems are often found to be \textit{short-range entangled}~\cite{ciracRenormalizationTensorProduct2009b}, so that the entropy of any finite region scales at most with the size of its boundary. States with such properties are highly desirable. They not only provide a more manageable subset of the state space for initiating searches for ground states, but also exhibit characteristics such as exponentially decaying correlations between regions, which mimic a local lightcone structure through Lieb-Robinson bounds~\cite{nachtergaeleLiebRobinsonBoundsQuantum2010,bravyiLiebRobinsonBoundsGeneration2006}.
	
	It is this type of local holographic behavior, in a quantum gravity context, that we are seeking in the following, identifying classes of quantum geometries that possess it. We focus on \textit{spin network states}, which define a complete basis of spatial quantum geometry. These states were originally envisioned by Penrose~\cite{penroseAngularMomentumApproach1971,penroseApplicationsNegativeDimensional1971} and later recovered to provide such a basis in the Loop Quantum Gravity (LQG) canonical quantisation programme~\cite{rovelliSpinNetworksQuantum1995,chircoThermallyCorrelatedStates2015}, in the \textit{spin foam }approach to (discrete) quantum gravity path integrals~\cite{perezSpinFoamApproach2013a}, as well as in Group Field Theories (GFT)~\cite{krajewskiGroupFieldTheories2012,oritiGroupFieldTheory2014,oritiMicroscopicDynamicsQuantum2011}.\\
	The goal of this work is to elaborate on criteria for holographic mappings between patches of a finite spatial region to exist. This is based on the earlier work~\cite{chircozhang,chircoRyuTakayanagiFormulaSymmetric2018,chircozhang2,colafranceschiQuantumGravityStates2021,colafranceschiHolographicMapsQuantum2021,chircoBulkAreaLaw2022} where spin network states were seen as tensor networks, and analysed via random tensor network techniques, which have been well-explored in the quantum information literature in connection to holography (and AdS/CFT correspondence)~\cite{yangBidirectionalHolographicCodes2016,haydenHolographicDualityRandom2016,pastawskiHolographicQuantumErrorcorrecting2015}. These works established the Ryu-Takayanagi entropy formula~\cite{ryuHolographicDerivationEntanglement2006,rangamaniHolographicEntanglementEntropy2017} for spin network states, as well as isometry conditions for bulk-to-boundary maps defined by the latter, with an important restriction: both combinatorial structures (the graph underlying the spin network states) and the algebraic data labelling them (the eigenvalues of quantum geometric operators) were held fixed. 
    We generalise those results by removing one of these substantial restrictions: the bulk region is modeled by a superposition of spin network states where we keep the graph structure fixed, but allow for general superpositions otherwise, i.e. we sum over the algebraic quantum numbers. This work thus represents a pivotal first step in studying local holography in its generality, within multiple quantum gravity approaches, and with possible application also to semiclassical states where a large number of different quantum geometries (spin network states) are superposed, and complements similar recent work generalising random tensor network holography~\cite{chengRandomTensorNetworks2022a,Dong:2023kyr,Akers:2024ixq}.\\
	The article is structured as follows. In section \ref{methodology} we detail the class of states under consideration, along with the criteria for holographic behavior. Within this section, we also introduce the methodology for assessing holography: starting from a state of quantum geometry, we express the purity of a reduced boundary state - our "holography measure" - through a random Ising model defined on the graph underlying the quantum geometry. In section \ref{conditionchapter} we present our results: after analysing the properties of the aforementioned Ising model, we derive a precise criterion for the superposition of spin network states to feature isometry of the induced map between network subregions. We conclude by discussing possible extensions of our work in section \ref{conclusion}. Some technical calculations are reported in the appendix. \\ 


\section{Methodology}
\label{methodology}
We start by explaining the issues we address and outlining the main features of our calculation. Several of these steps are analogous to previous results in the case of spin network states with fixed spins~\cite{colafranceschiHolographicEntanglementSpin2022} (referred to as single-sector case), here generalised to include superposition of such states. After giving examples for the types of quantum systems we would like to consider, we specify a class of tensor network states via their concrete construction. We then introduce an algebraic notion of holography, suitable for our context where the Hilbert space does not factorize over bulk and boundary regions. This notion of holography traces back to the bulk-to-boundary map defined by our states being an isometry, along the line of previous works on holographic tensor networks (see e.g.~\cite{haydenHolographicDualityRandom2016}), although the algebraic formulation represents a novel aspect of our analysis. We enquire as well about the isometric character of the maps that our states define between boundary subregions. As our work is statistical in nature, i.e. we ask questions about properties that may hold only `on average' with respect to the geometric data of our states, we then specify a choice of randomisation. Lastly, by adapting well-known techniques for random tensor networks to our general setting, we reframe the isometry condition in terms of the Rényi-2 entropy of the input region, and compute it via the analysis of a dual Ising model. A novel feature in this last step is the necessity to perform a cumulant expansion in terms of a \textit{different} statistical weight (distinct from but related to the uniform distribution over states)  in order to calculate the Rényi entropy.

\subsection{Questions and quantities of interest}
    Our goal is to enhance the characterisations of spin network states in terms of holography that were put forward in~\cite{colafranceschiHolographicMapsQuantum2021}. This serves to select subsets of states with favorable typical properties, as well as to make connections with research done in the context of tensor network holography. The main questions we address are the following: 
    \begin{itemize}
        \item Given a superposition of spin network states with fixed graph structure, consider the bulk-to-boundary map it defines, as well as the map between two subregions of the boundary; are these maps isometric?  
        \item Using the channel-state duality, properties of these maps can be traced back to properties of the corresponding states. In particular, isometry is related to maximised Rényi-2 entropy of the input region. We therefore ask: what is the Rényi-2 entropy of bulk and boundary subregions? 
    \end{itemize}
    \subsection{States under consideration}
	Throughout this work, we will study a restricted class of states made from superposing spin network states associated with the same (open) graph $\gamma$\footnote{We distinguish between open graphs, which are allowed to have a subset of links, called "semilinks" which do not end in another vertex (equivalently, one can see them as ending on $1$-valent vertices), and closed ones, in which every link ends in 2 ($D$-valent) vertices (with $D > 1$). 
 } and different assignments of spins on the its links. More specifically, we consider states constructed in analogy with projected entangled pair states (PEPS), i.e. obtained by contracting tensors associated to open spin network vertices according to $\gamma$, and therefore denoted as \textit{spin tensor network states}.
 
    Even though most of the constructions in this work are generic in terms of the relevant Hilbert space, let us be concrete in our choice of Hilbert spaces. Given a (compact) Lie group $G$, let $\hbb_x = L^2(G^D/G)$ be the Hilbert space of an open, $G-$invariant spin network vertex $x$ of valence $D$, whose links are identified by an index $\alpha=1,\dots,D$. For simplicity, we restrict the attention to the case in which all vertices of the graph possess the same valence, but the analysis can be easily generalised. Concretely, we will work with $ G= SU(2) $ spin networks, whose representations are labeled by a \textit{spin} $ j \in \frac{\mathbb{N}}{2} $, interpreted as quantised area and associated to the links of the spin network graph $\gamma$, as it is shown to be the case in canonical quantum gravity as well as simplicial quantum geometry. A link of $\gamma$ is generically indicated by $ e \in E_\gamma $ (with $E_\gamma$ being the total link set of $\gamma$), while vertices will be labeled by $x\in V_\gamma$ in the vertex set $V_\gamma$.  
    The single-vertex Hilbert space admits a decomposition
    \begin{equation}\label{SingleVertexHS}
        \hbb_x \cong \bigoplus_{\mathbf{j}^x} \ical_{\mathbf{j}^x}\otimes V_{\mathbf{j}^x}
    \end{equation}
    into a direct sum over the representation labels $\textbf{j}^x \coloneqq j^x_1, \dots , j^x_D$, where each summand is the tensor product of the (boundary) space $ V^{\textbf{j}^x} \cong \bigotimes_{\alpha=1}^{D} V^{j^x_{\alpha}} $, given by the tensor product of the representation spaces of dimension $ d_{j} = 2j +1 $ associated to the open links of vertex $ x $, and the (bulk) intertwiner space of vectors invariant under the diagonal $G$-action, $ \mathcal{I}^{\textbf{j}^x} = \text{Inv}_{G}(\bigotimes_\alpha V^{j^x_\alpha}) $.

    The main way in which the choice of Hilbert spaces on links and vertices affects the results is through the dimensionalities, in particular the dependence of $\dim(\ical_{\mathbf{j}^x})$ on the adjacent link spaces. Additional structures such as group actions play less of a role in this work.
    In particular, the dimension of the bulk input space on a vertex, $\dim(\ical_{\mathbf{j}^x})$ depends on the dimensions of its adjacent link spaces, $d_{j^x_e}$. Additionally, we will consider only finite dimensional vertex Hilbert spaces - in the aforementioned example, we will implement this through imposing cutoffs $\mathbb{j}<j^x_e<J$ on the values of spin labels.

    Importantly, the full Hilbert space of $ N $ distinguishable vertices decomposes into a direct sum of \textit{spin sectors}, each having as a basis the spin network states with fixed link spins:
	\begin{equation} \label{H_N}
		\hbb = \bigotimes_x \hbb_x \cong \bigoplus_{\ind{j}}\mathbb{H}_{\ind{j}} \qquad \mathbb{H}_{\ind{j}} \cong \mathcal{I}_{\ind{j}} \otimes V_{\ind{j}} = \bigotimes_x \mathcal{I}^{\textbf{j}^x} \otimes \bigotimes_x V^{\textbf{j}^x}, 
	\end{equation} 
	where $\ind{j}=\textbf{j}^1,\dots, \textbf{j}^N$ is the collection of spins over the whole set of semilinks.
    Our setup generalises immediately to the case where \textit{any} single-vertex Hilbert space is used, as long as it admits a decomposition of the same type as \ref{SingleVertexHS}. In particular, the dimension of the bulk space $\ical$ is allowed to depend on the representation labels in a nonlinear way. The crucial assumption is rather an initial factorisation of states over vertices, so that a PEPS-type construction of states is possible. This is not generally the case for states of quantum geometry in canonical or lattice quantum gravity (nor is the dependence on a single graph, even though it is more commonly assumed).
	
    The first step for the construction of the spin tensor network states consists of picking a state in $\hbb^N$ which factorizes per vertex:
	\begin{equation}\label{newstatesofpreparia}
		\ket{\Psi} = \bigotimes_x \ket{\Psi_x}.
	\end{equation}
	To turn this state into a spin network state with support on a graph $\gamma$, we apply a projection onto maximally entangled states of the spins living on the links (semilinks) forming the maximal closed subgraph $\Gamma$ of $\gamma$ consisting of all vertices with valence $>1$, according to the prescription outlined in e.g.~\cite{chircoRyuTakayanagiFormulaSymmetric2018,colafranceschiQuantumGravityStates2021}.
    More specifically, let 
	\begin{equation}\label{maxentstate}
		\ket{e_j} := \frac{1}{\sqrt{d_j}}\sum_m (-1)^{j+m}\ket{j ,m}_1\ket{j ,-m}_2
	\end{equation} 
    be a normalised singlet state of two semilinks (labeled here by 1 and 2) carrying the spin $j$, and consider a normalised superposition
    \begin{equation}
        \ket{e} = \bigoplus_{j_e \in \frac{\mathbb{N}}{2}} g_{j_e} \ket{e_{j_e}}
    \end{equation}
    with coefficients $ g_j $. The gluing of the $N$ vertices described by $\ket{\Psi}$ into the graph $\Gamma$ is then performed by projecting $\ket{\Psi}$ onto $ \ket{\Gamma} = \bigotimes_{e\in \Gamma} \ket{e} $, i.e. by applying to $\ket{\Psi}$ the operator (up to normalisation)
	\begin{equation}\label{linkprojector}
		\Pi_{\Gamma} = \bigotimes_{e\in \Gamma}\left(\bigoplus_{j_e}|g_{j_e}|^2 \ket{e_{j_e}}\bra{e_{j_e}}\right)
	\end{equation}
    The role of $\Pi_{\Gamma}$ is precisely that of entangling, in \textit{every spin sector}, the data of pairs of semilinks according to $\Gamma$. The result is a superposition of spin network states with support on $ \gamma $, the set of which we denote by $\hbb_\gamma = \Pi_\Gamma(\hbb)$.  Note that, when restricting the attention to vertex states picked on a single spin sector, the setting reduces to that of previous work~\cite{colafranceschiHolographicMapsQuantum2021,chircoBulkAreaLaw2022}.
    We thus constrained the link spins to be in (a superposition of) singlet states, in order to be glued according to $\Gamma$. 
    
    In principle, we can proceed to constrain the intertwiner degrees of freedom as well, and focus on the resulting boundary state. To do so, first notice that the graph Hilbert space splits into a sum over boundary link labels $E = j_\bnd$:
    \begin{equation}
        \hbb_\gamma \cong \bigotimes_{E} \ical_{E}\otimes V_{E}.
    \end{equation}
    Here, we introduced the boundary-fixed Hilbert spaces for vertices and boundary links, respectively
    \begin{equation}
        \ical_{E} = \bigoplus_{ \{ j_e: e\in \Gamma \} }\bigotimes_{x\in \gamma} \ical_{\mathbf{j}^x} 
        \qquad
         V_{E} = \bigotimes_{e\in\bnd}V^{j_e}
    \end{equation}
    in which the latter has a tensor product factorisation over boundary links, while the former retains a sum over bulk spins $j_e$. We can then choose pure states $\ket{\zeta_E}\in \ical_E = \bigoplus_{{j_e: e\in\Gamma}}\ical_{\ind{j}}$
    and define another projector
    \begin{equation}
            \Pi_B = \sum_{E} \frac{\ket{\zeta_E}\bra{\zeta_E}}{\langle \zeta_E\ket{\zeta_E}} \otimes \ibb_{V_E}
    \end{equation}
    which enforces the state to have certain intertwiner data in each sector. 

    Then, the projection
	\begin{equation}\label{projected} 
	       \ket{\phi_\bnd}= \bra{\zeta}\bra{\Gamma} \bigotimes_x \ket{\Psi_x}
	\end{equation}
	produces a state of the open (unglued) boundary semilinks, described by the Hilbert space
	\begin{equation}\label{BndSplits}
		\mathbb{H}_{\partial\gamma} \cong \bigoplus_{j_{\partial\gamma}} \bigotimes_{ e \in \bnd} V^{j_e}\cong \bigotimes_{ e \in \bnd}\bigoplus_{j_e}V^{j_e} \cong \bigotimes_{ e \in \bnd} V_e.
	\end{equation}
    Due to this factorisation, it is straightforward to speak of entanglement and measures of it in the boundary-reduced case. \\ In contrast, the fixed-graph Hilbert space $\hbb_\gamma$ with no restriction on the intertwiner data, thus with a generic superposition of them, has no obvious factorisation properties at all. In such a setting, the notion of entropy survives, but several subtleties arise in quantifying entanglement with it~\cite{lin_comments_2018,hollands_entanglement_2018,ma_entanglement_2016}. \\
    Nevertheless, holography can be characterised more directly than through entanglement scaling. We indeed use an alternative strategy, in which we can neglect this distinction and still make use of entropies as a computational tool to characterise holographic behavior; in particular, the latter will be traced back to isometric mappings between graph subregions. This is the task we tackle in the following. 
	
    \subsection{A notion of holography}
    We introduce here a simple but effective notion of information transport that allows us to make assertions about a form of holography. Details may be found in other work \cite{Project1Draft1}, in a more abstract context not focused on spin network states of quantum geometry. \\
    The main idea is to use a density matrix $\rho$ on the graph Hilbert space $\hbb_\gamma$, together with extension $i_I$ of bulk and partial trace maps $P\Tr_O$ of boundary operators, mirroring the operations on a product Hilbert space $\hbb_I\otimes \hbb_O$ of extending an operator on $\hbb_I$ $X\mapsto  X\otimes \ibb_O$, and $Y\mapsto \Tr_I[Y]$ as the partial trace returning operators on $\hbb_O$. We refer to the `input' and `output' systems by the same labels I and O from now on. With these objects, we may define a mapping between operator spaces $\acal_{I|O}$, more generally seen as input and output system, which we can use to turn operators, density matrices etc. in the bulk to equivalent ones in the boundary. The equivalence between the two systems is then the statement that the mapping between corresponding operator spaces is isometric.\\
    Such a statement may be translated into a calculable question about Rényi entropies of the bulk-reduced state $\rho_I$. This also connects the present framework to previous work. \\
    The pieces are combined as follows: we suppose that holography can be expressed by turning operators (or operations) $X_I$ on the bulk system into approximately equivalent operations on the boundary system. If the two are part of a larger system, we can extend bulk operators to the whole system by the operation $i_I$ . We could then evaluate $i_I(X_I)$ in the whole state $\rho$ of the system (here, our graph state), to get $\Tr[i_I(X)\rho]$, or instead reduce this to an effective operator on the boundary through the partial trace $P\Tr_O$. The algebras $\acal_{I|O}$ serve in this as a restriction to sets of operators where this `operator transport' suitably keeps all or most of the data of the system. \\
    This operator-focused approach not only allows for immediate transport of operators on top of Hilbert space states, but it is also necessary for the direct-sum Hilbert spaces we consider here, which, as we remarked, do not have straightforward factorization properties. Indeed, if we were to apply a Hilbert-space mapping paradigm, mapping bulk states to boundary states, then we would have to confront the question of which Hilbert space would be the bulk one, and which one is the boundary one, and how we can see these as subsystems; this question is highly ambiguous, absent factorization. \\
    To begin, we introduce the algebraic subsystems
    \begin{equation}\label{Bbndalgs}
        \bcal_I := \bigoplus_{E\in\mathcal{W}_I} B(\hbb_{I,E}:=\ical_E) \qquad \bcal_O := \bigoplus_{E\in\mathcal{W}_O} B(\hbb_{O,E}:=V_E)
    \end{equation}
    of the full algebra of operators on the graph Hilbert space
    \begin{equation}
        \acal = B(\hbb_\gamma) = \bigoplus_{E,E'} B(\ical_E\otimes V_E,\ical_{E'}\otimes V_{E'})
    \end{equation}
    where we choose the index sets $\mathcal{W}_{I|O}$ a posteriori to ensure isometric mapping between the two sets. We relate them to the full algebra through the associated extension and partial trace maps 
    \begin{equation}
        \bcal_{I|O}\stackrel{i_{I|O}}{\hookrightarrow} \acal_{I|O} \subset \acal \stackrel{P\Tr_{I|O}}{\rightarrow} \bcal_{I|O}
    \end{equation}
    given for example for the input/bulk algebra by
    \begin{align}
        i_I(X) = \sum_E X_E \otimes \ibb_{O,E} \qquad P\Tr_I[X] = \sum_E \Tr_{O,E}[X_E]
    \end{align}
    and whose images we name $\acal_{I|O} = \text{Im}(i_{I|O})$.\footnote{These $\acal_{I|O} $ are, technically speaking, the actual subsystems, and $i_{I|O} $ are the identifications allowing $\bcal_{I|O} $ to be treated as subsystems.} These are the `naive' partial trace and extension, and our choices of algebras can be motivated by this naive choice.\\
    However, additionally, we must make this choice for bulk and boundary algebras where we only have sector-diagonal operators, because any non-diagonally acting operator will not have a clear notion of `leaving the complement invariant'. If we understand a bulk operator $X = \sum_{E,F} X_{E,F} $ to leave the boundary $V_E$ invariant, its matrix elements with respect to a basis $\ket{E,i|j}$ of the bulk $\ical_E$ labeled by $i,j$, and $\ket{E,m|n}$ of the boundary $V_E$ labeled by $m,n$, should satisfy
    \begin{equation}
        \bra{E,j}\bra{E,n} X_{E,F} \ket{F,i}\ket{F,m} \sim \delta_{E,F}\delta_{m,n},
    \end{equation}
    but this is only a sensible equation if $E=F$ holds on the right hand side, and therefore must be enforced on the left, giving a sector-diagonal operator.\\
    On the other hand, if we instead defined bulk/boundary operators to be sector-mixing, such that
    \begin{equation}
        \bcal'_I = B(\bigoplus_E \ical_E) \qquad \bcal'_O = B(\bigoplus_E V_E) ,
    \end{equation}
    then there would be no obvious way to extend or realise the non-sector-diagonal ones to/in the algebra $\acal$. Said more succinctly, these sets are not subalgebras of $B(\hbb_\gamma)$ and therefore do not function as algebraic subsystems. We are therefore led to the choice of bulk and boundary algebras \ref{Bbndalgs} as the largest sensible one, and notice that they are not realised as operators on a single Hilbert space, but only a certain subset of those. In fact, they are the subalgebras $\bcal_{I|O} \subseteq \bcal_{I|O}' $ such that their \textit{center} is given by the boundary spin Casimirs $ \{ J^2_e : e \in \bnd \} $. This is, in fact, what we get if we remove from the full algebra $\acal$ all \textit{holonomy} operators $h_e$ on boundary links $e\in \bnd$~\cite{lin_comments_2018}. In this sense, because we are removing the boundary gauge field from the algebra (usually associated with magnetic degrees of freedom), we could interpret the bulk/boundary algebras \ref{Bbndalgs} as \textit{electric} algebras.\\
    Now, with these preliminary choices at hand, we can define the superoperator mapping bulk operators into boundary operators via the Choi-Jamiolkowski isomorphism~\cite{jiangChannelstateDuality2013,majewskiCommentChannelstateDuality2013}:
    \begin{equation}
        \tcal_\rho (X)  = K \, P\Tr_{O} [ i_I(X)\rho ] = \sum_E K c_E \, \Tr_{I_E} [ (X_E \otimes \ibb_{O_E})\rho_{E,E} ],
    \end{equation}
    where
     \begin{equation}
        \rho = \sum_{E,\Tilde{E}} \sqrt{c_E c_{\Tilde{E}}} \rho_{E,\Tilde{E}} \in \dcal(\hbb_\gamma)
    \end{equation}
    with $\Tr[\rho_{E,\Tilde{E}}] = \delta_{E,\Tilde{E}}$ and $c_E = \Tr_E[\rho] \geq 0$, $\sum_E c_E = 1$, $K>0$, and $\dcal(\hbb_\gamma)$ the set of density matrices on $\hbb_\gamma$. The sum is over boundary spin labels $E=j_\bnd$. \\
    In the following, we will work with proper subalgebras, so $\bcal_{I|O}\subset \bigoplus_{E} \mathbf{B}(\hbb_{I|O,E}) $. The main reason is that it is generically impossible to make the mapping on the full bulk input algebra isometric, just for dimensional reasons alone. We will make a choice of subalgebra by selecting a subindex set $\mathcal{W}$ of sectors $E$. The specific choice depends on the scenario of information transport we consider, and will be done a posteriori.\\
    Assume now that the dimension of $ \bcal_I $ does not exceed that of $ \bcal_O $ so that isometry between spaces of operators is possible in principle.
    We find the condition for $\mathcal{T}_\rho$ to be an \textit{isometry in the Hilbert-Schmidt inner product on the operator algebras} by equating the two expressions on the input (bulk) and output (boundary) sides
    \begin{align}
        \avg{\mathcal{T}_\rho(A),\mathcal{T}_\rho(B)}_O =\avg{A,B}_I .
    \end{align} 
    As illustrated in \cite{Project1Draft1}, this is equivalent to the requirement on the state $\rho$ that
    \begin{equation}\label{TrueIsomCond}
        c^2_E K^2 \Tr_{O^{\otimes 2}_E} [(\rho_{E,E}\otimes\rho_{E,E}) \mathcal{S}_{O_E}]  = \mathcal{S}_{I_E} \qquad \forall E\in \mathcal{W}.
    \end{equation}
    in which we use the \textit{swap operators} $\sw$ that exchange two identical copies of a Hilbert space.\\
    In the case where the states $\rho_{E,E}$ are \textit{pure}, the isometry condition above simplifies drastically to
    \begin{equation}\label{TracePreservation2}
        (\rho_{E,E})_I = \frac{\ibb_{I_E}}{D_{I_E}} \qquad c_E = \frac{D_{I_E}}{\sum_F D_{I_F}} \qquad \forall E \in \mathcal{W}.
    \end{equation}
    together with the condition $|K| = \sum_E D_{I_E}= D_I$. This, in fact, is just the requirement of trace preservation.
    Therefore, schematically, 
    \begin{equation}
    \rho_{E,E}\text{ pure }:\;\; \mathcal{T}_{\rho} \text{ Quantum channel} \implies \mathcal{T}_{\rho} \text{ isometry} 
    \end{equation}
    We will restrict to pure states $\rho_{E,E}$ in the following as results from~\cite{Project1Draft1} indicate that it is generically difficult to get isometry of $\tcal$ from mixed states.
	\textit{Therefore, in this work, we will establish when the transport superoperator is, typically, a quantum channel}, and this will be the sense in which we identify isometric and thus holographic states.  Furthermore, trace preservation is easy to convert into a calculable statement about entropies, which also directly connects to previous work. 
    
    Let us consider entropies for those isometry-inducing states. For any normalised input state $\sigma$, hermiticity of the channel $\mathcal{T}_\rho$ and isometry imply that the Rényi-2 entropy is left unchanged by the channel:
    \begin{equation}
        e^{-S_2(\tcal(\sigma))} 
        = \Tr[\tcal(\sigma)^2] 
        \stackrel{\tcal \text{ herm.}}{=} \Tr[\tcal(\sigma)^\dagger \tcal(\sigma)] 
        \stackrel{\tcal \text{ isom.}}{=}
        \Tr[\sigma^\dagger \sigma]
        =
        \Tr[\sigma^2]  
        = e^{-S_2(\sigma)}
    \end{equation}
    The Rényi-2 entropy of the reduced state $\rho_I$ itself is then also purely determined by the range of sectors and is maximal:
    \begin{equation}\label{EntropySplit}
        e^{-S_2(\rho_I)} = \sum_E c^2_E e^{-S_2((\rho_{E,E})_I)} = \sum_E \frac{\dim(\hbb_{I_E})^2 }{D_I^2} \frac{1}{\dim(\hbb_{I_E})} = \frac{1}{D_I}.
    \end{equation}
    We now shift our focus towards determining which classes of spin network states satisfy these last requirements.
    The method we use is entirely analogous to the one used in previous works~\cite{colafranceschiHolographicMapsQuantum2021,colafranceschiHolographicEntanglementSpin2022,haydenHolographicDualityRandom2016,qiHolographicCoherentStates2017} and relies on entropy calculations condition, written e.g. as $S_2((\rho_{E,E})_I) = \log(D_{I_E})$.
    Via the swap operator\footnote{Letting $ \mathcal{S}$ be the operator swapping two copies of a Hilbert space $\hbb$, we have $ \Tr_{\hbb^{\otimes 2}}[(A\otimes B)\mathcal{S} ] = \Tr_\hbb[AB] $ while $\Tr[(A\otimes B) ] = \Tr[A]\Tr[B] $.} we can rewrite the Rényi entropy as traces over two copies of the system:
	\begin{equation}\label{Rényi2}
		\frac{\Tr_{\hbb_I^2}[\rho_I^{\otimes 2} \mathcal{S}_I]}{\Tr_{\hbb_I^2}[\rho_I^{\otimes 2}]}
		=
		\frac{\Tr_{\hbb_{\bnd}^2}[(\ket{\phi}\bra{\phi})^{\otimes 2} \mathcal{S}_I]}{\Tr_{\hbb_{\bnd}^2}[(\ket{\phi}\bra{\phi})^{\otimes 2}]}.
	\end{equation}
    We have introduced the state $\ket{\phi}$ here as a stand-in for whichever pure state we assign to the full system. In the case of bulk-to-boundary or boundary-to-boundary maps, we will use pure states of different systems.
    For example, in the boundary-to-boundary case, by using the projectors $ \Pi_I , \Pi_\Gamma   $, we consider a state $\ket{\phi} = \Pi_I  \Pi_\Gamma \ket{\Psi}$ with $\ket{\Psi}\in \hbb^N$. We can then further express the numerator and denominator in terms of traces over $ \hbb^{\otimes 2}  $. For this, note that for product states $\ket{\Gamma}$ and $\ket{\zeta}$ for, respectively, links and intertwiners, the reduced state is
 \begin{equation*}
     \rho_I = Tr_O[\ket{\phi}\bra{\phi}] = \Tr_{\hbb_O\otimes\hbb_\Gamma}[\bra{{\Gamma},{\zeta}}\ket{\Psi}\bra{\Psi}\ket{{\Gamma},{\zeta}}]
     = 
     \Tr_{\hbb_O\otimes\hbb_\Gamma}[\ket{{\Gamma},{\zeta}}\bra{{\Gamma},{\zeta}}\ket{\Psi}\bra{\Psi}]
 \end{equation*}
 where we traced out the link and intertwiner spaces to impose the correct relations on the state, and projected out everything that is not in the boundary Hilbert space. We can thus equivalently apply the projectors $\Pi_I$ and $\Pi_\Gamma$ and trace over $ \hbb^{\otimes 2} $. With a product of vertex states as a starting point, the Rényi-2 entropy then takes the general form
	\begin{equation} 
		e^{-S_2(\rho_C)} = \frac{
			\Tr_{\hbb^{\otimes 2} }[ \Pi^{\otimes 2}_I\Pi^{\otimes 2}_{\Gamma} (\bigotimes_{ x} \ket{\Psi_x}\bra{\Psi_x})^{\otimes 2} \mathcal{S}_C]
		}{
			\Tr_{\hbb^{\otimes 2} }[ \Pi^{\otimes 2}_I\Pi^{\otimes 2}_{\Gamma} (\bigotimes_{ x} \ket{\Psi_x}\bra{\Psi_x})^{\otimes 2} ]	
		}
	\end{equation}
	Importantly, we have reduced the trace from $ \hbb_{\bnd}\otimes \hbb_I \otimes\hbb_\Gamma $, i.e. the tensor product of the Hilbert spaces for boundary links, intertwiners and internal links, to the `sector-diagonal' subspace $ \hbb $. This is possible since the vertex states we start from have intertwiner/link/boundary data whose sectors are matched, meaning $ \ket{\Psi_x} \in \hbb_x $.\\
    The case of bulk-to-boundary maps will be treated later in more detail.
	
    \subsection{Randomisation over vertex states}
	Instead of calculating the entropy for a particular state, we will make a typicity statement about our class of spin tensor networks. So, colloquially, we will ask \textit{"What is the average degree of isometry for states with specified graph structure?"}. The value of such a statement depends crucially on the deviation from the average. However, as was shown in previous work on random tensor networks~\cite{haydenHolographicDualityRandom2016}, the deviation is sufficiently small in a particular limit of bond dimensions. In our context, this limit must be taken on all bond dimension in the superposition. As we will discuss shortly, our construction requires the introduction of an upper cutoff $J$ on dimensions, which may be arbitrarily large, but in addition we require a lower cutoff $\mathbb{j}$. By taking this lower cutoff to be large, all bond dimensions involved in the superposition are large enough to suppress deviations from the average.\\
	In our class of states, then, we can write the average purity as a partition function of a randomised Ising model. To see this, we first average over a distribution of vertex states $ \ket{\Psi_x} = U_x \ket{\Psi_{ref}} $, where we choose some arbitrary pure reference state $\ket{\Psi_{ref}}$. This distribution is chosen to be uniform over the unitary group relating different single-vertex-states\footnote{The choice of uniform probability distribution is of course not the only possible one. At this stage, interpretative viewpoints as well as dynamical considerations can play an important role and suggest different choices. For example, in~\cite{chircozhang, chircozhang2}, this is where the group field theory dynamics of quantum geometry is inserted.}. More explicitly, we perform the integral 
	\begin{equation}
 \begin{aligned}
  \label{HaarMeasure}
		 \avg{(\ket{\Psi_x}\bra{\Psi_x})^{\otimes 2}}_{U_x} &\coloneqq \int_{\mathcal{U}(\hbb_x)} d\mu(U_x) (U_x \ket{\Psi_{ref}}\bra{\Psi_{ref}} (U_x)^\dagger)^{\otimes 2} \\&=: R_x\left((\ket{\Psi_x}\bra{\Psi_x})^{\otimes 2}\right)
   \end{aligned}
	\end{equation}
 where $d\mu(U_x)$ is the Haar measure on the unitary group $U_x$, for each vertex $x$ separately; the last line defines the operator $R_x$ implementing such an average. 
By linearity this average commutes with taking traces and we denote it by $ \avg{-}_U $ in the following.\\
	If all participating spins in the state are sufficiently large, say larger than some lower cutoff $ \mathbb{j} $,\footnote{What is required is that the lower cutoff scales in the number of vertices of the graph, as in $ \mathbb{j} >> N^k $ for some $ k > \frac{2}{\Delta E} $, with $\Delta E$ the spectral gap of the dual Ising model (see below); in fact, we need $N^2 << \mathbb{j}^{\ln(2\mathbb{j}+1)}$. This in turn implies $N > 15$ for the argument to make sense.} we can suppress fluctuations in the quotient
	\begin{equation}\label{AvgRényi1}
		\avg{e^{-S_2(\rho_I)}}_U =
		\avg{\frac{\Tr_{\hbb^2}[(\ket{\phi}\bra{\phi})^{\otimes 2} \mathcal{S}_I]}{\Tr_{\hbb^2}[(\ket{\phi}\bra{\phi})^{\otimes 2}]}}_U 
		\approx \frac{\avg{\Tr_{\hbb^2}[(\ket{\phi}\bra{\phi})^{\otimes 2} \mathcal{S}_I]}_U}{\avg{\Tr_{\hbb^2}[(\ket{\phi}\bra{\phi})^{\otimes 2}]}_U}
		=: \frac{Z_1}{Z_0}
	\end{equation}
	as has been shown in random tensor networks - the measure concentrates over the average if all spins are large. \\
    A short note on this regime is in order. Operationally, we are simply making a statement about a class of states which, in Peter-Weyl decomposition, consist only of certain representations. Physically, we may inspect this (still relatively large) class of states from several angles. For one, if no superposition is made and we deal with a single spin-network state, area operators associated with individual links take sharp values. This is already not the case once we superpose. In this large-j regime, the relative spacing in the spectrum of the area operator on any given link becomes arbitrarily small. However, this phenomenon already occurs at lower spins for \textit{sums} of area operators of different links, which are used for larger surfaces. Furthermore, while for a single sector the area values may be sharp, no sharpness is present ab initio for other quantities such as length operators or angles (spin networks are not eigenstates of those operators). Second, the states have a fixed entanglement pattern that is in no way erased by choosing large representation labels from the beginning. These quantum information properties are our focus, so the area values are of secondary importance. \\
    The operator $ R_x $ acting on two copies of the single-vertex Hilbert space has the property that it is invariant under unitary conjugation:
	\begin{equation} 
		V^{\otimes 2} R (V^\dagger)^{\otimes 2} = R
	\end{equation}
	by left-invariance of the Haar measure. \\
	Crucially, this requires the group to be a finite-dimensional Lie group  (this integral does not exist on the infinite unitary groups, so our Hilbert spaces must stay finite dimensional). Thus, we must require all spins of the state to be below some (arbitrarily large) upper cutoff $ J $. \textit{Therefore, all we can consider in our framework are subsets of the set of spin tensor network states which have only finitely many, sufficiently large spins in their superposition.}\\
	With this property, we can easily find what $ R_x $ is - the only two operators invariant under this action are the identity and the swap operator, and are combined in the form
	\begin{equation} 
		R_x = \frac{1}{\dim(\hbb_x)(\dim(\hbb_x)+1)} ( \mathbb{I}_x + \mathcal{S}_x).
	\end{equation}
	However, the dimensions here of course need to be the ones of the truncated Hilbert spaces, as otherwise the right hand side would vanish.
	Since we average over each vertex seperately, we really replace the initial random vertex states by
	\begin{equation} \label{MediumGrade}
		\frac{1}{\prod_x \dim(\hbb_x)(\dim(\hbb_x)+1)}\bigotimes_{x}  ( \mathbb{I}_x + \mathcal{S}_x).
	\end{equation}
    We will perform three types of average in this work, some finer than others. The `medium grade' is the one where each individual vertex Hilbert space $\hbb_x$ is averaged over. The coarser one averages instead over all of them simultaneously, so over $\bigotimes_x \hbb_x$, while the finer one performs an average over each sector $\hbb_{\mathbf{j^x}}$ of the single-vertex Hilbert spaces, and additionally allows for a probability weight $p_{\ind{j}}$ among the sectors\footnote{This weight is necessary to allow for states in the direct sum that are not active in all sectors. The states are of the form $\ket{\Psi} = \sum_{\ind{j}} \sqrt{p_{\ind{j}}} \otimes_x U_{\mathbf{j}^x}\ket{\Psi_{\mathbf{j}^x,\text{ref}}} $, with randomly selected unitaries.}. This distinction amounts to different granularities of typicality statements. A coarse average is easier to produce, has larger validity, but of course may not capture the finer details. A finer average, however, assumes more structure of the objects and that no mixing happens between the sectors averaged upon. In the coarser case, we replace the initial random vertex states by
    \begin{equation}\label{CoarseGrade}
        \frac{1}{\prod_x \dim(\hbb_x)(\prod_x\dim(\hbb_x)+1)}  ( \mathbb{I} + \mathcal{S}),
    \end{equation}
    whose operators act on all spin network vertices (equivalently, in the dual simplicial geometry picture, tetrahedra) simultaneously. On the other hand, the finer average produces
    \begin{equation} \label{FineGrade}
		\sum_{\ind{j},\ind{k}} p_{\ind{j}}p_{\ind{k}}
        \frac{\ibb_{\ind{j}} \otimes\ibb_{\ind{k}}
        }{\mathcal{D}_{\ind{j}} \mathcal{D}_{\ind{j}}}
        +
        \sum_{\ind{j}} p_{\ind{j}}^2
        \left(
        \frac{ \bigotimes_x (
            \ibb_{\hbb^{\otimes 2}_{\mathbf{j}^x}} + \sw_{\hbb^{\otimes 2}_{\mathbf{j}^x}} )
         }{
            \prod_x\mathcal{D}_{\mathbf{j}^x} (\mathcal{D}_{\mathbf{j}^x}+1)
         } - 
         \frac{\ibb^{\otimes 2}_{\ind{j}} 
        }{\mathcal{D}^2_{\ind{j}} }
        \right)
    \end{equation}
    which, in the regime of high spins, is well approximated by
    \begin{equation}\label{FineGradeHighRegimeExposition}
        \sum_{\ind{j}} p_{\ind{j}}^2 
        \frac{
        \bigotimes_x(
            \ibb_{\hbb^{\otimes 2}_{\mathbf{j}^x}} + \sw_{\hbb^{\otimes 2}_{\mathbf{j}^x}} ) - \ibb_{\hbb^{\otimes 2}_{\ind{j}}}
        }
        {\dim(\hbb_{\ind{j}})^2} 
        + \left( \sum_{\ind{j} } p_{\ind{j}} \frac{\mathbb{I}_{\ind{j}} }
        {\dim(\hbb_{\ind{j}})}  \right)^{\otimes 2} =: \mathcal{Q}^{(p)}+\mathbb{I}^{(p)}\otimes \mathbb{I}^{(p)}
    \end{equation}
    
	Then comes a crucial rewriting. To make working with the tensor product above tractable, we recognise that, when expanded as a sum, each term will have a number of swap operators, and identity operators do not matter. Each term can then be labeled by the set of vertices with swap operators on it, a $ -1 $ indicating a swap. \\
    The method introduced by Hayden \textit{et al.}~\cite{haydenHolographicDualityRandom2016} assigns to each vertex a $ \pm 1 $-valued \textit{Ising spin} $ \sigma_x $, which indicates whether a swap is on that vertex or not. This means the product (here for the medium grade average) turns into the sum over Ising configurations
	\begin{equation} 
		\prod_x \frac{1}{\dim(\hbb_x)(\dim(\hbb_x)+1)} \sum_{\vec{\sigma}} \bigotimes_x \mathcal{S}^{\frac{1-\sigma_x}{2}}_x
	\end{equation}
	To explain further this step, each term in the original sum is mapped to a unique Ising configuration such that the region of swap operators is the region of Ising spin-downs. Then, every configuration must be summed over.
	This turns the numerator and denominator of the average purity into \textit{Ising partition functions}:
	\begin{equation} 
		Z_{1|0} = \sum_{\vec{\sigma}} \Tr[\Pi_{}^{\otimes 2}\bigotimes_x (\mathcal{S}^{\frac{1-\sigma_x}{2}}_x) \mathcal{S}^{1|0}_I ]
		= 
		\sum_{\vec{\sigma}} e^{-H_{1|0}(\vec{\sigma})}
	\end{equation}
	and evaluation of the average purity is turned into a calculation of Ising-like partition sums. The projector $\Pi$ will depend on the application. In the case of large bond dimensions, we can approximate the sums by their ground state values as the lowest bond dimension plays the role of inverse temperature. The result is that achieving minimal purity of the reduced state, corresponding to having an isometry, depends on the size of the local input and output legs, along with the underlying graph structure. \\
    In the case of the coarser average, this sort of tensor product does \textit{not} appear and the expectation values can be directly computed. 
    Similarly, the fine average, in the high spin limit we are interested in, does not have the tensor product structure of the medium one. Instead, we have effectively a structure closer to the coarse average, with additional dependence in the weights $p$. Still, an Ising sum for each sector $\ind{j}$ is necessary, potentially complicating the calculations. We therefore delegate the calculations of the finer and coarser averages to appendices \ref{Appendix:CoarseAverage} and \ref{Appendix:FineAverage}, while the main body of this work is concerned with the medium average. As it can be seen in the appendix, the results are not drastically different.
 
    \subsection{Rewriting the Hamiltonian}
	To calculate the partition functions, we need to find a usable expression for the Ising Hamiltonian. This is straightforward when the Hilbert space factorises over vertices or links, but in the case of superposed spin sectors, this is less immediate. Because the Hilbert space does not factorise, we first have to split the trace into a sum over the spin sectors, in which we can then easily determine the Hamiltonian.\\
    More precisely, the total Hilbert space is the direct sum
    \begin{equation*}
        \hbb = \bigotimes_x \hbb_x = \bigoplus_{\ind{j}} \hbb_{\ind{j}}.
    \end{equation*}
    We can thus decompose the trace over $\hbb^{\otimes 2}$ into a sum over the spin sectors $\hbb_{\ind{j}}$, and rewrite the trace as follows:
	\begin{align}\label{MainTextPF2}
		Z_{1|0} &= \sum_{\vec{\sigma}} 
		\Tr[\Pi^{\otimes 2} \bigotimes_x \sw_x^{\frac{1-\sigma_x}{2}} (\mathcal{S}_I)^{1|0}]\\
		&= \sum_{(\ind{j},\ind{k},\vec{\sigma})} \Tr_{\hbb_{\ind{j}}\otimes\hbb_{\ind{k}}}
            [\Pi^{\otimes 2} \bigotimes_x \sw_x^{\frac{1-\sigma_x}{2}} (\mathcal{S}_I)^{1|0}] .
	\end{align}
	The individual summands depend, through the choice of the Hilbert space traced over, on the spin sets $\ind{j},\ind{k}$. The traces in each term are now over spaces that factorise over vertices and links, and accordingly the single-vertex swap operators do so, too: $ \sw_x = \sw_{\mathcal{I},x}\prod_\alpha\sw_{\alpha,x} $. The traces can then be evaluated over intertwiner, link and boundary parts separately.\\
	Then, the general form of decomposition we are looking for is as follows:
	\begin{equation} 
		\Tr_{j\times k} [ \Pi^{\otimes 2} \bigotimes_x \sw_x^{\frac{1-\sigma_x}{2}} \dots] = \Delta(\ind{j},\ind{k},\vec{\sigma}) K_{\ind{j}}K_{\ind{k}} e^{-H(\ind{j},\ind{k},\vec{\sigma})}.
	\end{equation}
	The three factors are non-unique, but fulfil specific functions: 
	\begin{itemize}
		\item The $ \Delta $-factor is boolean (or $[0,1]$-valued for the boundary-to-boundary mappings) and indicates whether a term vanishes - depending on the combination of spin sectors and Ising configuration, the term might be zero. Constraints arising from this are to be incorporated here.
		\item The $ K $-factor absorbs large contributions to the trace that depend only on the bond dimensions given through the spin sectors. They function as a normalising factor and will drop out of the calculation if one considers only a single factor. We can generally expect something of the form $K_{\ind{j}} = \Tr_{\hbb_{\ind{j}}}[\Pi] $.
		\item The Hamiltonian $ H $ is the main quantity of interest and contains all dependence on the Ising configuration. It also depends, in a normalised way, on the area spins of the spin sectors $ \ind{j},\ind{k} $. The function is designed such that it is nonvanishing only where the $ \Delta $-constraints are satisfied. This means there is no ambiguity which couplings the Ising model is subject to.
	\end{itemize}
	For example, in the following the decomposition for $ Z_0 $ will be chosen such that the Hamiltonian $ H_0 $ satisfies $ H_{0}(\ind{j},\ind{k},\vec{+1}) = 0 $, and the resulting choice of $ K_{\ind{j}} $ will be applied for $ Z_1 $ as well.\\
    Let us also define, for reference, the quantities 
	\begin{equation}\label{PFfixedSpins}
		Z^{(\ind{j},\ind{k})}_{1|0} = \sum_{\vec{\sigma}} \Delta_{1|0}(\ind{j},\ind{k},\vec{\sigma}) e^{-H_{1|0}(\ind{j},\ind{k},\vec{\sigma})}
	\end{equation}
	which enable us to phrase the discussion of the partition functions nicely. 
	By defining the normalised distribution over spin sectors
	\begin{equation}\label{IsingDistrib}
		P(\ind{j},\ind{k}) = \frac{K_{\ind{j}}K_{\ind{k}}}{Z_{0}} Z^{(\ind{j},\ind{k})}_{0}
	\end{equation}
	we see our quantity of interest as a probability average over Ising models
	\begin{equation}\label{ProbAvg1}
		\avg{e^{-S_2(\rho_I)}}_U \approx \frac{Z_1}{Z_0} = \sum_{(\ind{j},\ind{k})}  P(\ind{j},\ind{k}) 
		\sum_{\vec{\sigma}}
		\frac{e^{-H_{1}(\ind{j},\ind{k},\vec{\sigma})}}{Z^{(\ind{j},\ind{k})}_{0}}  \Delta_{1}(\ind{j},\ind{k},\vec{\sigma}).
	\end{equation} 
	\, \\
	If all spins in a given sector are large enough, we can perform a crucial approximation to the partition sums.
	In the Ising model, we may approximate the partition function by its ground state contribution if the excited states have very low weight. This is the case if the couplings of the model are very large, as any spin flip will increase the energy by an amount proportional to that coupling constant. When the spins are all large, we have that 
	\begin{equation} 
		Z^{(\ind{j},\ind{k})}_{0} \approx 1 , \qquad Z^{(\ind{j},\ind{k})}_{1} \approx \exp(-H_{1}(\ind{j},\ind{k},\vec{\sigma}_{GS}) )
	\end{equation}
	where $\vec{\sigma}_{GS}$ is the ground state configuration; this approximation massively simplifies the distribution $ P $ as well:
	\begin{equation} 
		Z_0 = \sum_{\ind{j},\ind{k}}K_{\ind{j}}K_{\ind{k}}Z^{(\ind{j},\ind{k})}_{0} \approx
		(\sum_{\ind{j}}K_{\ind{j}})^2
		\qquad P(\ind{j},\ind{k}) \approx p_{\ind{j}}p_{\ind{k}} \qquad p_{\ind{j}} = \frac{K_{\ind{j}}}{\sum_{\ind{k}}K_{\ind{j}}}
	\end{equation}
	In particular, given that $ Z_0 = \avg{Tr[\rho]^2}_U $, we can interpret the factorisation of the partition function as the statement $ \avg{Tr[\rho]^2}_U = \avg{Tr[\rho]}^2_U $ in the high-spin regime. This, in turn, simply reflects that in the high-spin regime, $\rho$ is on average a pure state, which is to be expected since we only work with such from the outset. \\
    In this high-spin regime, we may make general statements about the behaviour of entropies. Assume that the Ising sums in Eq.~\eqref{PFfixedSpins} have been calculated, and denote their approximate value by
    \begin{equation}
        Z^{(\ind{j},\ind{k})}_1 = e^{-X_{\ind{j},\ind{k}}}.
    \end{equation}
    Then, the full Rényi purity $e^{-S_2}$ takes the form of an expectation value in the probability density vectors $P = p\otimes p$ over pairs of spin sectors:
    \begin{equation}
        \avg{e^{-S_2}}_U \approx \sum_{\ind{j},\ind{k}} 
        p_{\ind{j}} p_{\ind{k}}
        e^{-X_{\ind{j},\ind{k}}} 
        =: \avg{e^{-X}}_P.
    \end{equation}
    We are therefore able to use a cumulant expansion for $X$ and write
    \begin{equation}
        \avg{e^{-X}}_P = \exp{\left(
        - \sum^\infty_{n=1}
        \frac{(-1)^{n-1}}{n!} \kappa_n(X)
        \right)
        }
    \end{equation}
    with cumulants of the random variable $X$, \begin{equation}
        \kappa_1(X) = \avg{X}_P \qquad \kappa_2(X) = \avg{X^2}_P-\avg{X}^2_P \qquad \text{etc.}
    \end{equation}\\
    Then, quite generally,
    \begin{equation}\label{CumulantExpansion}
        S_2(\rho_I) \approx \sum^\infty_{n=1}
        \frac{(-1)^{n-1}}{n!} \kappa_n(X) = 
        \avg{X}_P  - \frac{1}{2}\left(\avg{X^2}_P - \avg{X}_P^2\right) + \dots
    \end{equation}
    which shows that the overall entropy will not be `sharp' in the sectors, but be an average of the quantity $X$ that depends on the contributing sectors nontrivially. In the particular case that the partition sums evaluate, individually, to an `area' of a certain surface $ S_{\ind{j},\ind{k}} $ which bounds a bulk region $\Sigma_{\ind{j},\ind{k}}$ in the graph, such as
    \begin{equation}
        X_{\ind{j},\ind{k}} = \sum_{e\in S_{\ind{j},\ind{k}}} \log(d_{j_e}) \revcoloneqq \frac{1}{4} A_{\ind{j},\ind{k}}
    \end{equation}
    then this gives a Ryu-Takayanagi-type formula for the entropy which takes the $P$-expectation value of the area operators $\hat{A}_{\Sigma_{\ind{j},\ind{k}}}$ associated to the set of minimal surfaces $\Sigma_{\ind{j},\ind{k}}$:
    \begin{equation}\label{AveragedRTFormula}
        S_2(\rho_I) \approx \frac{1}{4} \avg{A}_P + \frac{1}{2\cdot 4^2} (\avg{A^2}_P-\avg{A}_P^2) + \dots
    \end{equation}
    where remaining terms capture higher cumulants of $A_\Sigma$.\\
    A well-known fact about random tensor networks with fixed bond dimensions (equivalent to the fixed-spin case here) is that they feature a Ryu-Takayanagi formula with a \textit{trivial area operator}. This means that the area operator $\hat{A}_\Sigma$ of the minimal surface $\Sigma$, appearing as
    \begin{equation}
        S_{vN}(\rho_I) \approx \frac{1}{4} \bra{\psi}\hat{A}_\Sigma \ket{\psi},
    \end{equation}
    is proportional to the identity operator on the graph Hilbert space,
    \begin{equation}
        \hat{A}_S = 4\left(\sum_{e\in S} \log(d_{j_e})\right) \ibb_{\hbb_\gamma}.
    \end{equation}
    This is also the case in our setting, because for just a single sector, there is only one possible value for the spin on each link. The `area' $a_e(j_e) := \log(d_{j_e})$ is then just a function of those c-number labels on the Hilbert space. This changes with multiple sectors as seen above: we have instead that the area operator is evaluated on a \textit{set} of minimal surfaces $\Sigma_{\ind{j},\ind{k}}$, and we can therefore write
    \begin{equation}
        S_2(\rho_I) \approx \frac{1}{4} \sum_{\ind{j},\ind{k}} P( \ind{j},\ind{k}) \bra{\psi}\hat{A}_{\Sigma_{\ind{j},\ind{k}}} \ket{\psi} \qquad \hat{A}_S := \sum_{e\in S} \hat{A_e} := \sum_{e\in S} \sum_{j_e} 4 \log(d_{j_e}) \ibb_{e,j_e}.
    \end{equation}
    The area operator on the right is not simply a multiple of the identity because it assigns different values to a surface $S$ depending on the state. In this sense, our area operator is nontrivial in a very similar sense to that of recent studies~\cite{Dong:2023kyr,Akers:2024ixq}. What is distinct is the possibility of multiple, distinct minimal surfaces which contribute to the entropy. However, this can be argued to be natural: if different sectors correspond to different sets of states with different metrics for spatial slices of a spacetime, then the condition of being the surface of minimal area depends on the metric in question, or more simply, on the sector. 
    Therefore, to obtain the entropy, one does not evaluate the area operator on a single given surface in the bulk, but a number of potentially different minimal surfaces determined by the state, and average their areas according to $P$. \\
    We note also that, while the area operator here is of the same form as the LQG area operator (for the diagonal series of eigenvalues)
    \begin{equation}
        \hat{A}^{LQG}_S := \sum_{e\in S} \hat{A}^{LQG}_e := \sum_{e\in S} \sum_{j_e} \sqrt{j_e(j_e+1)} \ibb_{e,j_e},
    \end{equation}
    it differs in the area values themselves: $a_{e,LQG}(s_e) = \sqrt{s_e(s_e+1)} $ does not even match in terms of scaling. Therefore, we can not naively identify the Ryu-Takayanagi graph-area operator above with the discrete geometric LQG area operator. If we want to do such an identification, there must be a difference between the `graph spins' $ j_e $ in our construction and the `LQG spins' $s_e$ such that the two match
    \begin{equation}
        \sqrt{s_e(s_e+1)} \approx \log(2j_e+1)
    \end{equation}
    at high values of the graph spins $j_e$. 
    An identification like this is routine in tensor network models of holography - a priori, to interpret the logarithms of bond dimensions as areas, a relation between the two must be stipulated. This can for example be done by embedding the tensor network graph in an ambient metric space and matching bond dimensions to the areas of dual surfaces in the space. Here, we can instead stipulate the matching through the microscopic LQG area operator and even give it a preliminary interpretation. We also note that this is independent from the question of superpositions and already features in the same way on a single link with fixed spin.\\
    This sort of matching makes sense if we see $d_{j_e}$ as an `effective bond dimension' of a system which subsumes many coarse-grained degrees of freedom, whereas the $s_e$ describe the `microscopic' geometry in terms of LQG's areas. The area operator appearing on the right hand side of the Ryu-Takayanagi formula should then be understood as a coarse-grained one. In fact, from this point of view it seems most plausible to interpret the graph states themselves as a sort of coarse-grained spin networks. \\
    
    It needs to be stressed that the nature of the averages $\avg{-}_U$ and $\avg{-}_P$ is quite different. The former is, effectively, a product of (uniform) averages over the Hilbert spaces $\hbb_x$ and a classical one, in that whatever quantity $\Hat{X}$\footnote{For clarity of exposition we use hats to denote operators in this argument to distinguish them from c-numbers.} we compute in it (i.e. any average expectation value) corresponds to an ensemble average of complex numbers
    \begin{equation}
        \avg{\Tr[\Hat{\rho} \Hat{X}]}_U = \sum_\psi P_\psi X_\psi\qquad X_\psi = \langle\psi|\Hat{X}|\psi\rangle = \Tr[\ket{\psi}\bra{\psi} \Hat{X}]
    \end{equation}
    \textit{each of which is a quantum expectation value}, and the probability $P_\psi$ is the uniform distribution. As far as the meaning of the average $\avg{-}_P$ is concerned, first note that we can write $p$ as
    \begin{equation}
        p_{\ind{j}} = \frac{K_{\ind{j}}}{\sum_{n}K_{\ind{n}}}\qquad K_{\ind{j}} = \Tr_{\hbb_{\ind{j}}}[\Hat{\Pi}] = \dim(\hbb_{\ind{j}})
    \end{equation}
    with the operator $\Hat{\Pi}$ that brings us from the full Hilbert space to the constrained one. For example, in the case of constraining onto states with definite graph pattern, $\Hat{\Pi}$ works as a projector that conditions the quantum probabilities on the given graph pattern. Then, $p_{\ind{j}}$ can be understood as a kind of combinatorial probability for a given, uniform randomly chosen vector $\ket{\psi}\in \Hat{\Pi}(\hbb)$ to be in the subsector $\hbb_{\ind{j}}$: the larger the dimension of the sector, the larger the chance for a 1-dimensional subspace to be a part of it.     The interpretation of $\avg{-}_P$ is then the following. Given some operator $\Hat{X}$ on the system, we can write its average in the same way as above, 
    \begin{equation}
        \avg{\avg{\Hat{X}}_\rho}_U = \frac{Z_1}{Z_0} \qquad Z_1 = \avg{\Tr_{\Hat{\Pi}(\hbb)}[\Hat{\rho} \Hat{X}]}_U = \sum_{\ind{j}} \Tr_{\hbb_{\ind{j}}}[\Hat{\Pi}  \Hat{X}],
    \end{equation}
    with the \textit{same} $Z_0$ as above. Then approximately, this can again be written as
    \begin{equation}
        \avg{\avg{\Hat{X}}_\rho}_U = \sum_{\ind{j}} p_{\ind{j}} X_{\ind{j}} = \avg{X}_p
        \qquad 
        X_{\ind{j}} = \frac{\Tr_{\hbb_{\ind{j}}}[\Hat{\Pi} \Hat{X}]}{\Tr_{\hbb_{\ind{j}}}[\Hat{\Pi}]} = \avg{\avg{\Hat{X}}_{\rho_{\ind{j}}}}_{U_{\ind{j}}}
    \end{equation}
    so to get the \textit{average expectation value of $\Hat{X}$ in a pure state}, we take the probability of the pure state to be in sector $\ind{j}$ and weigh by the average expectation value in that sector. For this interpretation, we see the object $X$ as a random variable on the space of events given by the sectors $\ind{j}$, with values $X_{\ind{j}}$. To any quantum operator $\Hat{X}$, we can associate such a classical random variable $X$. \\
    The extension to operators on two copies of the system brings with it the modification that the weights $X_{\ind{j},\ind{k}}$ are now dependent on a ground-state configuration $\sigma_{GS}$ of the Ising model, but otherwise the interpretation is the same. We can therefore see that the average $\avg{-}_\rho$ is quantum, $\avg{-}_U$ is classical and statistical and $\avg{-}_P$ is classical and combinatorial.\\
    
    We can also derive general necessary conditions for the purity to be minimal. To be more precise, consider that the condition on purity and therefore the isometry condition may be written as
    \begin{equation}\label{IsomRewriteTrick}
        \avg{p,M p} = \sum_{\ind{j},\ind{k}} p_{\ind{j}} M_{\ind{j}\ind{k}} p_{\ind{k}} = 0 \qquad M_{\ind{j}\ind{k}} = Z^{(\ind{j},\ind{k})}_{1} - \frac{1}{D_I}
    \end{equation}
    where $D_I $ is the total input dimension.  The special form of the matrix M allows us to get an idea of when this is the case. We may sketch the argument already at this stage, without referring to specific situations of interest. We can calculate the determinant using the matrix $Z_1$ with entries $Z^{(\ind{j},\ind{k})}_{1}$:
    \begin{align}
        \det(M) &= \det(Z_1  - \frac{1}{D_I}\Vec{1}^T \otimes \Vec{1} )\\
            &= \det(Z_1) (1 - \frac{1}{D_I}\Vec{1}^T (Z_1)^{-1} \Vec{1} )\\
            &= \det(Z_1) (1 - \frac{1}{D_I} \sum_{\ind{j},\ind{k}} (Z_1)^{-1}_{\ind{j}\ind{k}})
    \end{align}
    So assuming $\det(Z_1)\neq 0$, we write it as a diagonal, invertible part $W$ with entries $W^{\ind{j}} = Z^{\ind{j},\ind{j}}_1$ plus a matrix with empty diagonal. Then, factoring out the diagonal we obtain the form $Z_1 = W(\mathbb{I} + \alpha )$. We can then solve for the inverse of $Z_1$ and expand the above expression. At first order, $\det(\mathbb{I} + \alpha ) \approx e^{\Tr(\alpha)} = 1$ and $(Z_1)^{-1} \approx (W)^{-1} (\mathbb{I} - \alpha )$, meaning the above is
    \begin{align}
        \det(M) &= \det(W) (1 - \frac{1}{D_I} \sum_{\ind{j},\ind{k}} (W^{\ind{j}})^{-1} (\delta_{\ind{j},\ind{k}}-\alpha_{\ind{j}\ind{k}})) = \\
        &= \det(W) (1 - \frac{1}{D_I} \sum_{\ind{j}} (W^{\ind{j}})^{-1})
        + \frac{\det(W)}{D_I} \sum_{\ind{j},\ind{k}} (W^{\ind{j}})^{-1} \alpha_{\ind{j}\ind{k}}   
    \end{align}
           
    Then, the necessary condition at zeroth order in $\alpha$ is $\sum_{\ind{j}} (W^{\ind{j}})^{-1}) = D_I$. 
    If, schematically, $D_I = \sum_{\ind{j}} D_{I_{\ind{j}}} $, then $W^{\ind{j}} = \frac{1}{D_{I_{\ind{j}}}}$ fulfils the condition and the vector $p_{\ind{j}} = \frac{D_{I_{\ind{j}}}}{D_{I}}$ is a solution to Eq.~\eqref{IsomRewriteTrick}.
    Therefore, for sector-diagonal $Z_1$, we know a necessary and sufficient criterion on the elements of the partition sum to fulfil the isometry condition: 
    \begin{equation}
        p_{\ind{j}} = \frac{D_{{\ind{j}}}}{D} \stackrel{!}{=} \frac{D_{I_{\ind{j}}}}{D_{I}}
    \end{equation}
    which requires that the \textit{output dimensions} $D_{O_{\ind{j}}}$ are constant across all sectors, $D_{O_{\ind{j}}} \equiv D_{O}$ in order to factor out of the fraction. We will see this again later.
    \\


\section{Isometry conditions on superposed spin networks}\label{conditionchapter}
    \subsection{Bulk-to-boundary maps}
        In seeking holographic behaviour, we are concerned with the equivalence of a bulk and a boundary space. The setting for us is to consider a fixed graph connectivity $\gamma$ and a state in the Hilbert space
        \begin{equation}
            \hbb_{\gamma} = \bigoplus_{j_{\bnd}} \mathcal{I}_{j_{\bnd}}\otimes V_{j_{\bnd}}.
        \end{equation}
        We have a nontrivial center given by sector-diagonal operators $ \sum_{j_{\bnd}} \lambda_{j_{\bnd}} \mathbb{I}_{j_{\bnd}} $. Our interest will be in determining which choices of connectivity result in quantum channels that are isometric between the spaces $\bigoplus_{j_{\bnd}}\mathbf{B}(\mathcal{I}_{j_{\bnd}})$ and $\bigoplus_{j_{\bnd}}\mathbf{B}(V_{j_{\bnd}})$. In this bulk-to-boundary mapping case, we will call such a channel, and by proxy the state from which it arises, \textit{holographic}.
        The average purity of the reduced bulk state in the high spin regime is expressed as 
        \begin{equation}
            \avg{e^{-S_2(\rho_B)}}_U=\sum_{\ind{j},\ind{k},\Vec{\sigma}}p_{\ind{j}}p_{\ind{k}} \Delta_{1}(\ind{j},\ind{k};\Vec{\sigma}) e^{-H_{1} (\ind{j},\Vec{\sigma})}
        \end{equation}
        with all quantities defined in detail in appendix~\ref{Appendix:RIMBb}.\\
        We will now present a sufficient condition on the graph data, and set of input sectors $\mathcal{W}$, such that the isometry condition is fulfilled. Unless a sector $\ind{j}$ is excluded by having $\prod_{e\in\gamma} |g_{j_e}|^2 = 0$,
        \begin{enumerate}
            \item $\forall \ind{j}:\, \ind{j}_{\bnd}  \in \mathcal{W}$, \,\, $\Vec{\sigma} = +1$ is the minimum of $H_1(\ind{j},\Vec{\sigma})$.
            \item $\forall \ind{j}:\, \ind{j}_{\bnd}  \in \mathcal{W}$, \,\, $\prod_{e\in\gamma} |g_{j_e}|^2 \prod_{e\in\bnd} d_{j_e} = C  $ with $C$ independent of $ \ind{j} $.
        \end{enumerate}
        That this condition is sufficient can be checked directly.\\
        In general, $\Delta_1$ allows only terms labeled by $(\ind{j},\ind{k},\Vec{\sigma})$ such that $S_\uparrow \subseteq G_{\ind{j},\ind{k}} := \{ x\in \gamma | \mathbf{j}^x=\mathbf{k}^x \}$. In particular, there are no restrictions when the sectors are equal. If $\Vec{\sigma} = +1$ is the ground state of $H_1(\ind{j},\Vec{\sigma})$, then we approximate $Z^{\ind{j},\ind{k}}_1$ by the term corresponding to it. However, when $\ind{j} \neq \ind{k}$,  $\Delta_1=0$ for this Ising configuration, implying that $Z^{\ind{j},\ind{k}}_1 = Z^{\ind{j},\ind{j}}_1 \delta_{\ind{j},\ind{k}}$, with $Z^{\ind{j},\ind{j}}_1 = W^{\ind{j}} = \frac{1}{\dim(\mathcal{I}_{\ind{j}})}$. Additionally, the second condition implies for the probability weights $p_{\ind{j}}$ that
        \begin{equation}
            p_{\ind{j}} = \frac{\dim(\mathcal{I}_{\ind{j}})
                \prod_{e\in\gamma} 
                |g_{j_e}|^2 \prod_{e\in\bnd} d_{j_e} 
                }{
                \sum_{\ind{k}}  \dim(\mathcal{I}_{\ind{k}})
                \prod_{e\in\gamma} 
                |g_{k_e}|^2 \prod_{e\in\bnd} d_{k_e} 
                }
                =
                \frac{\dim(\mathcal{I}_{\ind{j}})
                C
                }{
                \sum_{\ind{k}}  \dim(\mathcal{I}_{\ind{k}})
                C
                } = \frac{\dim(\mathcal{I}_{\ind{j}})}{
                \sum_{\ind{k}}  \dim(\mathcal{I}_{\ind{k}})}
        \end{equation}
        which is, as we discussed before, sufficient to reach minimal purity. This condition is in a sense the obvious one. It requires that all sectors which appear in the class of states must themselves be able to support holographic transport, and also constrains a little the weights by which they are superposed. However, this does not include the case where some sectors are not holographic by themselves, but their superposition is. But in fact, if the choice of data given by the cutoffs, the glueing pattern and the coefficients $g$ are such that the matrix of partition sums $Z^{\ind{j},\ind{k}}_1$ is approximately diagonal, then the general argument given before enforces this sufficient set of conditions on the sectors and makes it necessary. Therefore, to find more general sufficient conditions, we need to inquire how diagonal said matrix is.\\
        Let us again use the language of nontrivial centers to investigate this issue. In the following, we denote by $E$ the set of boundary spin values $j_{\bnd}$ which are values of the spectrum of the center of the algebra in our Hilbert space. We decompose our unnormalised states according to them:
        \begin{align}\label{BndSpinDecomp}
            \ket{\phi} = \langle \Gamma \ket{\Psi}, \qquad (\rho_{E,E})_I = \frac{\Tr_{O_E}[\ket{\phi}\bra{\phi}]}{\Tr_{E}[\ket{\phi}\bra{\phi}]}, \qquad c_E = \frac{\Tr_{E}[\ket{\phi}\bra{\phi}]}{\Tr_{}[\ket{\phi}\bra{\phi}]}
        \end{align}
        We will study the impact of the trace preservation conditions \ref{TracePreservation2} on these objects and the subsequent constraints on $Z^{\ind{j},\ind{k}}_1$. To begin, we first match the entropy calculations of the state $\rho_I$ and its sector components in order to derive the right type of replica trick.\\
         Using formula \ref{EntropySplit}, we connect the following expressions:
         \begin{align}\label{BndSpinDecompEntropy}
             &&e^{-S_2((\rho_{E,E})_I)} = \frac{\Tr_{\hbb^2_{E}}[(\ket{\phi}\bra{\phi})^{\otimes 2}\sw_{I_E}]}{\Tr_{\hbb^2_E}[(\ket{\phi}\bra{\phi})^{\otimes 2}]}&& c^2_E = \frac{\Tr_{\hbb^2_E}[(\ket{\phi}\bra{\phi})^{\otimes 2}]}{\Tr_{\hbb^2}[(\ket{\phi}\bra{\phi})^{\otimes 2}]} \\
             &&e^{-S_2(\rho_I)} = \frac{\Tr_{\hbb^2_{}}[(\ket{\phi}\bra{\phi})^{\otimes 2}\sw_{I}]}{\Tr_{\hbb^2}[(\ket{\phi}\bra{\phi})^{\otimes 2}]} = 
             &&  \sum_E \frac{\Tr_{\hbb^2_{E}}[(\ket{\phi}\bra{\phi})^{\otimes 2}\sw_{I}]}{\Tr_{\hbb^2}[(\ket{\phi}\bra{\phi})^{\otimes 2}]}
         \end{align}
         which implies that we should use the sector-wise swap operator $\sw_I = \sum_E \sw_{I_E}$ for this calculation. While it is diagonal in boundary spins $E$, it is not so in the bulk ones. As we will now see, the same property holds for the matrix of partition sums. We can rewrite the purity $e^{-S_2(\rho_I)}$ in the same Ising-oriented fashion as before, $
             \avg{e^{-S_2(\rho_I)}}_U \approx \frac{Z_1}{Z_0}$ , but with sums over $E$: 
         \begin{equation}
            Z_{1|0} = \sum_{E,\Tilde{E}} \Tr_{\hbb_E \otimes \hbb_{\Tilde{E}}}[ (\ket{\phi}\bra{\phi})^{\otimes 2} \sw^{1|0}_I ],
         \end{equation}
         which are \textit{diagonal in} $E,\Tilde{E}$ due to the special swap operator here. \\
         We may therefore write $Z_{1|0} = \sum_E \Bar{Z}^{E,E}_{1|0}$, by defining the `boundary-fixed' partition sums
         \begin{equation}
             \Bar{Z}^{E,\Tilde{E}}_{1|0} = \Tr_{\hbb_E\otimes\hbb_{\Tilde{E}}}[
             \avg{(\ket{\phi}\bra{\phi})^{\otimes 2}}_U \sw^{1|0}_I
             ]
         \end{equation}
         which are related to the previous partition sums by 
         \begin{equation}
             \Bar{Z}^{E,\Tilde{E}}_{1|0} =
             \sum_{j_B, k_B} K_{\ind{j}} K_{\ind{k}} Z^{(\ind{j},\ind{k})}_{1|0}
         \end{equation}
         where the full spin sectors $\ind{j}=j_B\cup E,\ind{k}=k_B\cup E$ are comprised of the bulk spin sets $j_B,k_B$ and the boundary spins $E$,
         leading to
         \begin{equation}
             \avg{e^{-S_2(\rho_I)}}_U \approx \sum_E (\frac{\Bar{Z}^{E,E}_0}{Z_0}) (\frac{\Bar{Z}^{E,E}_1}{\Bar{Z}^{E,E}_0}).
         \end{equation}
         We can furthermore identify, through quick calculation, that
         \begin{align}\label{BndSpinDecompSectorIdent}
            \avg{e^{-S_2((\rho_{E,E})_I)}}_U = \frac{\Bar{Z}^{E,E}_1}{\Bar{Z}^{E,E}_0} && &&\avg{c^2_E}_U = \frac{\Bar{Z}^{E,E}_0}{Z_0}.
         \end{align} 
         The only assumption that goes into this is that the calculation of the two Rényi purity is the same if calculated over the full graph Hilbert space or over its individual sectors, which is true before averaging and which is true after averaging iff
         \begin{equation}\label{LocalisationOfSectors}
             \avg{ c^2_E e^{-S_2((\rho_{E,E})_I)} }_U \approx \avg{c^2_E}_U \avg{ e^{-S_2((\rho_{E,E})_I)} }_U.
         \end{equation}
         So a certain type of localisation of the average is required, but we expect it also naturally happens in the regime of large spins. We stress that, the partition sums need not be diagonal in the bulk spins.\\ 
         It is noteworthy that the average value of $c_E$ can easily be computed:
         \begin{equation}
             \avg{c_E}_U \approx
             \frac{\Tr_{E}[\Pi_\Gamma \avg{\ket{\Psi}\bra{\Psi}}_U]}
             {\Tr_{}[\Pi_\Gamma \avg{\ket{\Psi}\bra{\Psi}}_U]} = \frac{\Tr_{E}[\Pi_\Gamma ]}
             {\Tr_{}[\Pi_\Gamma ]} = \frac{D_E}{\sum_F D_F} = \sum_{j_B} p_{j_B\cup E}.
         \end{equation}
         Matching this with the required value necessitates that $D_{O_E}$ is in fact independent of $E$. In our case, this means a restriction to boundary spins $j_\bnd$ such that
         \begin{equation}
             D_{O_{j_\bnd}} = \prod_{e\in\bnd} d_{j_e} = D_O \; \forall j_\bnd.
         \end{equation}
        So, we can only use certain sectors on the output side once we fix this value $D_O$. \textit{In the following, we will assume such a value has been fixed once and for all.}\\
         We have much more precise control over the isometry condition now. We can check it in every sector $E$ separately and it amounts to
         \begin{align}
            \frac{\Bar{Z}^{E,E}_1}{\Bar{Z}^{E,E}_0} = \frac{1}{\dim(\hbb_{I_E})}
            \qquad
            \frac{\Bar{Z}^{E,E}_0}{Z_0} = \frac{\dim(\hbb_{I_E})^2}{(\sum_{\Tilde{E}}\dim(\hbb_{I_{\Tilde{E}}}))^2} \qquad \forall E \in \mathcal{W} .
         \end{align}
         Crucially, \textit{these conditions are necessary and sufficient}. The calculation of these `boundary-fixed partition sums' $\Bar{Z}^{E,E}_{1|0}$ can again be done through dual Ising models. These conditions can be further reformulated to yield useful constraints. \\
         Assuming the first condition, we can calculate what the second is: 
         \begin{equation}
             \frac{\Bar{Z}^{E,E}_0}{
             Z_0
             } =
             \frac{\Bar{Z}^{E,E}_1 D_{I_E}}{
             \sum_F \Bar{Z}^{F,F}_1 D_{I_F}
             }=
             D_{I_E}^2\frac{ 1}{
             \sum_F D_{I_F}D_{I_E}\frac{\Bar{Z}^{F,F}_1}{\Bar{Z}^{E,E}_1}
             }
         \end{equation}
         We can see from the expression in the denominator that we need $\frac{D_{I_E}}{\Bar{Z}^{E,E}_1} =\frac{1}{q}$, with $q$ constant, in order to achieve the second condition. This in turn means with the first condition
         \begin{equation}
             \Bar{Z}^{E,E}_1 = q D_{I_E} = k \frac{D_E^2}{D_{I_E}}  \qquad \Bar{Z}^{E,E}_0 = q D_{I_E}^2 = k D_E^2 \;\;\forall E\in \mathcal{W}, \qquad q = k D_O^2
         \end{equation}
         These are necessary conditions on all sectors which may be included in the input algebra $\acal_I$, so the set $ \mathcal{W}$. In fact, they are equivalent to the other set of conditions identified earlier.
         
        With inspiration from the single-vertex case (see \ref{appendix:SingleVertex} for details), we can make further clear the role of all these constraints. Define the new objects
        \begin{equation}
            Y^E_{1|0} = \frac{1}{D_E^2} \bar{Z}^{E,E}_{1|0} = \sum_{j_B,k_B} L^E_{j_B}L^E_{k_B} Z^{\ind{j},\ind{k}}_{1|0} \qquad K_{\ind{k}} = D_E L^E_{k_B}
        \end{equation}
        for which the isometry conditions become
        \begin{equation}
            Y^E_1 = k\frac{1}{D_{I_E}} \qquad Y^E_0 = k \;\;\forall E \in \mathcal{W}.
        \end{equation}
        Let us summarise the constraints we have from isometry or trace preservation. On the partition sums $Z^{\ind{j},\ind{k}}_{1|0}$, which represent the individual-sector data, there is the maximal entropy constraint for each Ising model. This can also pose constraints on the factors $K_{\ind{j}}$. For a single sector, this is all there is. On those factors as well as the dimensional data, there is the further restriction for multiple sectors that the output dimension must be fixed across all sectors, and possibly more subtle constraints.\\
        Quite luckily, if the output dimension must be manually fixed to be constant across all sectors of the problem, this reintroduces a notion of \textit{scale} into the discussion. What this means is that we can once again speak of low-energy limits for the Ising model in a sensible manner.\\
        For this purpose, divide all the couplings in the ising model by $\beta = \log(D_O)$. Then we can perform a universal low-temperature approximation on $Y^E_{1|0}$ by sending $\beta\rightarrow \infty$. \\
        To be more precise, let us write out the full dependence of the partition sums in terms of these quantities:
        \begin{equation}
            Z_{1|0} = \sum_{E} e^{2\beta} D_{I_E}^2 \sum_{j_B,k_B} L^E_{j_B}L^E_{k_B} \sum_{\Vec{\sigma}} \Delta_{1|0}(\ind{j},\ind{k};\Vec{\sigma}) e^{-\beta \Tilde{H}_{1|0}(\ind{j},\ind{k};\Vec{\sigma})}
        \end{equation}
        where we have rescaled $\Tilde{H}_{1|0} = \frac{H_{1|0}}{\beta}$ by rescaling the couplings $\Tilde{\lambda}_e = \frac{\lambda_e}{\beta},\Tilde{\Lambda}_x = \frac{\Lambda_x}{\beta}$. For the Ising models we consider for bulk-to-boundary and boundary-to-bulk, the $L$-factors are independent of $\beta$. Furthermore, there are subtle, indirect dependencies on $\beta$ in $D_{I_E},\Delta$ which come from  how many choices of $E$ there are for a given value of $\beta$, but we neglect these here, assuming that in the low-temperature limit these do not matter so much.\\
        Then, that limit is dominated by the lowest energy configurations of $\Tilde{H}_{1|0}$ and the combinations of $j_B,k_B$ which minimise it furthest. We can find estimates for these quite easily. First, take into account the constraints from $\Delta$. Let 
        \begin{equation}
            G_{\ind{j},\ind{k}} = \{x\in V_\gamma : \mathbf{j}^x = \mathbf{k}^x\}
        \end{equation}
        be the vertex set where the constraints do not change anything. Then for the different Hamiltonians, the constraints imply
        \begin{equation}
            \Tilde{H}_{1}: S_\uparrow \subseteq  G_{\ind{j},\ind{k}} \qquad 
            \Tilde{H}_{0}: S_\downarrow \subseteq  G_{\ind{j},\ind{k}}
        \end{equation}
        or else the configuration does not contribute.
        We can then compare the values of the Hamiltonians in the all-up or all-down configurations (subject to the constraints) to see which corresponds more to a minimum:
        \begin{align}
            \Tilde{H}_{1}(+) = \sum_{e \in \partial 
 G_{\ind{j},\ind{k}} \setminus \bnd}\Tilde{\lambda}_e +\sum_{x \in G_{\ind{j},\ind{k}}}\Tilde{\Lambda}_x &  &\Tilde{H}_{1}(-) = 1 \\
            \Tilde{H}_{0}(+) = 0&  &\Tilde{H}_{0}(-) = 1 + s_{\ind{j}}
        \end{align}
        where $s_{\ind{j}} = \frac{\log(D_{\ind{j}})}{\log(D_O)}$ and the reduced couplings are given as $\Tilde{\lambda}_e = \frac{\log(d_{j_e})}{\log(D_O)},\Tilde{\Lambda}_x = \frac{\log(D_{\mathbf{j}^x})}{\log(D_O)}$. We assumed here that the "all-up" configuration for $ \Tilde{H}_{1}$ is up on $G_{\ind{j},\ind{k}}$ and down elsewhere. This does not necessarily give the minimal energy configuration, but is a good approximation to it. In particular, when $\ind{j}=\ind{k}$, it reduces to $\Tilde{H}_{1}(+) = s_{\ind{j}}$. 
        We can think of the contribution from $\partial 
 G_{\ind{j},\ind{k}} \setminus \bnd$ as being the analogue of a \textit{bulk geometry-dependent minimal surface} - but we are calculating the entropy of the bulk here, so no such surface is attached to any boundary region and there is no meaning of RT formulas.\\
        Notably, the all-up configuration is always the minimum of $\Tilde{H}_{0}$, so the assumption $Z^{\ind{j},\ind{k}}_{0}\approx 1$ holds well.\\
        We can then approximate the partition functions as
        \begin{equation}
            Y^{E}_{1|0} =  \sum_{j_B,k_B} L^E_{j_B}L^E_{k_B} e^{-\beta E_{1|0}(\ind{j},\ind{k})}
        \end{equation}
        which for $Z_0$ gives the previous estimate and condition for isometry, and for $Z_1$ selects a certain type of contribution of minimal energy $E_1(j,k)$. Which combinations of $j_B,k_B$ give the lowest energy? This all depends a lot on the size, shape and values of spins in $G_{\ind{j},\ind{k}}$. However, we can think of two extreme cases for illustration. When the spin sets are equal, $G_{\ind{j},\ind{k}} = V_\gamma$ and so $E_1 (j,j) = r_E = \frac{D_{I_E}}{D_{O_E}}$. On the other hand, if the spin sets are nowhere equal, $G_{\ind{j},\ind{k}} = \emptyset$ and no configurations with up-spins are allowed. Instead, consider spin sets which are equal at a \textit{single} vertex $\{z\}= G_{\ind{j},\ind{k}}$, for which then 
        \begin{equation}
            E_1(j,k) = \Tilde{\Lambda}_z + \sum_{e \cap z, \slashed{\in} \bnd}\Tilde{\lambda}_e 
        \end{equation}
        so it once again depends on the values of the spins at hand at any given vertex. However, it seems feasible that such an `off-diagonal' contribution might be smaller than the `diagonal' one, if the values of the spins are not too large. In the following, we will simply assume that there is a number $g_E$ of combinations $(j_B,k_B,\Vec{\sigma})$ for which $\Tilde{H}_{1}$ is minimal at value 
        \begin{equation}
            E_{min,E} = \sum_{e\in S_{min,E}}\Tilde{\lambda}_e + \sum_{x\in \Sigma_{min,E}}\Tilde{\Lambda}_x
        \end{equation}
        for some bulk region $\Sigma_{min,E}$ and a boundary segment $S_{min,E}$ of it. Then 
        \begin{equation}
            Y^{E}_{1} \approx  g_E (2 - \delta_{j_{B,min},k_{B,min}}) L^E_{j_{B,min}}L^E_{k_{B,min}} e^{-\beta E_{min,E}}
        \end{equation}
        \begin{equation}
            Y^E_0 \approx (\sum_{j_B} L^E_{j_B})^2 = (\sum_{j_B} \prod_{e\in\gamma} |g_{j_e}|^2)^2 = (\prod_{e\in\gamma} 
            \sum_{j_e}|g_{j_e}|^2)^2 = 1
        \end{equation}
        and the next-to-leading term will be exponentially suppressed. As we can see, the condition on $Y^E_0$ is generically fulfilled in the bulk-to-boundary model because $|g_{j_e}|=1$ for all boundary links, making the value independent of $E$.\\
        So we are now in a position to give general conditions for isometry to happen:
        \begin{enumerate}
            \item First, the input and output algebras must be chosen such that the output dimension in each sector is a fixed $D_O$, which we take here to be quite large.
            \item We assume the localisation of Eq.~\eqref{LocalisationOfSectors}, which we conjecture to naturally happen in the regime of high spins.
            \item The input Rényi-2 entropy in each \textit{boundary} sector is maximal.
        \end{enumerate}
        This means we can, in this setting of fixed output dimension, check the degree of isometry purely by finding the minimiser $E_{min,E}$ and its degeneracy $g_E$, meaning we are just looking at an Ising model with extended set of variables $(j_B,k_B,\Vec{\sigma})$ and finding its ground state. The problem is particularly simple because the minimal value is most likely the one of configurations where the bulk spins are close to the lower cutoff . This of course introduces a tension between the approximations made. We need a high $D_O$ to perform the approximation, and require all spins to be large enough for the unitary average to localise, but still the minimal configuration will be the one with the smallest possible spins.\\
        As the single-bulk-link example in appendix~\ref{appendix:SingleBulkLink} shows, the third of the three conditions implies restrictions on the weights $g_j$ used to define the state $\rho$, of the form
        \begin{equation}
            \prod_{e\in\gamma}|g_j|^2 \sim \sqrt{D_{I_E}}.
        \end{equation}
Let us summarise the consequences of these results for the class of states themselves. In our setting, we consider PEPS-like spin network superpositions $\ket{\phi_\gamma}$ whose spins may take values between $\mathbb{j}$ and $J$. If we further consider the (sufficient) restrictions
\begin{enumerate}
    \item For \textit{any} sector $\ind{j}$ that features in  $\ket{\phi_\gamma}$, the boundary spins take the same \textit{total} value, i.e. $A_{\partial\gamma} = \sum_{e\in\bnd}\log(d_{j_E}) $ is independent of the sector.
    \item Fix any set of boundary spins $E=\{j_e\}$ in accordance to the above. There is a unique pair of spin sectors $(\ind{j},\ind{k})$ matching the boundary condition such that the Ising Hamiltonian $H_1$ achieves its minimum, and the ground-state energy gives $\log(D_{I_E})$. In terms of spins, this requires as much inhomogeneity in the spins in $\ind{j}$ over the graph $\gamma$ as possible. More generally, require that the state peaks on such sectors.

\end{enumerate}
Then, provided we restrict the input and output algebras sufficiently, the state $\ket{\phi_\gamma}$ induces an approximately holographic mapping $\tcal
_{\ket{\phi_\gamma} }$.

    \subsection{Boundary-to-boundary maps}
        In contrast to the usual question of holography, which is concerned with equivalence of a bulk and a boundary space, we may consider the system as a transport between two complementary boundary regions, $I, O=I^c\subset \bnd$ and ask for equivalence of the boundary regions. For this, as we will see, we will need to fix a state of the bulk intertwiners $\ical_E$ in each boundary sector. This means restricting the set of graph states that are averaged over to have prescribed data on intertwiners.\\
        We see that the full graph Hilbert space splits as
        \begin{equation}
            \hbb_\gamma \cong \bigoplus_{E_I,E_O} \hbb_{I,E_I}\otimes\hbb_{O,E_O}\otimes\mathcal{I}_{E}.
        \end{equation}
        We will see the first as an input space, the second as the output space and the third as a bath or background. This means we choose the algebras
        \begin{equation}
            \mathcal{B}_I = \bigoplus_{E_I} B(\hbb_{I,E_I}) \qquad \mathcal{B}_O = \bigoplus_{E_O} B(\hbb_{O,E_O})
        \end{equation}
        with the obvious extension and partial traces
        \begin{equation}
            i_I(\sum_{E_I} X_{E_I}) = \sum_{E_I,F_O} X_{E_I}\otimes \ibb_{F_O}\otimes \ibb_{\mathcal{I}_{E=E_I\cup F_O}} 
        \end{equation}
        \begin{equation}
            P\Tr_I[\sum_{E_I, E_O,F_I , F_O}X_{E_I\cup E_O;F_I\cup F_O}] = \sum_{E_I} \sum_{F_O}\Tr_{
            \hbb_{O,F_O}\otimes\mathcal{I}_{E_I\cup F_O}
            }[X_{E_I\cup F_O;E_I\cup F_O}]
        \end{equation}
        which are again adjoints. We routinely write $E=E_I\cup E_O$ as the sector whose component boundary spins are given by those in the region $I$ and $O$ together.\\
        Therefore, we can then again define a Choi map for any density matrix $\rho$ on the full system,
        \begin{equation}
            \tcal_\rho (X) = K P\Tr_O[i_I(X)\rho^{t_I}]  = K \sum_{E_O} \sum_{E_I}\Tr_{\hbb_{I,{E_I}}} [X_{E_I}\sigma_E]
        \end{equation}
        with the effective system density matrix in sector $E$
        \begin{equation}
            \sigma_E = \Tr_{\mathcal{I}_E}[\rho^{t_{\hbb_I}}] \in B(\hbb_E) = B(\hbb_{I,E_I})\otimes B(\hbb_{O,E_O}) .
        \end{equation}
        We can derive the applicable isometry condition here more easily because the effective system has a factorised Hilbert space
        \begin{equation}
            \hbb_{IO} = \hbb_I\otimes\hbb_O = \bigoplus_{E_I,E_O} \hbb_{I,E_I}\otimes\hbb_{O,E_O}.
        \end{equation}
        We simply see the effective system density as one on this system:
        \begin{equation}
            \sigma = \sum_{E_I,E_O} \sigma_{E_I\cup E_O} \in B(\hbb_{IO}) \cong \bigoplus_{E_I,E_O;F_I,F_O}B(
            \hbb_{I,E_I}\otimes\hbb_{O,E_O}
            ,\hbb_{I,F_I}\otimes\hbb_{O,F_O})
        \end{equation}
        This lets us write the Choi map as
        \begin{equation}
            \tcal_\rho (X) = K \Tr_{\hbb_I}[(X\otimes\ibb_O) \sigma]
        \end{equation}
        and we can apply the standard bipartite, centerless isometry condition
        \begin{equation}
            |K|^2 \Tr_{\hbb^{\otimes 2}_O}[ \sigma^{\otimes 2} \sw_{\hbb^{\otimes 2}_O}] = \sw_{\hbb^{\otimes 2}_I}.
        \end{equation}
        This becomes particularly easy for pure states $\sigma$ (equivalently, $\sigma_E$), where it is equivalent to trace preservation:
        \begin{equation}
            \frac{(\sigma)_I}{\Tr[\sigma]} = \frac{\ibb_{\hbb_I}}{\dim(\hbb_I)}\qquad K = \frac{\dim(\hbb_I)}{\Tr[\sigma]}.
        \end{equation}
        Of course, purity of $\sigma$ requires the state of the full system to factorise
        \begin{equation}
            \rho_{E,E} = \ket{\phi_E}\bra{\phi_E}\otimes\rho_{\mathcal{I},E}
        \end{equation}
        and is not very generic (as by Page's argument), unless either the effective system size or the environment size is very small compared to the other. Still, it is a generic result of a projective measurement on the effective system $\hbb_{IO}$ conditioned on some outcome, if one sees $\mathcal{I}$ as a kind of environment coupled to the effective system~\cite{breuerNonMarkovianDynamicsOpen2016}. In general, we know that for mixed states, trace preservation and isometry are \textit{mutually exclusive.} We will for now assume that such a factorisation holds and proceed.\\
        To restrict the average to states of this factorised form, we introduce another projector for the bulk information. First, notice that with the states we consider, $\rho_{E,E}\sim \ket{\psi_{\gamma,E}}\bra{\psi_{\gamma,E}}$ is always pure. Then, $\rho_{\mathcal{I},E}= \ket{\zeta_E}\bra{\zeta_E}$ must also be pure. We will \textit{fix a single pure state} $\ket{\zeta_E}$ in each sector, and the projector to this is then simply
        \begin{equation}
            \Pi_B = \sum_{E} \frac{\ket{\zeta_E}\bra{\zeta_E}}{\langle \zeta_E\ket{\zeta_E}} \otimes  \ibb_{IO,E}.
        \end{equation}
        The reason we fix a single one is that this reduces the effective dimension of the `environment' $\mathcal{I}_E$ for the channel to the minimum. This is one of the necessary requirements to find isometry for this tripartite case.\\
        So we project the randomised graph state $\ket{\psi_\gamma}$ into a fixed configuration of intertwiners $\zeta_E$ in each sector. This way, we obtain the pure state required to study isometry conditions the usual way. We can then write the second Rényi entropy as before, with
        \begin{equation}
            \avg{e^{-S_2(\sigma_I)}}_U \approx \frac{Z_1}{Z_0}.
        \end{equation}
        In this boundary-to-boundary mapping case, we will call such a channel, and by proxy the state from which it arises, \textit{transparent} if it is isometric.\\
        The purity of the reduced input state is, through calculations from appendix~\ref{Appendix:RIMbndbnd}, in the high-spin regime
        \begin{align}
            \avg{e^{-S_2(\sigma_I)}}_U &= \sum_{\ind{j},\ind{k},\Vec{\sigma}} P(\ind{j},\ind{k}) \Delta_1(\ind{j},\ind{k};\Vec{\sigma}) e^{-H_1(\ind{j},\ind{k};\Vec{\sigma})} \\
            &=\sum_{\ind{j},\ind{k}}\prod_{e\in\gamma }p_{j_e}p_{k_e}\sum_{\Vec{\sigma}} \cos{\theta(\ind{j},\ind{k};\Vec{\sigma})} e^{-\frac{1}{2}H_{\ind{j}}(\Vec{\sigma})} e^{-\frac{1}{2}H_{\ind{k}}(\Vec{\sigma})} \delta_{\ind{j}_\downarrow,\ind{k}_\downarrow}\delta_{j_{\partial S_\downarrow},k_{\partial S_\downarrow}}
        \end{align}
        which is a sum of nonnegative numbers and we take 
        \begin{equation}
            H_{\ind{j}} =  \sum_{e\in E_\gamma} \lambda_e \frac{1-\sigma_{s(e)}\sigma_{t(e)}}{2} + S_2(X(\ind{j}_{\bnd},\ind{j}_\uparrow))
        \end{equation}
        as an effective Hamiltonian. The form here is more complicated than that of the bulk-to-boundary case: the $\Delta$-factors are no longer in general boolean, but can take any value between $0$ and $1$. 
        We can take a limit of $\beta = \dim(\hbb_O)$ being very large, for fixed $r = \frac{\dim(\hbb_I)}{\dim(\hbb_O)}$. Then, we can drop all of the subdominant terms in the sum and approximate
        with the most dominant contribution $\Vec{\sigma}_{\text{min}} $. \\
        To begin, we study the behavior of the $\Delta$-factor. Of course, we must require as usual that
        \begin{equation}
            S_\downarrow \subseteq G_{\ind{j},\ind{k}}
        \end{equation}
        but the cosine factor also has an influence. For example, the all-up configuration $\Vec{\sigma} = \Vec{1}$ has $\cos{\theta(\ind{j},\ind{k};\Vec{\sigma})} = 1$ for all sectors. 
        For a given sector $\ind{j}$, we can evaluate the Hamiltonian on the extreme configurations for a hint on what the minimum might be:
        \begin{equation}
            H_{\ind{j}}(+1) = \sum_{e\in I} \lambda_e = \log(\dim(\hbb_{I,\ind{j}_I})
            \qquad 
            H_{\ind{j}}(-1) = \sum_{e\in O} \lambda_e + \log (
            \frac{\langle\zeta_{\ind{j}}\ket{\zeta_{\ind{j}}}}{\langle\zeta_E\ket{\zeta_E}}
            )
        \end{equation}
        
        We can study, for example, the case of a single vertex. This is very simple as there are no bulk spin sums involved, and the entropy-like quantities $\Sigma=0= S_2(X)$ vanish. This translates into the cosine $\cos{\theta}=1$ being 1 for all configurations and choices of spins, so the overall sum collapses into
        \begin{align}\label{btobSingleVertex}
            \avg{e^{-S_2(\sigma_I)}}_U 
            &= \sum_{\mathbf{j},\mathbf{k}} p_\mathbf{j} p_\mathbf{k} \sum_\sigma 
             e^{-\frac{1}{2}H_{\ind{j}}(\Vec{\sigma})} e^{-\frac{1}{2}H_{\ind{k}}(\Vec{\sigma})} \delta_{\ind{j}_\downarrow,\ind{k}_\downarrow}\\
             &= (\sum_{\mathbf{j}} p_\mathbf{j}e^{-\frac{1}{2}H_{\ind{j}}(+1)})^2 + \sum_{\mathbf{j}} p_\mathbf{j}^2e^{-H_{\ind{j}}(-1)}
        \end{align}
        
        This can be simplified by noting that the dimensionalities factorise and sum nicely:
        \begin{equation}
            D_{\mathbf{j}} = D_{I_{\mathbf{j}_I}} D_{O_{\mathbf{j}_O}} \implies \sum_\mathbf{j} D_\mathbf{j} = (\sum_{\mathbf{j}_I} D_{I_{\mathbf{j}_I}})(\sum_{\mathbf{j}_O}D_{O_{\mathbf{j}_O}})
        \end{equation}
        This leads to the expression
        \begin{align}
            \avg{e^{-S_2(\sigma_I)}}_U  
            &=
            \frac{1}{D_I} \left[
            \left(
            \sum_{E_I}\sqrt{\frac{D_{I_{E_I}}}{D_I}}
            \right)^2 + 
            \frac{1}{D_O} 
            \sum_{E_I} \left(\frac{D_{I_{E_I}}}{D_I}\right)^2
            \right]\\
            &= 
            \frac{1}{D_I} \left[
            e^{S_{\frac{1}{2}}(a) } + \frac{1}{D_O} e^{-S_2(a)}
            \right]
        \end{align}
        which is as simple as it gets. We can recognise the two terms as Rényi entropies $S_\alpha$ of the sequence $a_{\mathbf{j}_I} = \frac{D_{I_{\mathbf{j}_I}}}{D_I}$, allowing for the rewriting in the second line.
        We can see immediately that if the normalised sequence $a$ is constant, the term in the square brackets will be greater than 1 just from the first term. On the other hand, if it is peaked on exactly one sector, it returns $1+\frac{1}{D_O}$. This shows a general criterion for isometry:
        \begin{enumerate}
            \item we require large total output dimensions $D_O$;
            \item we require the sequence $a$ to be strongly peaked around one or at most a few input sector.
        \end{enumerate}
        For the single vertex, this is all we need. If we extend our study to a single internal link, we find constraints on the relative weights of different sectors in the bulk states, $\ket{\zeta_E}$, as seen in appendix~\ref{appendixExampleCalc}. We postpone a closer investigation of these constraints to future work.

To summarise once more: if we want to see the graph state $\ket{\phi_\gamma}$ as inducing an isometric boundary-to-boundary map, the following conditions must be satisfied (on top of the usual one on the spins size):
\begin{enumerate}
    \item the correlations between boundary data $V_E$ and intertwiner data $\ical_E$ in each boundary sector must be negligible;
    \item the graph state must be peaked on only a few sets of spin sectors on the input boundary region.  
\end{enumerate}


\section{Outlook}\label{conclusion}
In this work, we generalised earlier analyses on the holographic properties of quantum geometric states, i.e. spin networks, to the case of arbitrary superposition of algebraic (quantum geometric) data, for a fixed support graph. As in earlier work, we have done so via random tensor network techniques, taking advantage of the fact that spin networks are in fact generalised tensor networks themselves. This required extending appropriately such random tensor network techniques. In particular, while earlier work focused on unique, tunable bond dimensions, the true quantum gravity setting requires Hilbert spaces involving different combinations of bond dimensions on the same link. This superposition brings with it several new phenomena, which we addressed, using more general definitions of holographic behaviour, with respect to the standard Hilbert space-based one. These new phenomena are as follows:
\begin{itemize}
    \item Several new options for averaging procedures with different eases of computation.
    \item A multitude of dual Ising partition sums in place of a single one. These depend on pairs of bond dimensions $(\ind{j},\ind{k})$, and are subject to special constraints $\Delta$ depending on the problem.
    \item Appearance of a sum-over-sectors in the full expression of the purity, which may be interpreted as a kind of average in a distribution $p$ determined by the bond dimensions and other input data, through the $K$-factors.
    \item An `averaged' RT formula~\eqref{AveragedRTFormula} with a nontrivial area operator that is not sharp unless very special conditions are fulfilled. 
    \item A nontrivial center of the algebra given by the boundary area operators. This requires a generalised definition of bulk-to-boundary mappings via Choi maps and proper choices of input/output algebras.
\end{itemize}
We provided general conditions, both necessary and/or sufficient, for an average state of a discrete quantum geometry (described as a generalised tensor network) to give an isometric map between operator algebras. This is, in a clear sense, a kind of holographic behaviour. The conditions we found all entail two key ingredients:
\begin{itemize}
    \item We need the total boundary area to be constant across all involved sectors.
    \item We need the bulk Hilbert spaces to be comparably small.
\end{itemize}
While the latter may be expected from dimensional reasons or by the usual picture of bulk code subspaces~\cite{Jia:2020etj,Qi:2017ohu,Yang:2015uoa}, the former is quite surprising. It arises as a consequence of the split of the isometry condition into seperate conditions for each sector, which can only hold in parallel iff the total boundary area is a given fixed value across all sectors.\\
We also considered boundary-to-boundary transport of data in the presence of a bulk state, giving rise to the notion of a \textit{transparent} state. This is a dual viewpoint to holography: instead of the bulk being equivalent to the boundary, the bulk here functions as a realisation of a mapping between two boundary regions. \\
In future work, we would like to explore various ways to expand on this setup:
\begin{itemize}
    \item It is expected from several discrete quantum gravity formalisms that superpositions of different graph structures must arise. Up until now, our formalism has been adapted to a fixed graph, and thus a highly non-trivial extension is required to tackle the truly general case of quantum geometry states.
    \item In particular, the Group Field Theory formalism for quantum gravity points to an extension to a setting where graph vertices can be created or destroyed in a (bosonic) second-quantised language, and graph vertices are unlabelled (or must be identified by dynamical, physical data).
    \item The whole analysis should be extended to the dynamical and (in some approximation spatiotemporal level. The kinematical structures exposed here are generically subject to gravitational constraints and quantum gravity dynamics, be it encoded in a canonical constraint operator or in covariant, path integral-like amplitudes. They should be subject also, upon adoption of physical frames, to a relational temporal evolution.
    \item For a proper spacetime geometric interpretation and a rendezvous with the AdS/CFT correspondence, we need a better handle on how to reconstruct spacetime information from our data possibly combining techniques from quantum gravity literature (e.g. from GFT cosmology) and the AdS/CFT one~\cite{Neiman:2017zdr,Raju:2016vsu,Hubeny:2012wa,calcagniGroupFieldCosmology2012a,Gerhardt:2018byq,Pithis:2019tvp,Assanioussi:2020hwf,marchettiEffectiveDynamicsScalar2022,Oriti:2021rvm}.
    \item It would be interesting to take the renormalisation group flow perspective~\cite{Lunts:2015yua,Lee:2013dln,Akhmedov:1998vf,deBoer:1999tgo,Heemskerk:2010hk,Faulkner:2010jy} on the AdS/CFT correspondence seriously also in our finite context, and build coarse-graining maps using the boundary-to-boundary mappings we introduced. 
\end{itemize}

Our work, and the many possible extensions of it, prove that the interface between quantum gravity and quantum information is a very fertile research ground, and the search for a local, microscopic origin of holography is a powerful motivation to explore it.





\clearpage
\appendix
\appendixpage
\addappheadtotoc

    
    \section{Derivation of random Ising model}
        \subsection{Bulk-to-boundary maps}\label{Appendix:RIMBb}
            Here we derive the random Ising model components necessary for the bulk-to-boundary mapping analysis. Crucially, we will need the projector
            \begin{equation}
                \Pi_\Gamma = \sum_{j_B} \Pi_{\Gamma,j_B} \qquad \Pi_{\Gamma,j_B} = \bigotimes_{e\in\Gamma } |g_{j_e}|^2 \ket{e_{j_e}}\bra{e_{j_e}}
            \end{equation}
            with the maximally entangled state \ref{maxentstate} on each glued link, in each sector, with the weight $|g_{j_e}|^2$.\\
            Then each term in the random Ising model becomes
            \begin{align}
                &\Tr_{\hbb_{\ind{j}}\otimes \hbb_{\ind{k}}} 
                [ (\Pi_{\Gamma,\ind{j}_B}\otimes \Pi_{\Gamma,\ind{k}_B})  
                \sw_{S_\downarrow} \sw^{1|0}_B ]\\
                =
                &\Tr_{\hbb_{\Gamma,\ind{j}_B}\otimes \hbb_{\Gamma,\ind{k}_B}} [
                    (\Pi_{\Gamma,\ind{j}_B}\otimes \Pi_{\Gamma,\ind{k}_B}) \bigotimes_{e \in \Gamma} \sw^{\frac{1-\sigma_{s(e)} \sigma_{t(e)}}{2}}_{e}
                ]\times \\
                &\times\Tr_{\mathcal{I}_{\ind{j}}\otimes \mathcal{I}_{\ind{k}}} [
                    \bigotimes_x \sw^{\frac{1-b\sigma_x}{2}}_{B,x}
                ]
                \Tr_{\hbb_{\bnd,\ind{j}_{\bnd}}\otimes \hbb_{\bnd,\ind{k}_{\bnd}}} [
                     \bigotimes_{e \in \bnd} \sw^{\frac{1-\sigma_{s(e)}}{2}}_{e}
                ] 
            \end{align}
            where $b= (-1)^{1|0}$ in $Z_{1|0}$.
            We can then evaluate each part seperately.
            For the link factor, it factorises over each internal link:
            \begin{align}
                \prod_{e\in\Gamma} 
                |g_{j_e}|^2|g_{k_e}|^2\Tr_{\hbb_{j_e}\otimes \hbb_{k_e}} [
                    (\ket{e_{j_e}}\bra{e_{j_e}}\otimes \ket{e_{k_e}}\bra{e_{k_e}})  \sw^{\frac{1-\sigma_{s(e)} \sigma_{t(e)}}{2}}_{e}
                ]
            \end{align}
            where we define $\hbb_{j_e} := V_{s(e),j_e}\otimes V_{t(e),j_e}$. This evaluates to
            \begin{equation}
                \prod_{e\in\Gamma} 
                |g_{j_e}|^2|g_{k_e}|^2 \left( \delta_{j_e,k_e} d^{-1}_{j_e} \right)^{\frac{1-\sigma_{s(e)}\sigma_{t(e)}}{2}}.
            \end{equation}
            Then, the intertwiner factor is similarly
            \begin{equation}
                \Tr_{\mathcal{I}_{\ind{j}}\otimes \mathcal{I}_{\ind{k}}} [
                    \bigotimes_x \sw^{\frac{1-b\sigma_x}{2}}_{B,x}]
                    = 
                \prod_x \mathcal{D}_{\mathbf{j}^x} \mathcal{D}_{\mathbf{k}^x} \left(
                \delta_{\mathbf{j}^x,\mathbf{k}^x}\mathcal{D}^{-1}_{\mathbf{j}^x}
                \right)^{\frac{1-b\sigma_x}{2}}
            \end{equation}
            where $\mathcal{D}_{\mathbf{j}^x} = \dim(\mathcal{I}_{\mathbf{j}^x})$. Lastly, the boundary factor is
            \begin{equation}
                \Tr_{\hbb_{\bnd,\ind{j}_{\bnd}}\otimes \hbb_{\bnd,\ind{k}_{\bnd}}} [
                     \bigotimes_{e \in \bnd} \sw^{\frac{1-\sigma_{s(e)}}{2}}_{e}
                ]  
                = 
                \prod_{e\in \bnd} d_{j_e} d_{k_e} \left(
                \delta_{j_e,k_e} d^{-1}_{j_e}
                \right)^{\frac{1-\sigma_{s(e)}\sigma_{t(e)}}{2}}
            \end{equation}
            where we set $\sigma_{t(e)}=1$ for all boundary links.\\
            This means we have as our random Ising model data the choices:
            \begin{equation}
                \Delta_{1|0}(\ind{j},\ind{k};\Vec{\sigma}) = 
                \prod_x \delta_{\mathbf{j}^x,\mathbf{k}^x}^{\frac{1-b\sigma_x}{2}}
                \prod_{e\in\gamma} \delta^{\frac{1-\sigma_{s(e)}\sigma_{t(e)}}{2}}_{j_e,k_e} 
            \end{equation}
            \begin{equation}
                K_{\ind{j}} = \prod_x \mathcal{D}_{\mathbf{j}^x}
                \prod_{e\in\gamma} 
                |g_{j_e}|^2 \prod_{e\in\bnd} d_{j_e} = \Tr_{\hbb_{\ind{j}}}[\Pi_{\Gamma,j_B}]
            \end{equation}
            where we fix the convention $g_{j_e}=1$ for all $e\in \bnd$ to write this uniformly across all links of the graph. The Hamiltonian is
            \begin{equation}\label{BulksToBndHammy}
                H_{1|0} (\ind{j},\ind{k},\Vec{\sigma}) = \sum_{e\in\gamma} \lambda_e \frac{1-\sigma_{s(e)}\sigma_{t(e)}}{2} + \sum_x \frac{1-b\sigma_x}{2} \Lambda_x
            \end{equation}
            with couplings $\lambda_e= \log(d_{j_e}), \Lambda_x = \log(\mathcal{D}_{\mathbf{j}^x})$. It is clear that this Hamiltonian is bounded from above and below, with upper bound $\sum_e \lambda_e +\sum_x \Lambda_x $.\\
            The diagonal partition sums $Z^{\ind{j},\ind{j}}_1$, can be reinterpreted through
            \begin{equation}
               Z^{\ind{j},\ind{j}}_1 = \avg{\Tr_{\hbb^{\otimes 2}_{\ind{j}}} 
               [ \left( \frac{ \Pi_{\Gamma,j_B} }{ K_{\ind{j}} } \right)^{\otimes 2} \left(\ket{\Psi}\bra{\Psi}\right)^{\otimes 2} \sw_B] }_U
                = \avg{\Tr_{\mathcal{I}_{\ind{j}}}[ \Tr_{\hbb_{\bnd,j_{\bnd}}\otimes\hbb_{\Gamma,j_B}}[\frac{\Pi_{\Gamma,j_B}}{K_{\ind{j}}} \ket{\Psi}\bra{\Psi}]^2]}_U
            \end{equation}
            as the reduced bulk entropy of a state in the given sector, so as
            \begin{equation}
               Z^{\ind{j},\ind{j}}_1 = \avg{e^{-S_2(\rho_{I,\ind{j}})}}_U  \qquad \rho_{I,\ind{j}} = \Tr_{ \hbb_{\bnd,j_{\bnd}}\otimes\hbb_{\Gamma,j_B} }
               [\frac{\Pi_{\Gamma,j_B} }{ K_{\ind{j}} } \ket{\Psi}\bra{\Psi} ]
            \end{equation}
            Therefore, the diagonal partition sums are in fact in the interval $[\frac{1}{\dim(\mathcal{I}_{\ind{j}})}, 1]$.

        \subsection{Boundary-to-boundary maps}\label{Appendix:RIMbndbnd}
             We have two projectors in this case, one for the intertwiners in the bulk $(B)$ and one for the connected links in the bulk $(\Gamma)$. They split over spin sectors
            \begin{align}
                 \Pi_B = \sum_{E}\Pi_{B,E} &&   && \Pi_\Gamma = \sum_{j_B} \Pi_{\Gamma,j_B}
            \end{align}
            where the sums $j_B$ are only over bulk spins, as designated by the fixed graph pattern. In the following, the set of bulk spins of a sector $\ind{j}$ will be denoted $\ind{j}_B$ and we will indicate the decomposition into bulk and boundary spin values as $\ind{j} = \ind{j}_B \cup \ind{j}_{\bnd}$. \\
            The Ising partition sums we are dealing with in this case are (up to the usual constant factors)
            \begin{align}
                Z_{1|0} &= \Tr_{\hbb_{IO}^{\otimes 2}}[\avg{\sigma^{\otimes 2}}_U \sw^{1|0}_{I^2}] = \sum_{E,F} \Bar{Z}^{E,F}_{1|0} \\
                &= \sum_{E,F} \Tr_{(\hbb_{IO,E}\otimes\mathcal{I}_E)\otimes(\hbb_{IO,F}\otimes\mathcal{I}_F)}[
                (\Pi_{B,E}\otimes\Pi_{B,F}) \avg{ ( \ket{\psi_\gamma}\bra{\psi_\gamma} )^{\otimes 2} }_U
                \sw^{1|0}_{I_{E_I}\otimes I_{F_I}}
                ].
            \end{align}
            The traces are over the full sectors $E,F$ each, so we can expand this into the usual form with a trace over $\hbb_{\ind{j}}$. The $\ket{\psi_\gamma}\bra{\psi_\gamma}$-part becomes a trace over the internal link spaces, but the spin sets are identified with the ones coming from the overall trace, due to $\sw^{1|0}_{I_{E_I}\otimes I_{F_I}}$ being diagonal in the bulk spin sets. This makes the sum simplify to
            \begin{equation}
                Z_{1|0} = \sum_{\ind{j},\ind{k},\Vec{\sigma}} \Tr_{\hbb_{\ind{j}}\otimes\hbb_{\ind{k}}}
                [
                (\Pi_{B,E}\otimes\Pi_{B,F}) (\Pi_{\Gamma,\ind{j}_B}\otimes \Pi_{\Gamma,\ind{k}_B})
                \bigotimes_x \sw_x^{\frac{1-\sigma_x}{2}}
                \sw^{1|0}_{I_{\ind{j}_{\bnd,I}}\otimes I_{\ind{k}_{\bnd,I}}}
                ]
            \end{equation}
            Each term in the random Ising model becomes
            \begin{align}
                &\Tr_{\hbb_{\ind{j}}\otimes \hbb_{\ind{k}}} 
                [
                (\Pi_{B,E}\otimes\Pi_{B,F})
                (\Pi_{\Gamma,\ind{j}_B}\otimes \Pi_{\Gamma,\ind{k}_B}) 
                \sw_{S_\downarrow} \sw^{1|0}_I ]\\
                =
                &\Tr_{\hbb_{\Gamma,\ind{j}_B}\otimes \hbb_{\Gamma,\ind{k}_B}} [
                    (\Pi_{\Gamma,\ind{j}_B}\otimes \Pi_{\Gamma,\ind{k}_B}) \bigotimes_{e \in \Gamma} \sw^{\frac{1-\sigma_{s(e)} \sigma_{t(e)}}{2}}_{e}
                ]\times \\
                &\times\Tr_{\mathcal{I}_{\ind{j}}\otimes \mathcal{I}_{\ind{k}}} [
                    (\Pi_{B,E}\otimes\Pi_{B,F}) \bigotimes_x \sw^{\frac{1-\sigma_x}{2}}_{B,x}
                ]\\ 
                &\times\Tr_{\hbb_{\bnd,\ind{j}_{\bnd}}\otimes \hbb_{\bnd,\ind{k}_{\bnd}}} [
                    \sw^{1|0}_I \bigotimes_{e \in \bnd} \sw^{\frac{1-\sigma_{s(e)}}{2}}_{e}
                ] .
            \end{align}
            We can then evaluate each part seperately. Starting with the intertwiner factor,
            \begin{align}
                &\Tr_{\mathcal{I}_{\ind{j}}\otimes \mathcal{I}_{\ind{k}}} [
                    (\Pi_{B,E}\otimes\Pi_{B,F}) \bigotimes_x \sw^{\frac{1-\sigma_x}{2}}_{B,x}]\\
                &=
                \delta_{\ind{j}_\downarrow,\ind{k}_\downarrow}
                \Tr_{\bigotimes_{x\in S_\downarrow}\mathcal{I}_{\mathbf{j}^x}}
                [
                    X(E,\ind{j}_\uparrow)
                    X(F,\ind{k}_\uparrow)
                ]\\
                &=
                \delta_{\ind{j}_\downarrow,\ind{k}_\downarrow}
                \Tr_{\bigotimes_{x}\mathcal{I}_{\mathbf{j}^x}}
                [\Pi_{B,E}]
                \Tr_{\bigotimes_{x} \mathcal{I}_{\mathbf{k}^x}}
                [\Pi_{B,F}]
                e^{-\Sigma_B (\ind{j},\ind{k},\Vec{\sigma})}
            \end{align}
            where we denote
            \begin{equation}
                X(E,\ind{j}_\uparrow) = \Tr_{\bigotimes_{x\in S_\uparrow}\mathcal{I}_{\mathbf{j}^x}}[\Pi_{B,E}].
            \end{equation}
            with the entropy-like quantity
            \begin{equation}
                \Sigma_B (\ind{j},\ind{k},\Vec{\sigma}) := -\log\left[
                \frac{
                    \Tr_{\ind{j}_\downarrow}
                [
                    X(\ind{j}_{\bnd},\ind{j}_\uparrow) X(\ind{k}_{\bnd},\ind{k}_\uparrow)
                ]
                }{
                    \Tr_{\ind{j}_\downarrow}
                [X(\ind{j}_{\bnd},\ind{j}_\uparrow)]
                \Tr_{\ind{k}_\downarrow}
                [X(\ind{k}_{\bnd},\ind{k}_\uparrow)]
                }
                \right]
            \end{equation}
            which simplifies to the Rényi entropy of the reduced bulk `state' when the sectors coincide:
            \begin{equation}
                \Sigma_B (\ind{j},\ind{j},\Vec{\sigma}) = S_2(
                X(\ind{j}_{\bnd},\ind{j}_\uparrow)
                )
            \end{equation}  
            To be more concrete, for the choice
            \begin{equation}
                \ket{\zeta_E}=\sum_{j_B} \ket{\zeta_{\ind{j}}} \qquad
                \Pi_{B,E} = \frac{\ket{\zeta_E}\bra{\zeta_E}}{\langle\zeta_E\ket{\zeta_E}} = \sum_{j_B, k_B}\Pi_{E;j_B,k_B}
            \end{equation}
            from the main text, gives 
            \begin{equation}
                X(E,\ind{j}_\uparrow) = \sum_{a_\downarrow,b_\downarrow} \Tr_{\mathcal{I}_{\ind{j},\uparrow}}[\Pi_{E;\ind{j}_\uparrow\cup a_\downarrow,\ind{j}_\uparrow\cup b_\downarrow}] 
                =
                \sum_{a_\downarrow,b_\downarrow} 
                \Tr_{\mathcal{I}_{\ind{j},\uparrow}}
                [
                \frac{\ket{\zeta_{\ind{j}_\uparrow\cup a_\downarrow}}\bra{\zeta_{\ind{j}_\uparrow\cup b_\downarrow}}}{\langle\zeta_E\ket{\zeta_E}}
                ]
            \end{equation}
            which is a hermitean operator on intertwiners on the spin-down region between sectors like $a_\downarrow $ and $b_\downarrow$.\\
            We can make more general statements about the quantity by identifying the trace as a Hilbert-Schmidt inner product: let $X = X(E,\ind{j}_\uparrow), Y = X(F,\ind{k}_\uparrow)$, then
            \begin{equation}
                \avg{X,Y} = \text{Re}(\avg{X,Y}) = \Tr[XY] = ||X|| \, ||Y|| \cos{\theta^{HS}_{X,Y}}
            \end{equation}
            and with 
            \begin{equation}
                ||X|| = \sqrt{Tr[X^2]} = Tr[X] e^{-\frac{1}{2}S_2(X)}
            \end{equation}
            this gives
            \begin{equation}
                \Sigma_B (\ind{j},\ind{k},\Vec{\sigma}) = \frac{1}{2} S_2(X) + \frac{1}{2} S_2(Y) - \log\left(\cos{\theta^{HS}_{X,Y}}\right)
            \end{equation}
             which shows that we need to exclude the cosine part of this object from the Hamiltonian, as it can be $0$. By our convention, such factors are to be excluded. Alternatively, however, we can also write
             \begin{equation}
                \avg{X,Y} =\Tr[\sqrt{X}Y\sqrt{X}]
                =\Tr[\left(\sqrt{\sqrt{X}Y\sqrt{X}}\right)^2] =  e^{-S_2(\sqrt{\sqrt{X}Y\sqrt{X}})} \cos^2{\theta^F_{X,Y}} 
             \end{equation}
             where $ F(X,Y)= \cos^2{\theta^F_{X,Y}} = \left(\Tr[\sqrt{\sqrt{X}Y\sqrt{X}}]\right)^2$ is the \textit{Fidelity of states} $X,Y$. Due to its properties, and that of the Rényi entropy, we can see that the Hilbert-Schmidt inner product is actually nonnegative as long as $X,Y$ are states. Furthermore, due to the fidelity being a metric on the space of states, we know precisely that the cosine factors in both cases are $1$ iff $X=Y$. Orthogonality, however, so when the cosine vanishes, is possible for many different situations.\\
            The link factor is, in fact, identical to that of the bulk-to-boundary mapping.
            Lastly, the boundary factor is
            \begin{align}
                &\Tr_{\hbb_{\bnd,\ind{j}_{\bnd}}\otimes \hbb_{\bnd,\ind{k}_{\bnd}}} [
                    \sw^{1|0}_I \bigotimes_{e \in \bnd} \sw^{\frac{1-\sigma_{s(e)}}{2}}_{e}]\\
                &=
                \prod_{e\in\bnd}
                \Tr_{V_{s(e),j_e}\otimes V_{s(e),k_e}}
                [
                \sw^{
                \frac{1-\sigma_{s(e)}\sigma_{t(e)}}{2}
                }_e
                ]\\
                &=
                \prod_{e\in\bnd} d_{j_e}^2 \left( \delta_{j_e,k_e} d^{-1}_{j_e} \right)^{\frac{1-\sigma_{s(e)}\sigma_{t(e)}}{2}}
            \end{align}
            where we introduce boundary conditions for the Ising configurations on the boundary vertices ($t(e)$ in the above): for $Z_1$, on boundary vertices of region $I$, we set $\sigma_{t(e)} = -1$. For all other cases, $\sigma_{t(e)} = +1$. \\
            This means we have as our random Ising model data the choices:
            \begin{equation}
                \Delta_{1|0}(\ind{j},\ind{k};\Vec{\sigma}) = 
                \cos{\theta_{\ind{j},\ind{k}}}\delta_{\ind{j}_\downarrow,\ind{k}_\downarrow}
                \prod_{e\in\gamma} \delta^{\frac{1-\sigma_{s(e)}\sigma_{t(e)}}{2}}_{j_e,k_e} 
            \end{equation}
            \begin{equation}
                K_{\ind{j}} = \Tr_{\ind{j}}
                [\Pi_{B,\ind{j}_\bnd}] 
                \prod_{e\in\gamma} 
                |g_{j_e}|^2 \prod_{e\in\bnd} d_{j_e}  
            \end{equation}
            where we fix the convention $g_{j_e}=1$ for all $e\in \bnd$ to write this uniformly across all links of the graph. We also introduced the shorthand
            \begin{equation}
                \cos{\theta_{\ind{j},\ind{k}}} = \cos(\theta^{HS}_{
                X(\ind{j}_{\bnd},\ind{j}_\uparrow),
                X(\ind{k}_{\bnd},\ind{k}_\uparrow)
                })
            \end{equation}
            for the Hilbert-Schmidt angle between the objects $X$. It is not boolean but may be zero, which is why we include it in the $\Delta$-factor.
            The Hamiltonian is
            \begin{equation}\label{BndToBndHammy}
                H_{1|0} (\ind{j},\ind{k},\Vec{\sigma}) = \sum_{e\in\gamma} \lambda_e \frac{1-\sigma_{s(e)}\sigma_{t(e)}}{2} + \frac{1}{2} S_2(
                X(\ind{j}_{\bnd},\ind{j}_\uparrow)
                ) + \frac{1}{2} S_2(
                X(\ind{k}_{\bnd},\ind{k}_\uparrow)
                )
            \end{equation}
            with couplings $\lambda_e= \log(d_{j_e})$. It is clear that this Hamiltonian is bounded from above and below, with upper bound $\sum_e \lambda_e +\log(\dim(\mathcal{I}_{\ind{j}})) $.

        \section{Alternative averaging methods}
    \subsection{Coarse averaging}\label{Appendix:CoarseAverage}
	We can perform, to contrast our main approach, an average over all tetrahedral states as a whole, as opposed to averaging over individual tetrahedra's states. 
    This amounts to performing a unitary Haar average over the (truncated) Hilbert space
    \begin{equation}
    \hbb = \bigotimes_x \hbb_x
    \end{equation} 
    which does not treat different vertices seperately. Such an average does not keep information about localisation of properties on any given vertex and is in some sense `nonlocal'. \\
    We exemplify this here for the case of a boundary-to-boundary mapping by inserting both a bulk projector $\Pi_I$ as well as a projector for the input segment of the boundary, $\core$. This means we are calculating the entropy in a region $A$ of the `outer boundary' in the presence of an `input' or `core state' on the `inner boundary'. While in this average, the calculations simplify significantly, we lose the correspondence to a random Ising model and the bulk and boundary decouple as far as the boundary entropy is concerned. Still, we find that there is an `implied' transition of an entanglement shadow that is not seen explicitly.
	We choose as the state to be projected a general  
	\begin{equation}\begin{split}\label{generalf}
			\ket{\Psi}= U \ket{\Psi_0}=\bigoplus_{\ind{j}}\ket{\Psi_{\ind{j}}}  = \bigoplus_{\ind{j}}\sum_{\ind{k}}U_{\ind{j},\ind{k}}\ket{\Psi_{0,\ind{k}}} \in \hbb = \bigotimes_{ x} \hbb_x
		\end{split}
	\end{equation}
	where $ U $ is a general unitary $ \mathbb{H} \mapsto \mathbb{H} $ and $ \ket{\Psi_0} $ is a reference state. Split the boundary of choice into two parts - an \textit{inner and outer boundary} $ \bnd_{in},\bnd_{out} $. The inner will also receive data.
	Now, we can define the boundary state 
	\begin{equation}\label{boundarystatenew}
		\ket{\Phi} = \bra{\theta}\bra{\zeta}\bra{\Gamma} \Psi \rangle \in  \mathbb{H}_{\bnd_{out}} = \bigoplus_{j_{\bnd_{out}}} V_{j_{\bnd_{out}}}
	\end{equation} similar to.
	We have here introduced a core state $ \ket{\theta} = \sum_{j_{\text{in}}}\ket{\theta_{j_{\text{in}}}} $ on which we project the inner boundary.
	We are interested in the second Rényi entropy of this reduced state in dependence on the connectivity $ \Gamma $ and the presence of an interior $ \core = \ket{\theta}\bra{\theta} $ and intertwiner data $ \Pi_I $, from now on assumed to be a more general projector. For this, we need to use the replica trick again. We find 
	\begin{equation}\label{psumsnew2}
		Z_{1|0} = \Tr_{ \mathbb{H}^{\otimes 2} } \left[
		\core^{\otimes 2}
		\Pi_I^{\otimes 2}
		\left(\ket{\Gamma}\bra{\Gamma}\right)^{\otimes 2} \left(\ket{\Psi}\bra{\Psi}\right)^{\otimes 2}( \mathcal{S}_A | \mathbb{I}_{\mathbb{H}} ) \right].
	\end{equation}
	We now perform an average over the Haar measure on the unitary group $ \mathcal{U}(\mathcal{H}_J) $, where
	\begin{equation}\label{cutoffspace}	
		\mathcal{H}_J \coloneqq \bigoplus_{\ind{j},\mathbb{j} \leq j^x_\alpha \leq J} \mathbb{H}_{\ind{j}} \quad \text{with dimension } \mathcal{D}_J.
	\end{equation}
	For this case, we may use the same application of Schur's lemma as before. If $ \avg{f}_U $ denotes the average of a function $ f $ on $ U(\mathcal{H}_J) $ with respect to the Haar measure, we have that 
	\begin{equation}\label{cutoffSchur}
		\avg{\left(\ket{\Psi}\bra{\Psi}\right)^{\otimes 2}}_U = \frac{\mathbb{I}_{\mathcal{H}_J\otimes\mathcal{H}_J} + \mathcal{S}_{\mathcal{H}_J\otimes\mathcal{H}_J}}{\mathcal{D}_J(\mathcal{D}_J+1)}
	\end{equation}
	where S swaps the two copies of $ \mathcal{H}_J $.
	Now, we insert this into \ref{psumsnew2} and proceed as usual.
	
	\begin{align}\label{psumsnew3}
			\avg{Z_{1|0}}_U &= \Tr_{ \mathbb{H}^{\otimes 2} }
			\left[ 
			\core^{\otimes 2}
			\left(\ket{\zeta}\bra{\zeta}\right)^{\otimes 2}
			\left(\ket{\Gamma}\bra{\Gamma}\right)^{\otimes 2} \overline{\left(\ket{\Psi}\bra{\Psi}\right)^{\otimes 2}}( \mathcal{S}_A | \mathbb{I} ) \right]\\
			&=\Tr_{ \mathbb{H}^{\otimes 2} } \left[ 
			\core^{\otimes 2}
			\left(\ket{\zeta}\bra{\zeta}\right)^{\otimes 2}
			\left(\ket{\Gamma}\bra{\Gamma}\right)^{\otimes 2}
			\frac{\mathbb{I}_{\mathcal{H}_J\otimes\mathcal{H}_J} + \mathcal{S}_{\mathcal{H}_J\otimes\mathcal{H}_J}}
			{\mathcal{D}_J(\mathcal{D}_J+1)}( \mathcal{S}_A | \mathbb{I} ) \right]\\
			& =\frac{1}{\mathcal{D}_J(\mathcal{D}_J+1)}
			\Tr_{ \mathbb{H}^{\otimes 2} } \left[ 
			\core^{\otimes 2}
			\left(\ket{\zeta}\bra{\zeta}\right)^{\otimes 2}
			\left(\ket{\Gamma}\bra{\Gamma}\right)^{\otimes 2}
			\left(\mathbb{I}_{\mathcal{H}_J\otimes\mathcal{H}_J} + \mathcal{S}_{\mathcal{H}_J\otimes\mathcal{H}_J}\right)( \mathcal{S}_A | \mathbb{I} ) \right]
	\end{align}
	If we also neglect fluctuations in the low spin regime,\footnote{This requires that the lower cutoff must scale polynomially in the number of vertices of the graph, as in $ \mathbb{j} >> N^k $ for some $ k > \frac{2}{\Delta E} $ with the spectral gap of the later Ising model.} we may discard the $ J $-dependent prefactor  $\frac{1}{\mathcal{D}_J(\mathcal{D}_J+1)}$, which yields objects that we call $Y_{1|0}$, in the quotient and write 
    \begin{equation}\label{entropyfinalform1}
		\avg{e^{-S_2(A)}}_U = \avg{\frac{Z_1}{Z_0}}_U \approx \frac{\avg{Z_1}_U}{\avg{Z_0}_U} = \frac{Y_1}{Y_0} 
	\end{equation}
	However, we can no longer perform the conversion to an Ising model as before. The reason is that for that conversion, we need a tensor product $ \bigotimes_x\left(\mathbb{I}_{\hbb_x\otimes\hbb_x} + \mathcal{S}_{\hbb_x\otimes\hbb_x}\right) $ of operators acting on the vertices individually. Working with completely generic classes of states $ \Psi $ has removed the local structure from the problem entirely, and thus averaging over it will not allow us to recover that local data.\\
	Note that this is to be expected: an average \textit{removes} information from a distribution or random variable. The larger, or coarser, the average we perform, the more data we remove in the process. On the other hand, removing said data can pinpoint typical behaviour and allow for simpler calculations. In our case, the removal of local data clearly makes the calculation simpler - so simple, in fact, that we will not be able to talk about holographic surfaces or entanglement wedges or similar concepts, as those objects are not needed for the entropy calculation.\\
	In fact, we can perform the calculation as-is. For this, it is actually convenient to work in the form 
	\begin{equation} 
		Y_{1|0} =\Tr_{(\mathbb{H}_{\partial\gamma}\otimes\mathbb{H}_{b})^{\otimes 2}} \left[
		\core^{\otimes 2}
		\Pi_I^{\otimes 2}
		\left(\ket{\Gamma}\bra{\Gamma}\right)^{\otimes 2}
		\left(\mathbb{I}_{\mathbb{H}_{\partial\gamma}\otimes\mathbb{H}_{\partial\gamma}}\mathbb{I}_{\mathbb{H}_{b}\otimes\mathbb{H}_{b}} 
		+
		\mathcal{S}_{\mathbb{H}_{\partial\gamma}\otimes\mathbb{H}_{\partial\gamma}}\mathcal{S}_{\mathbb{H}_{b}\otimes\mathbb{H}_{b}}\right)
		( \mathcal{S}_A | \mathbb{I}) 
		\right]
	\end{equation}
    where the traces over bulk and boundary can be performed seperately. The reason that this is equal to the other sum is simply that the states $\ket{\psi}$ we put on the vertices initially were `diagonal' in the tensor product space $\mathbb{H}_{\partial\gamma}\otimes\mathbb{H}_{\text{bulk}}$ in the sense that the area spins agree between the two systems. This means we can trace over either $\hbb $ or $\mathbb{H}_{\partial\gamma}\otimes\mathbb{H}_{\text{bulk}}$. 
	The key fact is that the bulk state has no support on the boundary, so it does not matter for the trace over the boundary space at all. We study both parts seperately and find:
	\begin{align} 
		Y_{0} &=\Tr_{(\mathbb{H}_{\partial\gamma}\otimes\mathbb{H}_{b})^{\otimes 2}} \left[
		\core^{\otimes 2}
		\Pi_I^{\otimes 2}
		\left(\ket{\Gamma}\bra{\Gamma}\right)^{\otimes 2}
		\left(\mathbb{I}_{\mathbb{H}_{\partial\gamma}\otimes\mathbb{H}_{\partial\gamma}}\mathbb{I}_{\mathbb{H}_{b}\otimes\mathbb{H}_{b}} +
		\mathcal{S}_{\mathbb{H}_{\partial\gamma}\otimes\mathbb{H}_{\partial\gamma}}\mathcal{S}_{\mathbb{H}_{b}\otimes\mathbb{H}_{b}}\right) \right]\\
		&=\Tr_{(\mathbb{H}_{\partial\gamma})^{\otimes 2}} \left[
		\core^{\otimes 2} \mathbb{I}_{\mathbb{H}_{\partial\gamma}\otimes\mathbb{H}_{\partial\gamma}}  \right]
		\Tr_{(\mathbb{H}_{b})^{\otimes 2}} \left[ 
		\Pi_I^{\otimes 2}
		\left(\ket{\Gamma}\bra{\Gamma}\right)^{\otimes 2}  \right] \\
		&+\Tr_{(\mathbb{H}_{\partial\gamma})^{\otimes 2}} \left[   \core^{\otimes 2}
		\mathcal{S}_{\mathbb{H}_{\partial\gamma}\otimes\mathbb{H}_{\partial\gamma}} 
		\right]
		\Tr_{(\mathbb{H}_{b})^{\otimes 2}} \left[ 
		\Pi_I^{\otimes 2}
		\left(\ket{\Gamma}\bra{\Gamma}\right)^{\otimes 2}  \mathcal{S}_{\mathbb{H}_{b}\otimes\mathbb{H}_{b}} \right]\\
		&=\Tr_{(\mathbb{H}_{\partial\gamma})^{\otimes 2}} \left[
		\core^{\otimes 2} \right]
		\Tr_{(\mathbb{H}_{b})^{\otimes 2}} \left[ 
		\Pi_I^{\otimes 2}
		\left(\ket{\Gamma}\bra{\Gamma}\right)^{\otimes 2}  \right] \\
		&+\Tr_{(\mathbb{H}_{\partial\gamma})^{\otimes 2}} \left[   \core^{\otimes 2}
		\mathcal{S}_{\mathbb{H}_{\partial\gamma}\otimes\mathbb{H}_{\partial\gamma}} 
		\right]
		\Tr_{(\mathbb{H}_{b})^{\otimes 2}} \left[ 
		\Pi_I^{\otimes 2}
		\left(\ket{\Gamma}\bra{\Gamma}\right)^{\otimes 2}  \right]\\
		&=\left(
		\Tr_{(\mathbb{H}_{\partial\gamma})^{\otimes 2}} \left[
		\core^{\otimes 2} \right]
		+
		\Tr_{(\mathbb{H}_{\partial\gamma})^{\otimes 2}} \left[   \core^{\otimes 2}
		\mathcal{S}_{\mathbb{H}_{\partial\gamma}\otimes\mathbb{H}_{\partial\gamma}} 
		\right]
		\right)
		\Tr_{\mathbb{H}_{b}} \left[\Pi_I
		\ket{\Gamma}\bra{\Gamma}\right]^2 
	\end{align}
	While
	\begin{equation} 
		\begin{split}
			Y_{1} =\Tr_{(\mathbb{H}_{\partial\gamma}\otimes\mathbb{H}_{b})^{\otimes 2}} \left[
			\core^{\otimes 2}
			\Pi_I^{\otimes 2}
			\left(\ket{\Gamma}\bra{\Gamma}\right)^{\otimes 2}
			\left(\mathbb{I}_{\mathbb{H}_{\partial\gamma}\otimes\mathbb{H}_{\partial\gamma}}\mathbb{I}_{\mathbb{H}_{b}\otimes\mathbb{H}_{b}} +
			\mathcal{S}_{\mathbb{H}_{\partial\gamma}\otimes\mathbb{H}_{\partial\gamma}}\mathcal{S}_{\mathbb{H}_{b}\otimes\mathbb{H}_{b}}\right) 
			\mathcal{S}_A\right]\\
			=\left(
			\Tr_{(\mathbb{H}_{\partial\gamma})^{\otimes 2}} \left[
			\core^{\otimes 2} \mathcal{S}_A \right] 
			+
			\Tr_{(\mathbb{H}_{\partial\gamma})^{\otimes 2}} \left[   
			\core^{\otimes 2}
			\mathcal{S}_{\mathbb{H}_{\partial\gamma}\otimes\mathbb{H}_{\partial\gamma}} 
			\mathcal{S}_A
			\right]
			\right)
			\Tr_{\mathbb{H}_{b}} \left[\Pi_I
			\ket{\Gamma}\bra{\Gamma}\right]^2 
		\end{split}
	\end{equation}
	
	Which leads to the interesting result that the entropy does not depend on the bulk at all - in fact, the global average completely erased the information about the bulk combinatorics:
	\begin{equation} 
		\avg{e^{-S_2(A)}}_U =  \frac{Y_1}{Y_0} 
		=
		\frac
		{
			\Tr_{(\mathbb{H}_{\partial\gamma})^{\otimes 2}} \left[
			\core^{\otimes 2} \mathcal{S}_A \right] 
			+
			\Tr_{(\mathbb{H}_{\partial\gamma})^{\otimes 2}} \left[   
			\core^{\otimes 2}
			\mathcal{S}_{\mathbb{H}_{\partial\gamma}\otimes\mathbb{H}_{\partial\gamma}} 
			\mathcal{S}_A
			\right]
		}	
		{
			\Tr_{(\mathbb{H}_{\partial\gamma})^{\otimes 2}} \left[
			\core^{\otimes 2} \right]
			+
			\Tr_{(\mathbb{H}_{\partial\gamma})^{\otimes 2}} \left[   \core^{\otimes 2}
			\mathcal{S}_{\mathbb{H}_{\partial\gamma}\otimes\mathbb{H}_{\partial\gamma}} 
			\right]
		}
	\end{equation}
	We can simplify this further using the fact that $ \core $ only has support on the inner boundary, and $ \mathcal{S}_A $ only on the outer one. For example,
	\begin{equation} 
		\Tr_{(\mathbb{H}_{\partial\gamma})^{\otimes 2}} \left[   
		\core^{\otimes 2}
		\mathcal{S}_{\mathbb{H}_{\partial\gamma}\otimes\mathbb{H}_{\partial\gamma}} 
		\mathcal{S}_A\right]
		=
		\Tr_{(\mathbb{H}_{\partial\gamma,out})^{\otimes 2}} \left[   
		\mathcal{S}_{\mathbb{H}_{\partial\gamma,out}} 
		\mathcal{S}_A\right]
		\Tr_{(\mathbb{H}_{\partial\gamma,in})^{\otimes 2}} \left[   
		\core^{\otimes 2}
		\mathcal{S}_{\mathbb{H}_{\partial\gamma,in}} \right]		
	\end{equation}
	which, when used in the above, yields
	\begin{equation} 
		\begin{split}
			\avg{e^{-S_2(A)}}_U =  \frac{Y_1}{Y_0} =
			\frac
			{
				\Tr_{(\mathbb{H}_{\partial\gamma,out})^{\otimes 2}} \left[
				\mathcal{S}_A \right] 
				+
				e^{-S_2(\core)}
				\Tr_{(\mathbb{H}_{\partial\gamma},out)^{\otimes 2}} \left[   
				\mathcal{S}_{\mathbb{H}_{\partial\gamma,out}} 
				\mathcal{S}_A
				\right]
			}	
			{
				\Tr_{(\mathbb{H}_{\partial\gamma,out})^{\otimes 2}} \left[
				\mathbb{I} \right]
				+
				e^{-S_2(\core)}
				\Tr_{(\mathbb{H}_{\partial\gamma,out})^{\otimes 2}} \left[  
				\mathcal{S}_{\mathbb{H}_{\partial\gamma,out}} 
				\right]
			}\\
			=
			\frac
			{
				\dim(\mathbb{H}_{A}) \dim(\mathbb{H}_{\bar{A}})^2
				+
				e^{-S_2(\core)}
				\dim(\mathbb{H}_{A})^2 \dim(\mathbb{H}_{\bar{A}})
			}	
			{
				\dim(\mathbb{H}_{\partial\gamma,out})^2
				+
				e^{-S_2(\core)}
				\dim(\mathbb{H}_{\partial\gamma,out})
			} 
			=
			\frac
			{
				\delta^{||\bar{A}||} 
				+
				e^{-S_2(\core)}
				 \delta^{||A||}
			}	
			{
				 \delta^{||\bnd_{out}||}
				+
				e^{-S_2(\core)}
			}
		\end{split}
	\end{equation}
	where we defined the truncated single-link Hilbert space dimension
        \begin{equation}
            \delta = \dim(\bigoplus_{\mathbb{j} \leq j^x_\alpha \leq J} V^{j^x_\alpha}) = \frac{d_J(d_J+1)-d_{\mathbb{j}}(d_{\mathbb{j}}+1)}{2}.
        \end{equation}
	So, the average Rényi-2 entropy is given approximately as
	\begin{equation} 
		-\log(\avg{e^{-S_2(A)}}_U ) =
		-\log(\frac
		{
			 \delta^{||\bar{A}||} 
			+
			e^{-S_2(\core)}
			 \delta^{||A||}
		}	
		{
			 \delta^{||\bnd_{out}||}
			+
			e^{-S_2(\core)}
		})
	\end{equation}
	which, for large enough $  \delta $, has limiting behaviour:
	\begin{equation} 
		\avg{S_2(A)}_U\approx -\log(\avg{e^{-S_2(A)}}_U ) \approx \min\{S_2(\core) + ||\bar{A}||\ln( \delta),||A||\ln( \delta)\}
	\end{equation}
	and in particular $ \avg{S_2(\bnd)}_U = \min\{||\bnd_{out}||\ln( \delta),S_2(\core)\}$.
	Numerically, we can find that this approximation is very good for even low values of $  \delta $.\\
	It appears that the entropy will only depend on the size of the region A, the outer boundary and the entropy of the core. The crucial fact is that after total randomisation, no connection between bulk and boundary exists anymore.\\
	Furthermore, as the region A grows, its entropy will at some point make a transition to count the number of links in the \textit{complement} instead - It shows Page curve behaviour with an offset. \\
    A simple interpretation connected to the other averaging procedures is that there is a bulk region akin to an entanglement wedge bounded by A and a minimal surface which grow as A grows. This surface eventually `envelops' the core, but does not enter it,  upon which the minimal surface wraps around and now contains the core as well as A's complement in the boundary.\\ 
    In this figurative sense, a minimal surface is still virtually present, though no longer accessible as the Ising model is no longer present. 
    
    \subsection{Fine averaging}\label{Appendix:FineAverage}
    Here, we will study a different kind of averaging procedure over the class of spin tensor networks with stricter control over the participation of different spin sectors. To be more precise, we write the full state $\ket{\Psi} \in \bigotimes_x \hbb_x$ as
    \begin{equation}
        \ket{\Psi} = \sum_{\ind{j}} \sqrt{p_{\ind{j}}} \otimes_x U_{\mathbf{j}^x}\ket{\Psi_{\mathbf{j}^x,\text{ref}}}
    \end{equation}
    where the unitaries $U_{\mathbf{j}^x}\in \mathcal{U}(\hbb_{\mathbf{j}^x})$ are picked at random with the Haar distribution, but the weights $p_{\ind{j}}\in [0,1]$, $\sum_{\ind{j}} p_{\ind{j}} = 1$ are \textit{fixed} and act as `manual dials' for us to manipulate. This average allows us to make a typicality statement about superpositions where, for example, high weight $p$ lies on sectors with the right input and output dimensionality to support isometries. It stands in contrast to the medium grade average, where the weights $p_{\ind{j}}$ were not free but instead determined by the set of considered sectors in the input and output algebra alone.\\
    To proceed, first let $\mathfrak{j},\mathfrak{k}$ denote full collections of (unglued) spins $\{\mathbf{j}^x:x\in \gamma\}$ over all vertices.
    Notice that the average state in this procedure is
    \begin{equation}
        \avg{\ket{\Psi}\bra{\Psi}}_U = \sum_{\mathfrak{j},\mathfrak{k}} \sqrt{p_{\mathfrak{j}}p_{\mathfrak{k}}} \bigotimes_x \avg{
        \left(
        U_{\mathbf{j}^x}\ket{\Psi_{\mathbf{j}^x,\text{ref}}}
        \bra{\Psi_{\mathbf{k}^x,\text{ref}}}U_{\mathbf{k}^x}^\dagger
        \right)
        }_{U}
    \end{equation}
        We will use the identities
    \begin{align}
        \avg{
        \left(
        U_{\mathbf{j}^x}
        \ket{\Psi_{\mathbf{j}^x,\text{ref}}}
        \bra{\Psi_{\mathbf{k}^x,\text{ref}}}
        U_{\mathbf{k}^x}^\dagger
        \right)
        }_U = \delta_{\mathbf{j}^x,\mathbf{k}^x} \frac{\ibb_{\mathbf{j}^x}}{\mathcal{D}_{\mathbf{j}^x}} = W(\mathbf{j}^x,\mathbf{k}^x)
    \end{align}
    as well as
    \begin{align}
        &\avg{
        \left(
        U_{\mathbf{a}^x}\otimes U_{\mathbf{c}^x}
        \ket{\Psi_{\mathbf{a}^x,\text{ref}}}\ket{\Psi_{\mathbf{c}^x,\text{ref}}}
        \bra{\Psi_{\mathbf{b}^x,\text{ref}}}\bra{\Psi_{\mathbf{d}^x,\text{ref}}}
        U_{\mathbf{b}^x}^\dagger\otimes U_{\mathbf{d}^x}^\dagger
        \right)
        }_U \\
        = W(\mathbf{a}^x,\mathbf{b}^x)&\otimes W(\mathbf{c}^x,\mathbf{d}^x)
        + \delta_{\mathbf{a}^x, \mathbf{c}^x}\delta_{\mathbf{b}^x, \mathbf{d}^x} \delta_{\mathbf{a}^x, \mathbf{b}^x} \left(
         \frac{
            \ibb_{\hbb^{\otimes 2}_{\mathbf{a}^x}} + \sw_{\hbb^{\otimes 2}_{\mathbf{a}^x}}
         }{
            \mathcal{D}_{\mathbf{a}^x} (\mathcal{D}_{\mathbf{a}^x}+1)
         }
         -
         W(\mathbf{a}^x,\mathbf{a}^x)^{\otimes 2}
        \right)
    \end{align}
    which, in simpler terms, means that if $a=b, c=d, a\neq c$, it evaluates to $W(\mathbf{a}^x,\mathbf{b}^x)\otimes W(\mathbf{c}^x,\mathbf{d}^x)$, but if instead $a=b=c=d$, it evaluates to $\frac{
            \ibb_{\hbb^{\otimes 2}_{\mathbf{a}^x}} + \sw_{\hbb^{\otimes 2}_{\mathbf{a}^x}}
         }{
            \mathcal{D}_{\mathbf{a}^x} (\mathcal{D}_{\mathbf{a}^x}+1)
         }$.\\
    Therefore, the average state is
    \begin{equation}
        \avg{\ket{\Psi}\bra{\Psi}}_U = \sum_{\mathfrak{j}} p_{\mathfrak{j}} \bigotimes_x
        W(\mathbf{j}^x,\mathbf{j}^x) = 
        \sum_{\mathfrak{j}} p_{\mathfrak{j}} 
        \frac{\ibb_{\mathfrak{j}}}{\mathcal{D}_{\mathfrak{j}}}
    \end{equation}
    which really shows that the average state in this setting is completely agnostic in each sector up to its relative weight in contribution to the full state. \\
    On the other hand, the average state on two copies is more involved:
    \begin{align}
        \avg{\left( \ket{\Psi}\bra{\Psi}\right)^{\otimes 2}}_U
        &= \sum_{\ind{j}\neq\ind{k}} p_{\ind{j}}p_{\ind{k}}
        W(\mathbf{j}^x,\mathbf{j}^x)\otimes W(\mathbf{k}^x,\mathbf{k}^x)
        +
        \sum_{\ind{j}} p_{\ind{j}}^2
        \frac{\bigotimes_x(
            \ibb_{\hbb^{\otimes 2}_{\mathbf{j}^x}} + \sw_{\hbb^{\otimes 2}_{\mathbf{j}^x}} )
         }{
           \prod_x \mathcal{D}_{\mathbf{j}^x} (\mathcal{D}_{\mathbf{j}^x}+1)
         }\\
         &=
         \sum_{\ind{j},\ind{k}} p_{\ind{j}}p_{\ind{k}}
        \frac{\ibb_{\ind{j}} \otimes\ibb_{\ind{k}}
        }{\mathcal{D}_{\ind{j}} \mathcal{D}_{\ind{j}}}
        +
        \sum_{\ind{j}} p_{\ind{j}}^2
        \left(
        \frac{ \bigotimes_x (
            \ibb_{\hbb^{\otimes 2}_{\mathbf{j}^x}} + \sw_{\hbb^{\otimes 2}_{\mathbf{j}^x}} )
         }{
            \prod_x\mathcal{D}_{\mathbf{j}^x} (\mathcal{D}_{\mathbf{j}^x}+1)
         } - 
         \frac{\ibb^{\otimes 2}_{\ind{j}} 
        }{\mathcal{D}^2_{\ind{j}} }
        \right)
    \end{align}
    which, in the regime of high spins, is well approximated by
    \begin{equation}\label{FineGradeHighRegime}
        \sum_{\ind{j}} p_{\ind{j}}^2 
        \frac{
        \bigotimes_x(
            \ibb_{\hbb^{\otimes 2}_{\mathbf{j}^x}} + \sw_{\hbb^{\otimes 2}_{\mathbf{j}^x}} ) - \ibb_{\hbb^{\otimes 2}_{\ind{j}}}
        }
        {\dim(\hbb_{\ind{j}})^2} 
        + \left( \sum_{\ind{j} } p_{\ind{j}} \frac{\mathbb{I}_{\ind{j}} }
        {\dim(\hbb_{\ind{j}})}  \right)^{\otimes 2} =: \mathcal{Q}^{(p)}+\mathbb{I}^{(p)}\otimes \mathbb{I}^{(p)}
    \end{equation}
    which is similar in structure to the coarse average, but with a `weighted identity'
    \begin{equation}
        \mathbb{I}^{(p)}
        :=
        \sum_{\ind{j} } p_{\ind{j}} \frac{\mathbb{I}_{\ind{j}} }
        {\dim(\hbb_{\ind{j}})}.  
    \end{equation}
    We may proceed from here by introducing an Ising spin configuration for each sector, $\sigma_{\ind{j},x}$. The average can then be split into two parts, the first involving $\mathcal{Q}$ and the other $\mathbb{I}^{(p)}$, which can be evaluated exactly. Then each term in the first part is its own, \textit{seperate}, Ising sum with the all-up state removed. Due to this removal we can expect for some cases the dominant contribution to come from the second, exactly calculable part.\\

    Using this exactly calculable, second part in the bulk-to-boundary calculations has two effects:
    \begin{itemize}
        \item It replaces the $K$-factors by $ K_{\ind{j}} \mapsto p_{\ind{j}} \frac{K_{\ind{j}}}{D_{\ind{j}}} = p_{\ind{j}} \prod_e |g_{j_e}|^2 = \Tilde{p}_{\ind{j}} $
        \item It removes all contributions from Ising configurations that are not all-up, so the only contributing configuration is $\Vec{\sigma} = +1$.
    \end{itemize}
    This simplifies calculations dramatically. In fact, it allows an immediate, exact expression for the Rényi purity:
    \begin{equation}
        \avg{e^{-S_2(\rho_I)}}_U \approx \sum_{\ind{j}} \Tilde{p}_{\ind{j}}^2 \frac{1}{\dim(\ical_{\ind{j}})}.
    \end{equation}
    So a tuning of the weights $p$ will allow for holography quite easily, e.g. by 
    \begin{equation}
        \Tilde{p}_{\ind{j}} = \frac{\dim(\ical_{\ind{j}})}{D_I} .
    \end{equation} 
    This is a solution if, as before, we fix the boundary spins.\\
    For another example, the maximum entropy sequence $p_n$ on a set of sectors $\ind{j}_n$ subject to the constraint of isometry is then schematically of the form
    \begin{equation}
        p_n \sim \frac{\tau}{c_n}\mathcal{W}(\frac{c_n}{\tau})
    \end{equation} 
    for the sequence $c_n = \frac{g_n}{\dim(\ical_{\ind{j}_n})} $, and some constant $\tau$ in the Lambert $\mathcal{W}$ function. So while there may still be constraints on the weights, the situation is now perfectly controllable.


\section{Example calculations}
 \label{appendixExampleCalc}
    \subsection{Boundary-to-Boundary: single internal link}	
    \begin{figure}
		\centering
		\includegraphics[width=0.5\linewidth]{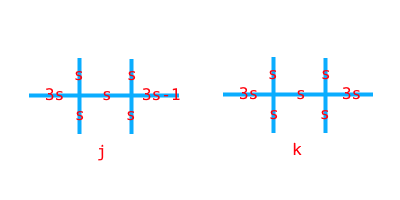}
		\caption{ We consider a superposition of two spin sectors which are only different on a single link. The values are chosen as to give a controllably low intertwiner space dimension of 1 or 2.}
		\label{fig:example1}
	\end{figure}
    To illustrate the complexity of the calculations involved, we shall give an innocuous example. We will not impose the known necessary conditions for isometry and work in the boundary-to-boundary case, where we include a matrix $\rho^I$ for bulk/intertwiner degrees of freedom in the trace expressions for $Z_{1|0}$. We do not impose that it is a projector from the beginning, but this restriction and others may of course be imposed at the end. \\
    The graph in question, see figure \ref{fig:example1}, is the simplest nontrivial one, and we only consider two distinct spin sectors, which agree on all but one link. This means that there are only four Ising configurations. We choose the area spins such that only one intertwiner space is nontrivial. We will also refer to the vertices just by L(eft) or R(ight) for convenience. In spin sector $\ind{k}$, all intertwiner spaces are 1-dimensional, while in $\ind{j}$, the right one is 2-dimensional. We can then give the intertwiner state
	\begin{equation}
		\rho^I= \begin{bmatrix}
			\rho^I_{\ind{j},\ind{j}} & \rho^I_{\ind{j},\ind{k}} \\
			\rho^I_{\ind{k},\ind{j}} & \rho^I_{\ind{k},\ind{k}}
		\end{bmatrix}\qquad
		\rho^I_{\ind{j},\ind{j}} = \begin{bmatrix}
			a & b \\
			\bar{b} & d
		\end{bmatrix}
		\qquad   \rho^I_{\ind{j},\ind{k}} = \begin{bmatrix}
			u \\
			v
		\end{bmatrix} \qquad \rho^I_{\ind{k},\ind{k}} = w = 1-(a+d)
	\end{equation}
	From this, we can derive the reduced entropies 
	\begin{equation}
		S_2((\rho^I_{\ind{j},\ind{j}})_R) =S_2((\rho^I_{\ind{j},\ind{j}})) = -\log\left[ \frac{a^2+d^2+|b|^2}{(a+d)^2} \right] \qquad S_2((\rho^I_{\ind{j},\ind{j}})_L) = 0
	\end{equation}
	while the ones for $\ind{k}$ vanish.\\
    We choose the input area $ C $ to be the rightmost link, where the two sectors disagree. The individual sectors induce, at least for large spins, an isometry, making the only question what parameters need to be chosen to make their superposition induce an isometry.\\
    We show some of the calculations for this case here, but some are omitted for brevity and can be easily reproduced in Mathematica code.\\
    We first discuss the constraints on configurations. The case chosen here has intertwiner constraints which can be ignored, so we have the following:
	\begin{enumerate}
		\item $\Delta_{0}(\ind{j},\ind{j},\vec{\sigma}) =\Delta_{0}(\ind{j},\ind{k},\vec{\sigma}) = 1$  ,  $\Delta_{0}(\ind{j},\ind{k},(+_L,+_R)) = 1 $  and  $\Delta_{0}(\ind{j},\ind{k},(-_L,+_R))  = 1 $.
		\item $\Delta_{0}(\ind{j},\ind{k},(-_L,-_R)) = \prod_{e: \sigma_x = -1} \delta_{j_e,k_e} = 0 = \Delta_{0}(\ind{j},\ind{k},(+_L,-_R))$
	\end{enumerate}
	Essentially, as the spins on the right vertex do not agree between the two sectors, the Ising spin may not point down on it. The only difference in the numerator $\Delta$-factors is that the boundary pinning field flips around where spins have to agree. We take the pinning field to be $-1$ on the \textit{rightmost link} where the areas disagree:
	\begin{enumerate}
		\item $\Delta_{1}(\ind{j},\ind{k},(+_L,+_R)) = 0 = \Delta_{1}(\ind{j},\ind{k},(-_L,+_R))$
		\item $\Delta_{1}(\ind{j},\ind{k},(-_L,-_R)) = 1 = \Delta_{1}(\ind{j},\ind{k},(+_L,-_R))$
	\end{enumerate}
	In other words, the right Ising spin must be down. This is a manifestation of the general rule that configurations where the value of the Hamiltonian would be ambiguous need to be excluded. Overall, even in this innocuous example we see that there can be a fair reduction of possible configurations contributing to the mixed partition sums. 
	Then, we can calculate the Hamiltonian values individually.\\
    \begin{tabular}{ |p{1.3cm}||p{1.2cm}|p{2.8cm}|p{3.5cm}|p{4cm}|  }
		\hline
		\multicolumn{5}{|c|}{Hamiltonian values for the 4 configurations} \\
		\hline
		& (+,+) &(+,-)&(-,+)&(-,-)\\
		\hline
		$H_0(\ind{j},\ind{j})$   & $ 0 $    &$3L_2+L_{6,-}+ S_2 $&   $ 3L_2+L_{6,-} $ & $4L_2+L_{6,-}+L_{6,+}+ S_2 $\\
		$H_0(\ind{k},\ind{k})$&   0  & $3L_2+L_{6,+} $  &$3L_2+L_{6,+}$& $4L_2+2L_{6,+}$\\
		$H_0(\ind{j},\ind{k})$ &0 & \text{-}&  $2L_2+L_{6,+}+ \Sigma_{(-,+)} $& \text{-}\\
		\hline
		$H_1(\ind{j},\ind{j})$    &$ L_{6,-} $ & $3L_2+ S_2 $&  $ 3L_2+L_{6,+}+L_{6,-} $& $4L_2+L_{6,+}+ S_2 $\\
		$H_1(\ind{k},\ind{k})$&   $L_{6,+}$  & $3L_2$&  $3L_2+2L_{6,+} $ & $ 4L_2+L_{6,+} $\\
		$H_1(\ind{j},\ind{k})$& -  & $3L_2+\Sigma_{(+,-)}$   & \text{-} & $ 3L_2+L_{6,+} $+$\Sigma_{(-,-)}$\\
		\hline
	\end{tabular}\\
    We have used the shorthands
        \begin{equation}
            L_2 = \ln(2s+1) \qquad L_{6,\pm} = \ln(6s\pm 1)
        \end{equation}
	\begin{align*}
		\Sigma_{(\sigma_L,\sigma_R)} = \Sigma(\ind{j},\ind{k};(\sigma_L,\sigma_R))
		\qquad 
		S_2 = S_2(\Pi^I_{\ind{j},\ind{j}})
	\end{align*}(set these to zero in the following)
	We thus find the six partition sums
	\begin{enumerate}
		\item $Z^{\ind{j},\ind{j}}_0 = 1+ e^{-S_2-3L_2-L_{6,-}}
		+e^{-3L_2-L_{6,-}}
		+e^{-S_2-4L_2-L_{6,-}-L_{6,+}}$
		\item $Z^{\ind{k},\ind{k}}_0 = 1+ e^{-3L_2-L_{6,+}}
		+e^{-3L_2-L_{6,+}}
		+e^{-4L_2-2L_{6,+}}$
		\item $Z^{\ind{j},\ind{k}}_0 = 1 + e^{-\Sigma_{(-,+)}-2L_2 - L_{6,+}}$
		\item $Z^{\ind{j},\ind{j}}_1 = e^{-L_{6,-}}+ e^{-S_2-3L_2}
		+e^{-3L_2 - L_{6,-}-L_{6,+}}
		+e^{-S_2-4L_2-L_{6,+}}$
		\item $Z^{\ind{k},\ind{k}}_1 = e^{-L_{6,+}}+ e^{-3L_{2}}+e^{-2L_{6,+}-3L_2}
		+ e^{-L_{6,+}-4L_2}$
		\item $Z^{\ind{j},\ind{k}}_1 =  e^{-\Sigma_{(+,-)} - 3 L_2}
		+e^{-\Sigma_{(-,-)}  - 3L_2 - L_{6,+}}$
	\end{enumerate}
	and we note that as expected, the values of the $Z_0$ partition sums approach 1 as we increase the areas. Each term in the $Z_1$ sums also decays with some power of the area. Additionally, the decay in the mixed sums is stronger than the others due to suppression of certain configurations.\\
	It is instructive to see the single-sector results at this point. For the first sector, the purity is
	\begin{align} 
		\frac{Z^{\ind{j},\ind{j}}_1 }{	Z^{\ind{j},\ind{j}}_0 }
		&= 
		\frac{e^{-L_{6,-}}+ e^{-3L_2}e^{-S_2}
			+e^{-3L_2}e^{-L_{6,-}}e^{-L_{6,+}}
			+e^{-4L_2}e^{-L_{6,+}}e^{-S_2}}
		{1+ e^{-3L_2}e^{-L_{6,-}}e^{-S_2}
			+e^{-3L_2}e^{-L_{6,-}}
			+e^{-4L_2}e^{-L_{6,-}}e^{-L_{6,+}}e^{-S_2}}\\
		&\approx e^{-L_{6,-}} \text{ for large }s
	\end{align}
	which is the reciprocal dimension of the boundary input space in the large-spin limit.
	The same thing happens with the other sector, where $ \frac{Z^{\ind{k},\ind{k}}_1 }{Z^{\ind{k},\ind{k}}_0 } \approx e^{-L_{6,+}} $. Thus both sectors individually yield an isometric map when spins are large enough.
	Now, with the two weights 
	\begin{equation}
		K_{\ind{j}} = e^{4L_2+L_{6,+}+L_{6,-}}(a+d) \qquad K_{\ind{k}} = e^{4L_2+2L_{6,+}}w
	\end{equation}
	and the entropies of the intertwiner state
	\begin{equation}
		e^{-S_2} = t\coloneqq \frac{a^2+d^2+|b|^2}{(a+d)^2} 
	\end{equation}
	\begin{equation}
		e^{-\Sigma_{(-,-)}} = e^{-\Sigma_{(+,-)}} = q \coloneqq
		\frac{|u^2|+|v|^2}{w(a+d)}
		\qquad
		e^{-\Sigma_{(+,+)}}
		=
		e^{-\Sigma_{(-,+)}}
		= 1
	\end{equation}
	this gives us $\frac{Z_1}{Z_0}$. The full expression is rather unilluminating, but we give a few special cases of interest. When the spin is taken to be asymptotically large:
	
	\begin{align}\label{3rdOrderExpr}
		\frac{Z_1}{Z_0} 
		&=
		\frac{2 w^2-2 w+1}{6 s}
		+
		\frac{(1-2 w)^3}{36 s^2}\\
		&+
		\frac{27 (w-1)^2 e^{-S_2}+(61-54 q) w^2+(54 q-10) w+24 w^4-48 w^3+1}{216 s^3}
		+
		O\left(\left(\frac{1}{s}\right)^4\right).
	\end{align}
	This is particularly simple in that the $b$ and $u,v$ parameters do not contribute up to second order. Thus, in the high spin limit, these tend to not matter as much. Studying only the leading 2 orders, we can ask which parameters give us the most mixing.
	These are easily found to be, for fixed s, $w\in \{\frac{1}{2}, \frac{2s+1}{2}\}$, which means that only $ w=\frac{1}{2} $ is a valid minimum. \\
	If we take the third order into account, we instead have as minima
	\begin{equation}
		(w,q) = (1,\frac{2}{27} \left(18 s^2-9 s+16\right)) \approx
		(1,\frac{4}{3} s^2)
	\end{equation}
	irrespective of the value of the entropy $e^{-S_2}$. Further orders preserve this minimum.
	A simple numerical study of the minimum of the purity for fixed $ s $ reveals that the overall value decreases with spin inversely as expected:
	\begin{equation}
		G(s):=\min_{a,d,w,t,q}(\frac{Z_1}{Z_0}(a,d,w,t,q,s)) \approx \frac{1}{12s}.     
	\end{equation}
	So, the maximal achievable entropy for fixed spin $s$ is then
	\begin{equation}
		S_2 \approx  \ln(12s)
	\end{equation}
	We can then estimate the maximal dimension for a transparent subspace by $G^{-1} \approx 12s $, which is precisely the dimension of $\hbb_C$ in the studied case. Numerically, we find that this is achieved when $(a,d,w)\approx(\frac{1}{4},\frac{1}{4},\frac{1}{2})$ in the large-spin limit, with no significant dependence on $q$ or $ t $. This, in particular, implies $ c_j = c_k = \frac{1}{2} $ and has the minimum $ w=\frac{1}{2} $ seen before. The result agrees qualitatively with the second order result from \ref{3rdOrderExpr}, confirming that the large-spin regime is well approximated by it.\\
	\,\\
	If we instead choose the region $C$ to be the upper right link, we get the same type of result: there is a state of maximal entropy which makes the induced map into an isometry, with $G(s)= \frac{1}{2s+1}$, and which is the minimum of $\frac{Z_1}{Z_0}$ for fixed spin sectors. The parameters of the minimum are, however, different: $(a,d,w)\approx(\frac{1}{3},\frac{1}{3},\frac{1}{3})$. Here, $ c_j = 2c_k $. This shows that whether an isometry exists or not can depend sensitively on the boundary region under consideration and the state of bulk degrees of freedom.\\

    \subsection{Bulk-to-boundary: single vertex}\label{appendix:SingleVertex}
        We give a full analytical calculation of the partition sums for bulk-to-boundary mappings for a single spin network vertex. While this example may appear trivial, it illustrates the complexity of the calculations that already appear without tensor network contractions when superposition is allowed. \\ 
        We begin with the full partition sums
        \begin{equation}
            Z_{1|0} = 
            \sum_{\vec{\sigma}}\Tr_{\hbb^{\otimes 2}}[ \Pi^{\otimes 2} \bigotimes_x \sw_{\hbb_x}^{\frac{1-b\sigma_x}{2}}] 
            = \sum_{E,\Tilde{E}} Z^{E,\Tilde{E}}_{1|0}
        \end{equation}
        with boundary-fixed partition sums
        \begin{align}
            Z^{E,\Tilde{E}}_{1|0} &= \sum_{\vec{\sigma}}\Tr_{\hbb_E\otimes \hbb_{\Tilde{E}}}[ \Pi^{\otimes 2} \bigotimes_x \sw_{\hbb_x}^{\frac{1-b\sigma_x}{2}}] \\
            &= \sum_{\vec{\sigma}}\sum_{j_B,k_B} \Tr_{\hbb_{E\cup j_B}\otimes \hbb_{\Tilde{E}\cup k_B}}[ \Pi^{\otimes 2} \bigotimes_x \sw_{\hbb_x}^{\frac{1-b\sigma_x}{2}}] \\
            &= \sum_{\vec{\sigma}}\sum_{j_B,k_B} K_{E\cup j_B} K_{\Tilde{E}\cup k_B} Z^{E\cup j_B,\Tilde{E}\cup k_B}_{1|0}.
        \end{align}
        Now each term in this is in general hard to compute already, but here it completely trivialises. Due to absence of internal links,  $\Pi$ and the sums over $j_B,k_B$ disappear entirely. Furthermore, the sums are diagonal in boundary spins $E$, meaning we only need to consider
        \begin{equation}
            Z^{E,E}_{1|0} = K^2_{E}  Z^{E\cup\emptyset,E\cup\emptyset}_{1|0}.
        \end{equation}
        where the right partition sum is defined through the Ising model. Using \ref{Appendix:RIMBb}, this is
        \begin{equation}
            Z^{E\cup\emptyset,E\cup\emptyset}_{1|0} = \sum_{\sigma_x} e^{-\frac{1-\sigma_{x}}{2}\sum_{e\in\bnd} \lambda_e  - \frac{1-b\sigma_x}{2} \Lambda_x}
            =  D_{I_E}^{-(1|0)}
            +  D^{-1}_{O_E} D_{I_E}^{-(0|1)}
        \end{equation}
        and 
        \begin{equation}
            K^2_E = D^2_{I_E} D^2_{O_E} = D^2_E
        \end{equation}
        and therefore the Ising sums are
        \begin{align}
            Z_0 &= \sum_E (D^2_{E} + D_{E}) \\
            Z_1 &= \sum_E D^2_{I_E} D^2_{O_E}( D_{I_E}^{-1} +  D^{-1}_{O_E}   ) 
            = \sum_E D_{I_E} D_{O_E}( D_{I_E} +  D_{O_E}   ).
        \end{align}
        Let us introduce $r_E = \frac{D_{I_E}}{D_{O_E}}$, which allows us to rewrite everything in terms of it as an expansion parameter $r$. Isometry requires $r\leq1$, and this holds for any gauge invariant spin network vertex by definition of intertwiner spaces. \\
        \begin{align}
            Z_0 &= \sum_E ( \frac{D^2_{I_E}}{r_E} (\frac{D^2_{I_E}}{r_E}+1) \\
            Z_1 &= \sum_E D^3_{I_E} \frac{1}{r_E}(1 +  \frac{1}{r_E}   ).
        \end{align}
        We can establish under which conditions we have holography by calculating the purity and weight seperately in each sector. These are given, respectively, by \ref{BndSpinDecomp},\ref{BndSpinDecompEntropy},\ref{BndSpinDecompSectorIdent}:
        \begin{align}
            \frac{Z^{E,E}_1}{Z^{E,E}_0} &= \frac{( D_{I_E} +  D_{O_E}   )D_E}{(1+D_E)D_E} 
            = D_{I_E}\frac{\frac{1}{r_E}(1+\frac{1}{r_E})}{\frac{1}{r_E}(1+\frac{D_{I_E}^2}{r_E})}\\
            &= 
            \frac{1}{D_{I_E}} \frac{1+r_E}{1+\frac{r_E}{D_{I_E}^2}} = \frac{1}{D_{I_E}} \left( 1 + (1-\frac{1}{D_{I_E}^2})r_E + \mathcal{O}(r^2_E) \right)
        \end{align}
        which is generically close to the isometric value for small $r_E$, and in fact for 1-dimensional input spaces only differs at second order.

        The cross-sector conditions can be understood as restrictions on which combinations of $E$ we may have in the input algebra. We must demand for some constant $q$ that
        \begin{equation}
            D_E(D_{I_E}+ D_{O_E} ) = q D_{I_E} \qquad D_E (D_E+1) = q D_{I_E}^2
        \end{equation}
        which already includes the condition from before that $r_E$ is small. 
        We can just solve the two conditions directly for some conditions by multiplying the first by $D_{I_E}$ and solving for $r_E$: 
        \begin{equation}
            D_{I_E}^2+D_E = D_E +1 \implies r_E = \frac{1}{D_E}
        \end{equation}
        Inserting this solution into the first equation yields
        \begin{equation}
            D_E+1 = q = \text{ const}
        \end{equation}
        so in fact \textit{the dimensions must be independent of the sector label $E$}. It also yields more trivially that 
        \begin{equation}
            D_{I_E}^2 = r_E D_E = 1 \implies D_{I_E} = 1, D_E = D_{O_E} = q-1 = \text{ const}.
        \end{equation}
        So although we saw before that generically for small $r_E$, the entropy condition is fulfilled easily, this is not the case for the cross-sector condition: it requires that the output dimension may also not depend on the sector label. In fact, it also requires, far stronger,  the restriction to 1-dimensional inputs in each sector. Therefore, we must restrict ourselves to a fixed output dimension $D_{O}$ and select only those $E$ such that $D_{I_E}=1, D_{O_E} = D_O$. Then we can find holographic behaviour.\\
        For the trivalent case, the input condition is always fulfilled; But in the 4-valent case already, we have to restrict ourselves: \textit{either}, at least one of the dimensions on the boundary links is $1$, \textit{or} the largest dimension is
        \begin{align}
            d_{max} = d_1+d_2+d_3 -2.
        \end{align}
        This means that there are overall 2 constraints on the 4 free variables we can choose. However, the overall message is clear: while the fixing of the output dimension is generically necessary, there are also strong restrictions on the input dimension. Incidentally, the restriction to dimension $1$ inputs also makes corrections to the entropy vanish, simply because the minimal entropy is also the maximal entropy.\\
        We note though that in the high-$D_O$ approximation, holography is generic: 
        \begin{equation}
            Y^E_1 = \frac{1}{D_O}+\frac{1}{D_{I_E}} = e^{-\beta}+\frac{1}{D_{I_E}} \approx\frac{1}{D_{I_E}}  \qquad Y^E_0 = 1+\frac{1}{D_E} = 1+ \frac{1}{D_{I_E}} e^{-\beta} \approx 1
        \end{equation}
        which trivially fulfils holography. We should therefore think of the restriction to intertwiner dimension 1 perhaps more as having very low input dimension compared to the output $D_O$.
        
        \subsection{Bulk-to-boundary: single internal link} \label{appendix:SingleBulkLink}
        In the case of a single link, we are also able to perform most of the calculations analytically. We label for convenience the endpoints by x and y and the set of boundary spins on x or y as $E_x$ and $E_y$, respectively. The spins of the single bulk link will be labeled by u or v, and we denote intertwiner dimensions on a vertex $x$ depending on link spins by $\mathcal{D}_{\{j^x_e\}}$.\\
        Once again we have
        \begin{align}
            Z^{E,E}_{1|0} &= \Tr_{\hbb_E\otimes \hbb_{E}}[ \Pi^{\otimes 2} \bigotimes_x \sw_{\hbb_x}^{\frac{1-b\sigma_x}{2}}] \\
            &= \sum_{u,v} \Tr_{\hbb_{E\cup u}\otimes \hbb_{E\cup v}}[ \Pi^{\otimes 2} \bigotimes_x \sw_{\hbb_x}^{\frac{1-b\sigma_x}{2}}] \\
            &= \sum_{u,v} K_{E\cup u} K_{E\cup v} Z^{E\cup u,E\cup v}_{1|0}.
        \end{align}
        and we can split
        \begin{equation}
                K_{E\cup u} = D_{O_E} \mathcal{D}_{E_x\cup u}\mathcal{D}_{E_y\cup u}
                 |g_{u}|^2  = L_{E\cup u} D_{O_E}
        \end{equation}
        \begin{align}
            H^{E\cup u}_{1|0}(\Vec{\sigma})
            &=
            \frac{1-\sigma_{x}}{2}\sum_{j_e \in E_x} \log(2j_e+1) 
            +\frac{1-\sigma_{y}}{2}\sum_{j_e \in E_y} \log(2j_e+1)
            + \frac{1-b\sigma_x}{2} \Lambda_x + \frac{1-b\sigma_y}{2} \Lambda_y\\
            &+ \frac{1-\sigma_x \sigma_y}{2} \log(2u+1)
        \end{align}
        into its bulk contribution $L_{E\cup u}$, which stays in the sum, and the pure boundary contribution $D_{O_E}$. Next, we consider the effect of the constraints. The constraints, in general, state that for $Z_1$ ($Z_0$), the spin-down (spin-up) region must be contained in the region $G_{\ind{j},\ind{k}}$ of vertices where all the incident spin labelings agree. This restricts allowed Ising configurations and leads to a suppression of sector-off-diagonal partition sums, because the usual lowest energy configuration is typically disallowed.\\
        In our case, for $u\neq v$, $G_{\ind{j},\ind{k}} = \emptyset$ (not counting the boundary vertices), meaning that only the spin-down (spin-up) configuration can contribute to off-diagonal sums. Let 
        \begin{equation}
            a_{x|y} = \mathcal{D}_{E_{x|y}\cup u}^{-1}
            \qquad 
            b_{x|y} = \prod_{j_e \in E_{x|y}} d_{j_e}^{-1}
        \end{equation}
        , then
        \begin{align}
            Z^{E\cup u,E\cup v}_{1} &= e^{-H^{E\cup u}_{1}(\Vec{-1})} = 
            b_x b_y \\
            Z^{E\cup u,E\cup v}_{0} &= e^{-H^{E\cup u}_{0}(\Vec{+1})}
            = 1
        \end{align}
        This means that the $u-v$ Matrices $Z^{E\cup u,E\cup v}_{1|0}$ have the form of a constant contribution in each entry as well as a diagonal part. The diagonal terms are again
        \begin{align}
            Z^{E\cup u,E\cup u}_{1} &= a_x a_y + d^{-1}_u a_x b_y + d^{-1}_u a_y b_x + b_x b_y \\
            Z^{E\cup u,E\cup u}_{0} &= 1 + d^{-1}_u a_x b_x + d^{-1}_u a_y b_y + a_x a_y b_x b_y 
        \end{align}
        and so we have obtained the full partition sums for fixed boundary spins.

        We can use the other objects defined in the main text by
        \begin{equation}
            L_{\ind{j}}=|g_u|^2\qquad Y^E_{1|0}= \sum_{u,v} |g_u|^2|g_v|^2 Z^{E\cup u,E\cup v}_{1|0}
        \end{equation}
        and fix the output dimension
        \begin{equation}
            D_O = D_{O_E} = (b_x b_y)^{-1}
        \end{equation}
        and so we find:
        \begin{align}
            Y^E_{1|0} = \sum_{u} |g_u|^4 Z^{E\cup u,E\cup v}_{1|0} + D_O^{(-1|0)} \left[
            (\sum_u |g_u|^2)^2 - \sum_u |g_u|^4
            \right]
        \end{align}
        Now by noting the normalisation condition $\sum_u |g_u|^2 = 1$ for all internal links, the term in the brackets becomes a Rényi-2 entropy of the sequence of coefficients
        \begin{equation*}
            (\sum_u |g_u|^2)^2 - \sum_u |g_u|^4
             = 1 - \exp( - S_2( (|g_u|^2)_u ).
        \end{equation*}
        We can evaluate for example $Y^E_1$ in the special case where the two vertices are 4-valent and all the boundary spins are the same, at value $j = \frac{n-1}{2}$, meaning $D_O = n^6$. Then, we can explicitly calculate the intertwiner dimension in closed form, and find that $D_{I_E} = n^2$, and the sum splits into 3 parts: the constant part above, the part where $m := 2 u +1$ is less than $n$ and the part where $m>n$. The first part is as above, while the second and third are, respectively,
        \begin{equation}
            \sum_{m=1}^n |g_{m}|^4 (\frac{1}{n^6} + \frac{1}{m^2}(1 + \frac{2}{n^3})) 
        \end{equation}
        \begin{equation}
            \sum_{m = n+2k, k\in [1:n]} |g_{n+2k}|^4 \left[\frac{2}{n^3 (n-k) (2 k+n)}+\frac{1}{(k-n)^2}+\frac{1}{n^6}\right]
        \end{equation}
        This is already quite complex, but because the sums are finite, they of course converge. In fact, for $g=1$ they even have a closed form each:
        \begin{equation*}
            \frac{n^5 H_n^{(2)}+2 n^2 H_n^{(2)}+1}{n^5}
        \end{equation*}
        \begin{equation*}
            \frac{-\pi ^2 n^5+6 n^5 \psi ^{(1)}(1-n)+4 \gamma  n-4 n \psi ^{(0)}\left(\frac{n}{2}+1\right)+4 n \psi ^{(0)}\left(\frac{3 n}{2}+1\right)+4 n \psi ^{(0)}(1-n)+6}{6 n^5}
        \end{equation*}
        where $H_n^{(2)}$ gives the $n$th Harmonic number of order 2, $\psi^{(k)}$ are the polygamma functions and $\gamma$ is the Euler-Mascheroni constant. In the limit of large $n$, they have simple expansions
        \begin{equation*}
            \frac{\pi ^2}{6}-\frac{1}{n}+O\left(\left(\frac{1}{n}\right)^2\right)
        \end{equation*}
        \begin{equation*}
           \frac{5 \pi ^2}{6} -\frac{1}{n}+\pi ^2 \cot ^2(\pi  n)
            +O\left(\left(\frac{1}{n}\right)^2\right).
        \end{equation*}
        In general, for varying $n$ the sum shows oscillatory behaviour around any integer.\\
        As for $Y^E_0$, we find the same split with exact values
        \begin{equation*}
            \frac{2 n^3 H_n^{(2)}+H_n^{(2)}+n^7}{n^6} = n  +O\left(\left(\frac{1}{n}\right)^2\right)
        \end{equation*}
        \begin{align*}
            &\frac{6 n^7+4 \gamma  n^2-4 n^2 \psi ^{(0)}\left(\frac{n}{2}+1\right)+4 n^2 \psi ^{(0)}\left(\frac{3 n}{2}+1\right)+4 n^2 \psi ^{(0)}(1-n)+6 \psi ^{(1)}(1-n)-\pi ^2}{6 n^6} \\
            &= n +O\left(\left(\frac{1}{n}\right)^2\right)
        \end{align*}
        As a result, we see that indeed, with $|g|=1$ we cannot fulfil the isometry conditions, and that
        \begin{equation}
            \frac{Y^E_1}{Y^E_0} \approx \frac{\pi^2}{2 n}
        \end{equation}
        which does not scale like $\frac{1}{D_{I_E}} = \frac{1}{n^2}$. This shows that the choice $g=1$ hinders holographic behaviour.
        Now, we can also find what choice for $g$ we \textit{must} make. In a high-$\beta$ approximation, the sums are generally ($v_n = |g_{\frac{n-1}{2}}|^2 $):
        \begin{equation}
            Y^E_1 \approx \sum_n v_n^2 \frac{1}{D_{I_E}}, \qquad Y^E_0 \approx (\sum_n a_n)^2
        \end{equation}
        So again, in this limit the only condition we need to check is the entropy one, which takes the form
        \begin{equation}
            \frac{Y^E_1}{Y^E_0} \approx \sum_n r_n a_xa_y(n) \stackrel{!}{=} \frac{1}{\sum_n (a_xa_y(n))^{-1}} \qquad r_n = \frac{v_n^2}{(\sum_n v_n)^2}
        \end{equation}
        which has as valid solutions for $r_n$, if there are $M$ values for the spins,
        \begin{equation}
            r_n = \frac{1}{a_xa_y(n)} \left( \frac{1}{M \sum_m (a_xa_y(m))^{-1} } +c_n \right)\qquad \sum_n c_n = 0.
        \end{equation}
        Ultimately, solving for $a_n$ here is irrelevant, but what matters is the scaling:
        \begin{equation}
            |g_u|^2 \sim \frac{1}{\sqrt{a_xa_y(u)}}
        \end{equation}
        So demanding isometry puts strong constraints on the scaling of the coefficients $g$ we use to define the state $\rho$ and map $\tcal_\rho$.
        This is interesting, for one because it is consistent with the assumption that large spins must dominate for the approximations to work, but also because this is the only constraint we had to put on the problem to get isometry. This is because for the setting here, it was quite natural to assume that the all-up configuration is the ground state of $\Tilde{H}_1$. Under this assumption, and the constancy of $D_{O_E}$, though, we are however already almost at isometry, as given in the main section.\\
        The most intriguing part of this scaling, though, is that the left hand side does not depend on $E$. This implies at least that
        \begin{equation}
            \frac{(a_xa_y(n))^{-1}}{\sum_m (a_xa_y(m))^{-1}}
        \end{equation}
        is independent of $E$ for all $n$.\\
        We can therefore understand that the full isometry condition boils down to 3 essential ingredients:
        \begin{enumerate}
            \item Constancy of $D_{O_E} = D_O$.
            \item Knowledge of the minima of $\Tilde{H}_1$ and their proximity to the all-up configuration.
            \item Constraints on the $g_{j_e}$ coefficients that relate them to the input dimensions.
        \end{enumerate}


	\bibliographystyle{JHEP}
	\bibliography{ReferencesCorrected.bib}  

\providecommand{\href}[2]{#2}\begingroup\raggedright\begin{thebibliography}{10}

\bibitem{PhysRevD.7.2333}
J.D.~Bekenstein, \emph{Black holes and entropy},
  \href{https://doi.org/10.1103/PhysRevD.7.2333}{\emph{Phys. Rev. D} {\bfseries
  7} (1973) 2333}.

\bibitem{Hawking:1975vcx}
S.W.~Hawking, \emph{{Particle Creation by Black Holes}},
  \href{https://doi.org/10.1007/BF02345020}{\emph{Commun. Math. Phys.}
  {\bfseries 43} (1975) 199}.

\bibitem{eisertAreaLawsEntanglement2010}
J.~Eisert, M.~Cramer and M.B.~Plenio, \emph{Area laws for the entanglement
  entropy - a review},  \href{https://arxiv.org/abs/0808.3773}{{\ttfamily
  0808.3773}}.

\bibitem{witten1998anti}
E.~Witten, \emph{Anti de sitter space and holography},  1998.

\bibitem{maldacenaLargeLimitSuperconformal1999}
J.M.~Maldacena, \emph{The {{Large N Limit}} of {{Superconformal Field
  Theories}} and {{Supergravity}}},
  \href{https://arxiv.org/abs/hep-th/9711200}{{\ttfamily hep-th/9711200}}.

\bibitem{Araujo_Regado_2023}
G.~Araujo-Regado, R.~Khan and A.C.~Wall, \emph{Cauchy slice holography: a new
  ads/cft dictionary},
  \href{https://doi.org/10.1007/jhep03(2023)026}{\emph{Journal of High Energy
  Physics} {\bfseries 2023} (2023) }.

\bibitem{donnellyGravitationalEdgeModes2021}
W.~Donnelly, L.~Freidel, S.~Moosavian and A.~Speranza, \emph{Gravitational edge
  modes, coadjoint orbits, and hydrodynamics}, .

\bibitem{freidelEdgeModesGravity2020}
L.~Freidel, M.~Geiller and D.~Pranzetti, \emph{Edge modes of gravity. {{Part
  I}}. {{Corner}} potentials and charges}, .

\bibitem{freidelExtendedCornerSymmetry2021}
L.~Freidel, R.~Oliveri, D.~Pranzetti and S.~Speziale, \emph{Extended corner
  symmetry, charge bracket and {{Einstein}}’s equations}, .

\bibitem{ciracRenormalizationTensorProduct2009b}
J.I.~Cirac and F.~Verstraete, \emph{Renormalization and tensor product states
  in spin chains and lattices},
  \href{https://arxiv.org/abs/0910.1130}{{\ttfamily 0910.1130}}.

\bibitem{nachtergaeleLiebRobinsonBoundsQuantum2010}
B.~Nachtergaele and R.~Sims, ``Lieb-{{Robinson Bounds}} in {{Quantum Many-Body
  Physics}}.''.

\bibitem{bravyiLiebRobinsonBoundsGeneration2006}
S.~Bravyi, M.B.~Hastings and F.~Verstraete, \emph{Lieb-{{Robinson}} bounds and
  the generation of correlations and topological quantum order},
  \href{https://arxiv.org/abs/quant-ph/0603121}{{\ttfamily quant-ph/0603121}}.

\bibitem{penroseAngularMomentumApproach1971}
R.~Penrose, \emph{Angular momentum: an approach to combinatorial space-time}, .

\bibitem{penroseApplicationsNegativeDimensional1971}
R.~Penrose, ``Applications of negative dimensional tensors..''

\bibitem{rovelliSpinNetworksQuantum1995}
C.~Rovelli and L.~Smolin, \emph{Spin networks and quantum gravity}, .

\bibitem{chircoThermallyCorrelatedStates2015}
G.~Chirco, C.~Rovelli and P.~Ruggiero, \emph{Thermally correlated states in
  {{Loop Quantum Gravity}}}, .

\bibitem{perezSpinFoamApproach2013a}
A.~Perez, \emph{The {{Spin Foam Approach}} to {{Quantum Gravity}}},
  \href{https://arxiv.org/abs/1205.2019}{{\ttfamily 1205.2019}}.

\bibitem{krajewskiGroupFieldTheories2012}
T.~Krajewski, ``Group field theories.''.

\bibitem{oritiGroupFieldTheory2014}
D.~Oriti, ``Group {{Field Theory}} and {{Loop Quantum Gravity}}.''.

\bibitem{oritiMicroscopicDynamicsQuantum2011}
D.~Oriti, ``The microscopic dynamics of quantum space as a group field
  theory.''.

\bibitem{chircozhang}
G.~Chirco, D.~Oriti and M.~Zhang, \emph{{Group field theory and tensor
  networks: towards a Ryu\textendash{}Takayanagi formula in full quantum
  gravity}}, \href{https://doi.org/10.1088/1361-6382/aabf55}{\emph{Class.
  Quant. Grav.} {\bfseries 35} (2018) 115011}
  [\href{https://arxiv.org/abs/1701.01383}{{\ttfamily 1701.01383}}].

\bibitem{chircoRyuTakayanagiFormulaSymmetric2018}
G.~Chirco, D.~Oriti and M.~Zhang, \emph{Ryu-{{Takayanagi Formula}} for
  {{Symmetric Random Tensor Networks}}}, .

\bibitem{chircozhang2}
G.~Chirco, A.~Goe\ss{}mann, D.~Oriti and M.~Zhang, \emph{{Group field theory
  and holographic tensor networks: dynamical corrections to the
  Ryu\textendash{}Takayanagi formula}},
  \href{https://doi.org/10.1088/1361-6382/ab7bb9}{\emph{Class. Quant. Grav.}
  {\bfseries 37} (2020) 095011}
  [\href{https://arxiv.org/abs/1903.07344}{{\ttfamily 1903.07344}}].

\bibitem{colafranceschiQuantumGravityStates2021}
E.~Colafranceschi and D.~Oriti, \emph{Quantum gravity states, entanglement
  graphs and second-quantized tensor networks},
  \href{https://arxiv.org/abs/2012.12622}{{\ttfamily 2012.12622}}.

\bibitem{colafranceschiHolographicMapsQuantum2021}
E.~Colafranceschi, G.~Chirco and D.~Oriti, ``Holographic maps from quantum
  gravity states as tensor networks.''.

\bibitem{chircoBulkAreaLaw2022}
G.~Chirco, E.~Colafranceschi and D.~Oriti, \emph{Bulk area law for boundary
  entanglement in spin network states: Entropy corrections and horizon-like
  regions from volume correlations},
  \href{https://arxiv.org/abs/2110.15166}{{\ttfamily 2110.15166}}.

\bibitem{yangBidirectionalHolographicCodes2016}
Z.~Yang, P.~Hayden and X.-L.~Qi, \emph{Bidirectional holographic codes and
  sub-{{AdS}} locality},  \href{https://arxiv.org/abs/1510.03784}{{\ttfamily
  1510.03784}}.

\bibitem{haydenHolographicDualityRandom2016}
P.~Hayden, S.~Nezami, X.-L.~Qi, N.~Thomas, M.~Walter and Z.~Yang,
  \emph{Holographic duality from random tensor networks},
  \href{https://arxiv.org/abs/1601.01694}{{\ttfamily 1601.01694}}.

\bibitem{pastawskiHolographicQuantumErrorcorrecting2015}
F.~Pastawski, B.~Yoshida, D.~Harlow and J.~Preskill, \emph{Holographic quantum
  error-correcting codes: {{Toy}} models for the bulk/boundary correspondence},
   \href{https://arxiv.org/abs/1503.06237}{{\ttfamily 1503.06237}}.

\bibitem{ryuHolographicDerivationEntanglement2006}
S.~Ryu and T.~Takayanagi, \emph{Holographic {{Derivation}} of {{Entanglement
  Entropy}} from the anti--de {{Sitter Space}}/{{Conformal Field Theory
  Correspondence}}}, .

\bibitem{rangamaniHolographicEntanglementEntropy2017}
M.~Rangamani and T.~Takayanagi, ``Holographic {{Entanglement Entropy}}.''.

\bibitem{chengRandomTensorNetworks2022a}
N.~Cheng, C.~Lancien, G.~Penington, M.~Walter and F.~Witteveen, \emph{Random
  tensor networks with nontrivial links},  June, 2022.
\newblock 10.48550/arXiv.2206.10482.

\bibitem{Dong:2023kyr}
X.~Dong, S.~McBride and W.W.~Weng, \emph{Holographic {Tensor} {Networks} with
  {Bulk} {Gauge} {Symmetries}}, .

\bibitem{Akers:2024ixq}
C.~Akers and A.Y.~Wei, \emph{Background independent tensor networks}, .

\bibitem{colafranceschiHolographicEntanglementSpin2022}
E.~Colafranceschi and G.~Adesso, ``Holographic entanglement in spin network
  states: A focused review.''.

\bibitem{lin_comments_2018}
J.~Lin and D.~Radicevic, \emph{Comments on {Defining} {Entanglement}
  {Entropy}},  Sept., 2018.
\newblock 10.48550/arXiv.1808.05939.

\bibitem{hollands_entanglement_2018}
S.~Hollands and K.~Sanders, \emph{Entanglement measures and their properties in
  quantum field theory},  May, 2018.

\bibitem{ma_entanglement_2016}
C.-T.~Ma, \emph{Entanglement with {Centers}},
  \href{https://doi.org/10.1007/JHEP01(2016)070}{\emph{Journal of High Energy
  Physics} {\bfseries 2016} (2016) 70}.

\bibitem{Project1Draft1}
S.~Langenscheidt and D.~Oriti, ``Channel-state duality with centers.''.

\bibitem{jiangChannelstateDuality2013}
M.~Jiang, S.~Luo and S.~Fu, \emph{Channel-state duality},
  \href{https://doi.org/10.1103/PhysRevA.87.022310}{\emph{Physical Review A}
  {\bfseries 87} (2013) 022310}.

\bibitem{majewskiCommentChannelstateDuality2013}
W.A.~Majewski and T.I.~Tylec, \emph{Comment on “{Channel}-state duality”},
  \href{https://doi.org/10.1103/PhysRevA.88.026301}{\emph{Physical Review A}
  {\bfseries 88} (2013) 026301}.

\bibitem{qiHolographicCoherentStates2017}
X.-L.~Qi, Z.~Yang and Y.-Z.~You, \emph{Holographic coherent states from random
  tensor networks}, .

\bibitem{breuerNonMarkovianDynamicsOpen2016}
H.-P.~Breuer, E.-M.~Laine, J.~Piilo and B.~Vacchini, \emph{Non-{Markovian}
  dynamics in open quantum systems},
  \href{https://doi.org/10.1103/RevModPhys.88.021002}{\emph{Reviews of Modern
  Physics} {\bfseries 88} (2016) 021002}.

\bibitem{Jia:2020etj}
H.F.~Jia and M.~Rangamani, \emph{Petz reconstruction in random tensor
  networks}, .

\bibitem{Qi:2017ohu}
X.-L.~Qi, Z.~Yang and Y.-Z.~You, \emph{Holographic coherent states from random
  tensor networks},
  \href{https://doi.org/10.1007/JHEP08(2017)060}{\emph{Journal of High Energy
  Physics} {\bfseries 08} (2017) 060}.

\bibitem{Yang:2015uoa}
Z.~Yang, P.~Hayden and X.-L.~Qi, \emph{Bidirectional holographic codes and
  sub-{AdS} locality},
  \href{https://doi.org/10.1007/JHEP01(2016)175}{\emph{Journal of High Energy
  Physics} {\bfseries 01} (2016) 175}.

\bibitem{Neiman:2017zdr}
Y.~Neiman, \emph{Towards causal patch physics in {dS}/{CFT}},
  \href{https://doi.org/10.1051/epjconf/201816801007}{\emph{EPJ Web of
  Conferences} {\bfseries 168} (2018) 01007}.

\bibitem{Raju:2016vsu}
S.~Raju, \emph{Smooth {Causal} {Patches} for {AdS} {Black} {Holes}},
  \href{https://doi.org/10.1103/PhysRevD.95.126002}{\emph{Physical Review D}
  {\bfseries 95} (2017) 126002}.

\bibitem{Hubeny:2012wa}
V.E.~Hubeny and M.~Rangamani, \emph{Causal holographic information},
  \href{https://doi.org/10.1007/JHEP06(2012)114}{\emph{Journal of High Energy
  Physics} {\bfseries 06} (2012) 114}.

\bibitem{calcagniGroupFieldCosmology2012a}
G.~Calcagni, S.~Gielen and D.~Oriti, \emph{Group field cosmology: a
  cosmological field theory of quantum geometry},
  \href{https://doi.org/2020030911294324}{\emph{Classical and Quantum Gravity}
  {\bfseries 29} (2012) 105005}.

\bibitem{Gerhardt:2018byq}
F.~Gerhardt, D.~Oriti and E.~Wilson-Ewing, \emph{Separate universe framework in
  group field theory condensate cosmology},
  \href{https://doi.org/10.1103/PhysRevD.98.066011}{\emph{Physical Review D}
  {\bfseries 98} (2018) 066011}.

\bibitem{Pithis:2019tvp}
A.G.A.~Pithis and M.~Sakellariadou, \emph{Group {Field} {Theory} {Condensate}
  {Cosmology}: {An} {Appetizer}},
  \href{https://doi.org/10.3390/universe5060147}{\emph{Universe} {\bfseries 5}
  (2019) 147}.

\bibitem{Assanioussi:2020hwf}
M.~Assanioussi and I.~Kotecha, \emph{Thermal quantum gravity condensates in
  group field theory cosmology},
  \href{https://doi.org/10.1103/PhysRevD.102.044024}{\emph{Physical Review D}
  {\bfseries 102} (2020) 044024}.

\bibitem{marchettiEffectiveDynamicsScalar2022}
L.~Marchetti and D.~Oriti, \emph{Effective dynamics of scalar cosmological
  perturbations from quantum gravity},
  \href{https://doi.org/10.1088/1475-7516/2022/07/004}{\emph{Journal of
  Cosmology and Astroparticle Physics} {\bfseries 2022} (2022) 004}.

\bibitem{Oriti:2021rvm}
D.~Oriti and X.~Pang, \emph{Phantom-like dark energy from quantum gravity},
  \href{https://doi.org/10.1088/1475-7516/2021/12/040}{\emph{Journal of
  Cosmology and Astroparticle Physics} {\bfseries 12} (2021) 040}.

\bibitem{Lunts:2015yua}
P.~Lunts, S.~Bhattacharjee, J.~Miller, E.~Schnetter, Y.B.~Kim and S.-S.~Lee,
  \emph{Ab initio holography},
  \href{https://doi.org/10.1007/JHEP08(2015)107}{\emph{Journal of High Energy
  Physics} {\bfseries 08} (2015) 107}.

\bibitem{Lee:2013dln}
S.-S.~Lee, \emph{Quantum renormalization group and holography},
  \href{https://doi.org/10.1007/JHEP01(2014)076}{\emph{Journal of High Energy
  Physics} {\bfseries 01} (2014) 076}.

\bibitem{Akhmedov:1998vf}
E.T.~Akhmedov, \emph{A remark on the {AdS}/{CFT} correspondence and the
  renormalization group flow},
  \href{https://doi.org/10.1016/S0370-2693(98)01270-2}{\emph{Physics Letters B}
  {\bfseries 442} (1998) 152}.

\bibitem{deBoer:1999tgo}
J.~de~Boer, E.~Verlinde and H.~Verlinde, \emph{On the {Holographic}
  {Renormalization} {Group}},
  \href{https://doi.org/10.1088/1126-6708/2000/08/003}{\emph{Journal of High
  Energy Physics} {\bfseries 08} (2000) 003}.

\bibitem{Heemskerk:2010hk}
I.~Heemskerk and J.~Polchinski, \emph{Holographic and {Wilsonian}
  {Renormalization} {Groups}},
  \href{https://doi.org/10.1007/JHEP06(2011)031}{\emph{Journal of High Energy
  Physics} {\bfseries 06} (2011) 031}.

\bibitem{Faulkner:2010jy}
T.~Faulkner, H.~Liu and M.~Rangamani, \emph{Integrating out geometry:
  {Holographic} {Wilsonian} {RG} and the membrane paradigm},
  \href{https://doi.org/10.1007/JHEP08(2011)051}{\emph{Journal of High Energy
  Physics} {\bfseries 08} (2011) 051}.

\end{thebibliography}\endgroup

\end{document}